\documentclass[11pt,a4paper]{article}
\usepackage{jheppub}
\usepackage{xcolor,,hyperref}
\usepackage{amsmath}
\usepackage{physics}
\usepackage{graphicx}
\usepackage{svg}
\usepackage{subcaption}
\usepackage{array}
\usepackage{tikz}
\usepackage{url}
\usepackage{multirow}
\usepackage{tabularx}
\newcolumntype{Y}{>{\centering\arraybackslash}X}
\usepackage{makecell}

\def\black{\color{black}}

\def\mC{{\mathcal{C}}}

\def\mc{{\mathbf{c}}}

\def\hefto{{\mathbf{o}}}

\def\tempC{{C}}

\def\tempO{{\textbf{o}}}

\def\cataO{{\mathbf{o}}}

\def\smeftO{{\mathcal{O}}}

\def\l{\langle}
\def\r{\rangle}

\usepackage{tikz,xcolor,hyperref}

\definecolor{lime}{HTML}{A6CE39}
\DeclareRobustCommand{\orcidicon}{\hspace{-4pt}
\begin{tikzpicture}
	\draw[lime, fill=lime] (0,0) 
	circle [radius=0.16] 
	node[white] {\hspace{0.1mm}{\fontfamily{qag}\selectfont \tiny ID}};
	\draw[white, fill=white] (-0.07,0.1) 
	circle [radius=0.01];
\end{tikzpicture}
\hspace{-3.2mm}
}

\foreach \x in {A, ..., Z}{\expandafter\xdef\csname 
orcid\x\endcsname{\noexpand\href{https://orcid.org/\csname orcidauthor\x\endcsname}
	{\noexpand\orcidicon}}
}

\def\gs{\mathrel{
	\rlap{\raise 0.511ex \hbox{$>$}}{\lower 0.511ex \hbox{$\sim$}}}}
\def\ls{\mathrel{
	\rlap{\raise 0.511ex \hbox{$<$}}{\lower 0.511ex \hbox{$\sim$}}}}

\title{SMEFT predictions for  semileptonic processes}

\author[]{Siddhartha Karmakar\orcidA{}}
\author[]{, Amol Dighe\orcidC{}}
\author[]{and Rick S. Gupta\orcidB{}}

\affiliation[]{Tata Institute of Fundamental Research, Homi Bhabha Road, Colaba, Mumbai 400005, India}

\emailAdd{siddhartha@theory.tifr.res.in}
\emailAdd{amol@theory.tifr.res.in}
\emailAdd{rsgupta@theory.tifr.res.in}

\preprint{TIFR/TH/24-3}
\abstract{The  $SU(2)_L\times U(1)_Y$ invariance of the Standard Model Effective Field Theory (SMEFT) predicts multiple restrictions in the space of Wilson coefficients of  $U(1)_{em}$ invariant effective lagrangians such as the  Low-energy Effective Field Theory (LEFT), used for low-energy flavor-physics observables, or the Higgs Effective Field Theory (HEFT) in unitary gauge,  appropriate for weak-scale observables. In this work, we derive and list all such predictions for semileptonic operators up to dimension 6. We find that these predictions can be expressed as 2223 linear relations among the HEFT/LEFT Wilson coefficients, that are completely independent of any assumptions about the alignment of the mass and flavor bases. These relations connect diverse experimental searches such as rare meson decays, high-$p_T$ dilepton searches, top decays, $Z$-pole observables, charged lepton flavor violating observables and non-standard neutrino interaction searches. We demonstrate how these relations can be used to derive strong indirect constraints on multiple Wilson coefficients that are currently either weakly constrained from direct experiments or have no direct bound at all. These relations also  imply, in general,  that evidence for new physics  in a particular search channel must be accompanied by correlated anomalies in other channels.}


\begin{document}

\maketitle
\flushbottom

\section{Introduction}\label{sec: intro}
The Standard Model Effective Field Theory (SMEFT)~\cite{Buchmuller:1985jz, Grzadkowski:2010es, Jenkins:2013zja, Isidori:2023pyp} is a model-independent way to incorporate the effects of beyond Standard Model (BSM) physics at low energies.  It modifies the Standard Model SM lagrangian by the addition of all possible higher dimensional operators respecting  the SM symmetries:
 \begin{align}
	\mathcal{L}&= \mathcal{L}_{SM}+\frac{1}{\Lambda^2}\sum_{i}{\mC}_i^{(6)}{\cal O}_i^{(6)}+\cdots , \label{SMEFTexpansion}
\end{align}
where $\Lambda$ is the cut-off scale, typically of the order of TeV or higher. Here, ${\mathcal{O}}_i^{(d)}$ represent the $d$-dimensional BSM operators and $\mC_i^{(d)}$ represent the corresponding Wilson coefficients (WCs). We assume here that the new physics preserves baryon and lepton numbers and therefore do not include dimension-5 operators. The ellipsis represents higher order operators with dimension $> 6$.

SMEFT is manifestly invariant under $SU(3)_C\times SU(2)_L\times U(1)_Y$, the SM gauge symmetry. As a consequence, there are specific relationships among different flavor observables. For instance, the SMEFT requirement that the up-type and down-type left-handed fermionic fields should arise from $SU(2)_L$ doublets implies relations among flavor observables probing the up sector and those probing the down sector.  In this work, we initiate a systematic derivation of such relations, beginning with the semileptonic processes in this article.

In flavor physics, effective field theories (EFT) have long served as a standard framework to parameterize the effects of heavy new physics. However,  for most flavor physics processes, the experimental energy scale is at or below the mass of the $b$ quark; this includes weak decays of mesons, neutral meson mixing, $\tau$ decays, etc. The relevant EFT at these energies is the so called Low-energy Effective Field Theory (LEFT)\footnote{LEFT is sometimes referred to as weak effective field theory (WET or WEFT) in literature~\cite{Jenkins:2017jig, Aebischer:2017gaw, Aebischer:2017ugx, London:2021lfn}.}~\cite{Buchalla:1995vs}, which assumes only the $SU(3)_C\times U(1)_{em}$ invariance and not the full $SU(3)_C\times SU(2)_L\times U(1)_Y$ invariance of SM. 
 
The flavor structure of new physics (NP) can also be probed at higher scales, for instance, in flavor-violating decays of the $Z, W^\pm$, and the Higgs boson $h$,  via flavor-violating production or decay of the top quark $t$, or by constraining the Drell-Yan processes initiating from a flavor off-diagonal diquark state.  
In order to include both high-energy and low-energy observables, one of course needs to write all possible $SU(3)_C\times U(1)_{em}$  invariant operators, as in LEFT, but terms involving the top quark, Higgs boson and electroweak bosons also need to be included. An appropriate framework that can encompass both, low-energy flavor observables as well as this second class of processes involving heavier SM states,  is the so-called Higgs Effective Field Theory  (HEFT)~\cite{ Alonso:2012px, Buchalla:2013rka, Pich:2016lew}. This is a more general framework than SMEFT and also includes scenarios where  the EW symmetry is realized non-linearly. In the unitary gauge, it leads to a lagrangian involving all possible $SU(3)_C\times U(1)_{em}$ invariant operators. Given the HEFT lagrangian, it is possible to derive the corresponding  LEFT  lagrangian by simply integrating out the heavier SM states $W, \,Z, \,h$ and $t$.

\begin{figure}[t]
	\centering
	\includegraphics[width=0.9\textwidth]{"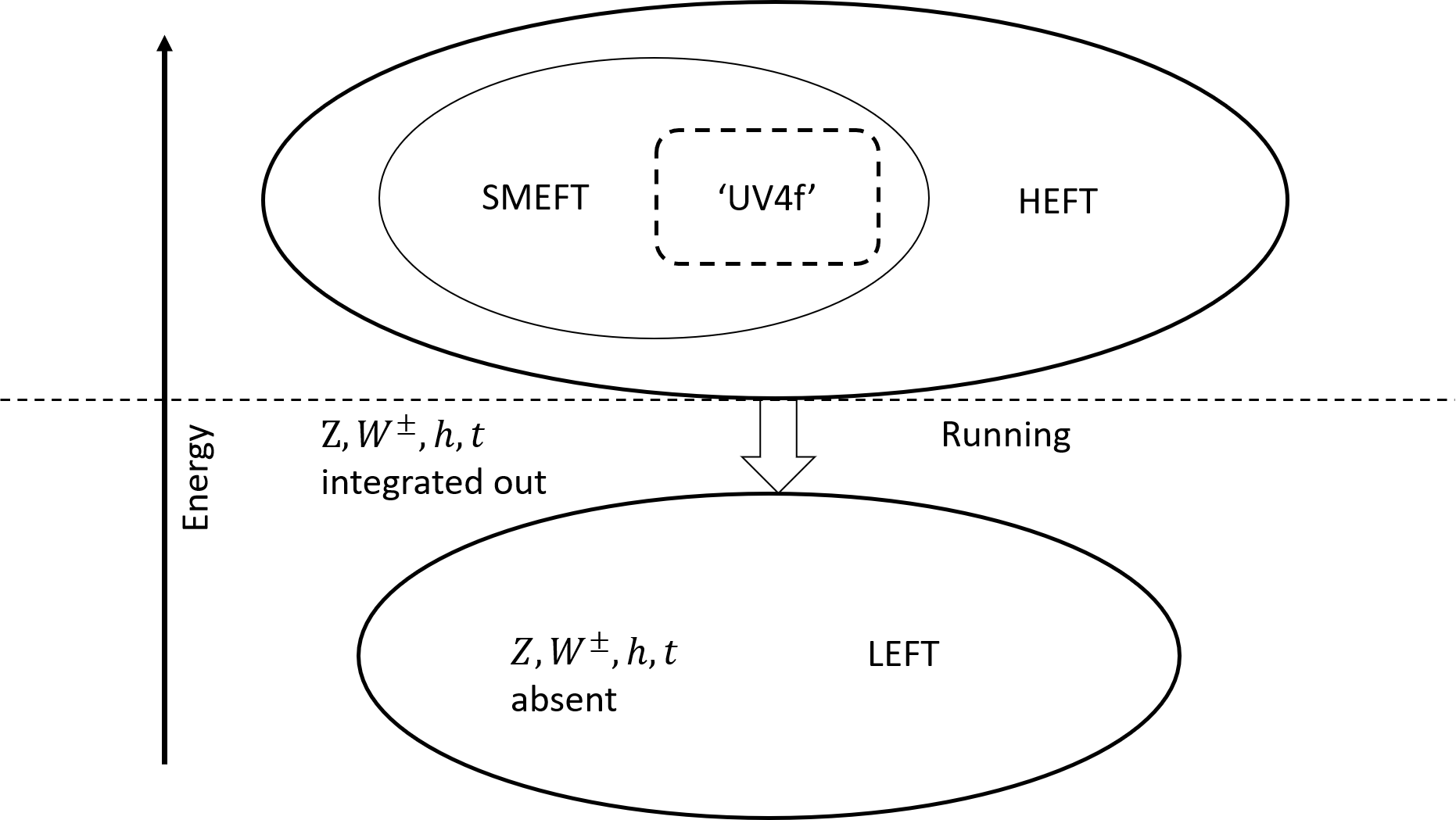"}
	\caption{\label{vendiag} Schematic representation of EFTs above and below the electroweak scale. UV4f represents the subset of SMEFT where the BSM physics only has four-fermion operators.}
\end{figure}

For a given set of processes, a general parametrization of possible BSM deviations assuming only $SU(3)_C\times U(1)_{em}$ invariance gives rise to many more free parameters up to a given order than the number of SMEFT WCs to that order. This is simply because the former does not assume the full $SU(3)_C\times SU(2)_L\times U(1)_Y$ invariance of SM. This situation has been schematically presented in Fig.~\ref{vendiag} where SMEFT can be seen to be a subset of the more general HEFT. Within this region satisfying SMEFT assumptions,  the smaller number of free parameters  implies relationships among the  WCs of HEFT. These relationships can be thought of as predictions of SMEFT at a certain order; these predictions can be broken only by violating the basic underlying assumptions of SMEFT.

An apparent obstacle in deriving these relations is that, while SMEFT is written in the flavor basis, HEFT or LEFT operators have to be written in the mass basis if we wish to connect them to physical observables. The equations connecting HEFT Wilson coefficients in the mass basis to SMEFT Wilson coefficients in the flavor basis, thus, contain elements of the rotation matrices of the left-handed and right-handed up-type and down-type fermions, which cannot be fixed by experiments.   We show, however, that only the measurable elements of the Cabbibo-Kobayashi-Maskawa (CKM) quark-mixing matrix and the  Pontecorvo-Maki-Nakagawa-Sakata (PMNS) lepton-mixing matrix appear in the final relations among HEFT WCs. This allows us to derive the implications of SMEFT on flavor physics observables in a way that is completely independent of assumptions about the alignment of the flavor and the mass bases, often referred to as UV flavor assumptions. 

In this work, we consider the 3240 semileptonic four-fermion operators in HEFT that get contributions from the 1053 SMEFT operators, giving rise to 2187 constraints. In addition, we consider 144 HEFT operators  that can contribute to low-energy flavor observables via the exchange of $Z, W^\pm$ and $h$ bosons.  In SMEFT, these arise from 108  independent operators, thus implying 36 constraints in the HEFT space.  We derive all these 2223 constraints and express them as analytic relations independent of any UV flavor assumptions.

Some other recent studies have also considered the implications of the $SU(2)_L\times U(1)_Y$ invariance of SMEFT on flavor observables~\cite{Alonso:2014csa, Cata:2015lta, Fuentes-Martin:2020lea, Bause:2020auq, Bause:2020xzj, Bissmann:2020mfi, Bause:2021cna, Bause:2021ihn, Bause:2022rrs, Sun:2023cuf, Grunwald:2023nli, Greljo:2023bab, Fajfer:2012vx, Bause:2023mfe, Chen:2024jlj}. 
To the best of our knowledge,  however,  the present work is the first study to comprehensively derive and list all the 2223 analytic relations relevant for semileptonic processes (see, however, Ref.~\cite{Bause:2020auq, Bause:2021cna, Bause:2021ihn} where a subset of the above  relations has been presented.) Our approach also makes it clear that these implications can be obtained and presented in a way that is free from all UV flavor assumptions.  A similar approach has been used to derive SMEFT predictions in Higgs physics in Ref.~\cite{gupta, LHCHiggs}.

The SMEFT predictions derived in this work are expressed as linear relationships among $SU(3)_C\times U(1)_{em}$ invariant BSM couplings in the mass basis. These relationships can be directly translated to exact relations among experimental observables. As we will see, these relations connect diverse experimental searches: low-energy flavor observables in different sectors such as kaon, B-meson, charm and $\tau$-decays, the Drell-Yan process at high-$p_T$, top production and decay channels, $Z$-decays, and searches for non-standard neutrino interaction. These relationships thus allow us to utilize experimental limits on a set of well-constrained observables to put bounds on other, otherwise poorly constrained, observables.
Our work demonstrates that indirect constraints on many WCs --- such as those related to $d\bar d\to\nu\bar \nu$, $u_i\to u_j\nu \bar \nu$ and top decays --- obtained in the above manner, would surpass direct bounds. 

Another crucial implication of these relations among WCs is that, in general they disallow an isolated non-vanishing WC.  This is because a nonzero WC will, via the linear relations, imply a nonzero value for multiple other WCs. This indicates that deviations from SM would typically not appear in isolated channels. For instance, it is known that the observed excess in $B\to K \nu \nu$ branching fraction can be explained by a nonzero WC for the operator involving the transition $b\to s \nu \nu$. We show that this would imply non-vanishing values for WCs involving processes such as $b\to c \ell\nu_\ell$, $b\to u \ell\nu_\ell$ , $t \to c \mu e$, $t \to u \mu e$, etc.

While the SMEFT predictions we derive are completely independent of UV flavor assumptions, we find that as far as phenomenological implications are concerned, the sharpest conclusions can be drawn in an important class of models where the dominant new physics effects come from four-fermion operators and not from modifications of $Z, W^\pm$ and $h$ couplings. We call these models `UV4f' models and represent them by the dashed rectangle in Fig.~\ref{vendiag}.  This is a highly motivated class of UV completions that encompasses a majority of the models proposed to explain the flavor anomalies. These include minimal leptoquark models~\cite{Hiller:2014yaa, Gripaios:2014tna, deMedeirosVarzielas:2015yxm, Sahoo:2015wya} and many  $Z'$ models~\cite{Altmannshofer:2014cfa, Bonilla:2017lsq, Bian:2017rpg, Alonso:2017uky, Cline:2017ihf}  proposed in the literature.

The plan of this paper is as follows. In Sec.~\ref{sec: SMEFTtoHEFT}, we present the list of relevant operators in SMEFT, HEFT and LEFT and provide the relations among the WCs.  
We discuss the phenomenological applications of these relations in Sec.~\ref{sec: indirect_bounds}, where we derive the indirect bounds on WCs associated with left-handed quarks and leptons. In Sec.~\ref{sec: phenoevidence}, we discuss possible directions of NP searches suggested by the relations among the WCs, given some of the current observed deviations from SM.  
We present concluding remarks in Sec.~\ref{sec: conclusion}.
In Appendix~\ref{app:HEFTbasis}, we write the HEFT operators used in the text in the $SU(2)_L \times U(1)_Y$ invariant form, with the electroweak symmetry non-linearly realized, and compare our list with the previous literature. In Appendix~\ref{app: SMEFTbasis}, we briefly discuss some details of the SMEFT basis used and the rationale for our choice. In Appendix~\ref{app:left-heft}, we present all the analytic relations in terms of semileptonic LEFT WCs and WCs that modify the $Z$, $W^{\pm}$ and Higgs couplings to fermions. In Appendix~\ref{app: numbers}, we present tables of $90\%$ C.L limits on the LEFT WCs.

\section{SMEFT predictions {\black for semileptonic operators}}\label{sec: SMEFTtoHEFT}

In this section, we present all possible semileptonic operators respecting the $U(1)_{em}$ symmetry (as the $SU(3)_C$ symmetry is always respected, we will not mention it separately from here on), and derive the analytic relations among them that are predicted by SMEFT. We consider the following lagrangian terms at the weak scale:
\begin{align}
    \mathcal{L}_{\rm HEFT} &\supset \mathcal{L}^{\rm SM} +\sum_{f,i,j} \left[{\mc}_{fZ}\right]^{ij} \,(\bar f_i\gamma^\mu \,f_j)\,Z_\mu +  \sum_{f_u, f_d,i,j}\left[{\mc}_{f_u f_{d}  W}\right]^{ij} \,(\bar f_{u_i} \gamma^\mu\,f_{d_j})\,W^+_\mu \nonumber\\
    & ~+ \sum \left[{\mc}_{fh}\right]^{ij}\,(\bar f_{i} \,P_{R}\, f_{j})\,h+ \frac{1}{\Lambda^2}\,\sum_{i}{\mc}_{i}\,{\hefto}^{4f}_{sl,i}+h.c.,\label{heftlag}
\end{align}
where, in addition to the SM lagrangian $\mathcal{L}^ {SM}$ and  the term  containing all possible semileptonic four-fermion operators ${\hefto}^{4f}_{sl} $, we also include corrections to the couplings of $Z$, $W^{\pm}$ and Higgs boson $h$ to fermions.\footnote{ We have not considered four-quark operators and electroweak dipole operators. Although these can contribute to semileptonic processes, they do not get matched to semileptonic operators at the tree level. Furthermore,  these operators are constrained from processes which are not semileptonic. The four-quark operators can get constraints from nonleptonic decays, whereas the dipole operators are bounded by measurements such as the precise observations of dipole moments of elementary particles,  the $b \to s \gamma$ process etc.} This is because the diagrams with Higgs, $W^\pm$ and $Z$ exchange can generate four-fermion effective operators at the low energies relevant to semileptonic flavor observables. Here $f \in \{u_L, u_R, d_L, d_R,  e_L, e_R, \nu_L\}$, $f_{u}$ denotes neutrinos and up-type quarks (both left-handed and right-handed) whereas $f_{d}$ denotes down-type quarks and charged leptons. A lagrangian containing all these operators with independent coefficients is equivalent to the HEFT lagrangian, $\mathcal{L}_{\rm HEFT}$, in the unitary gauge. This is because, although formally  $SU(2)_L\times U(1)_Y$ invariance is non-linearly realized in the HEFT  lagrangian, in the unitary gauge HEFT reduces to a lagrangian with all possible $U(1)_{em}$-invariant terms. As we show in Appendix~\ref{app:HEFTbasis}, our list of operators can be rewritten in an invariant form with non-linearly realized electroweak symmetry as in Ref.~\cite{Buchalla:2012qq}.  The HEFT basis of  Appendix~\ref{app:HEFTbasis}  excludes some redundant operators that appeared in the HEFT bases presented in earlier literature~(e.g. \cite{Buchalla:2012qq, Cata:2015lta}) and also includes some operators that were missed in previous work. Further note that in the UV4f scenario discussed in the Sec.~\ref{sec: intro}, the coupling modifications of $Z,W^{\pm}$ and $h$ are absent, i.e. the second, third and fourth terms on the RHS of eq.\,(\ref{heftlag}) vanish.

The semileptonic four-fermion operators ${\hefto}^{4f}_{sl}$  can be directly probed by high-energy processes such as the Drell-Yan process $\bar{q}_i q_j \to ll$,  top production and decay processes, etc. We consider these operators in Sec.~\ref{four-fermionic}  and list the dimension-6 (dim-6) SMEFT operators that contribute to them. We find that the number of HEFT operators is larger than the number of dim-6 SMEFT operators, which results in SMEFT predictions for these HEFT WCs. These predictions are in the form of linear relations among the HEFT WCs; we explicitly derive these relations in Sec.~\ref{four-fermionic}.

Next we consider the corrections to the SM couplings of $Z,W^{\pm}$ and $h$ to fermions, indicated by the second, third and fourth terms in the RHS of eq.\,(\ref{heftlag}).  Although our reason for inclusion of these operators is that they contribute to low-energy semileptonic processes via $Z,W^\pm$ and $h$ exchange,  these couplings can be probed independently by studying decays of the  $Z, W^{\pm}$ and $h$. We list the SMEFT operators contributing to these in Sec.~\ref{sec:ZWH}. We find that, while the number of SMEFT operators is the same as the number of HEFT operators for  $h$ coupling corrections, the number of contributing SMEFT operators in the case of gauge boson coupling corrections is smaller. This results in relations among the corrections to $Z$ and $W^\pm$ couplings; we derive these in Sec.~\ref{sec:ZWH}.

Finally, in Sec.~\ref{sec: leftcor}  we rewrite the analytic relations derived in  Sec.~\ref{four-fermionic} and Sec.~\ref{sec:ZWH} in terms of WCs at the low scale relevant for most of the important flavor observables, such as those connected to meson mixing, rare decays, etc.  The lagrangian relevant at these scales is the sum of the LEFT neutral-current and charged-current lagrangians\footnote{Note that, to distinguish different EFTs, we denote the Wilson coefficients by `${\mC}$' for SMEFT, by `${\mc}$' for HEFT and by `$C$' for LEFT. The corresponding operators are denoted by `$\smeftO$', `$\hefto$' and '$O$' respectively.}
\begin{align}
	\mathcal{L}_{\textrm{LEFT}}^{\rm NC} &= \mathcal{L}^{\rm NC}_{\textrm{SM}} +\frac{4 G_F}{\sqrt{2}}\sum_{i}^{\textrm{NC only}}\,{\tempC}_i \, O_i^{\rm NC}~,\label{LEFT1}\\
	\textrm{and} \quad \mathcal{L}_{\textrm{LEFT}}^{\rm CC} &=  \mathcal{L}^{\rm CC}_{\textrm{SM}} + \frac{4 G_F}{\sqrt{2}}\sum_{i}^{\textrm{CC only}}\,{\tempC}_i \, O_i^{\rm CC}~, \label{LEFT2}
\end{align}
where the first terms on the RHS arise from the first term in eq.\,(\ref{heftlag}) by integrating out $Z, W^\pm$ and $h$, assuming SM couplings.\footnote{ A loop factor of $e^2/(16\pi^2)$ is usually included for the NC Lagrangian for LEFT in literature~\cite{Aebischer:2015fzz,Bause:2020auq}. In our convention, we have not included this factor in order to have uniformity in our analytic relations to be presented later.}  Here `${\rm NC}$' and `${\rm CC}$' stand for neutral-current and charged-current, respectively.  In order to obtain the SMEFT predictions for relations among the LEFT WCs, we need to match the LEFT coefficients above to the HEFT ones including the effect of  $Z$, $W^\pm$, and $h$ exchange diagrams. These matching relations can then be inverted to write the HEFT WCs and the relations among them in terms of the LEFT ones. We carry out this procedure in Sec.~\ref{sec: leftcor}.

\begin{table}
	\begin{center}
		\small
		\begin{minipage}[t]{0.49\textwidth}
			\renewcommand{\arraystretch}{1.5}
			\begin{tabularx}{\textwidth}[t]{|c|Y c|}
				\hline
				\multicolumn{3}{|c|}{Vector operators $ LLLL $} \\
				\hline
				& NC & Count\\
				$[{\mc}_{e_L d_L} ^{V}]^{\alpha \beta i j}$        & $(\bar e_L^\alpha \gamma_\mu e_L^\beta)(\bar d_L^i \gamma^\mu d_L^j)$ & 81 (45)\\
				$[{\mc}_{euLL} ^{V}]^{\alpha \beta i j}$        & $(\bar e_L^\alpha \gamma_\mu e_L^\beta)(\bar u_L^i \gamma^\mu u_L^j)$ &81 (45)\\
				$[{\mc}_{\nu dLL} ^{V}]^{\alpha \beta i j}$        & $(\bar \nu_L^\alpha \gamma_\mu \nu_L^\beta)(\bar d_L^i \gamma^\mu d_L^j)$ &81 (45)\\
				$[{\mc}_{\nu uLL} ^{V}]^{\alpha \beta i j}$        & $(\bar \nu_L^\alpha \gamma_\mu \nu_L^\beta)(\bar u_L^i \gamma^\mu u_L^j)$ &81 (45)\\
				& CC &\\
				$[{\mc}_{LL} ^{V}]^{\alpha \beta i j}$        & $(\bar e_L^\alpha \gamma_\mu \nu_L^\beta)(\bar u_L^i \gamma^\mu d_L^j)$&162 (81)\\[3ex]
                \hline
				\multicolumn{3}{|c|}{Vector operators $ RRRR $} \\
				\hline
				& NC & Count\\
				$[{\mc}_{edRR} ^{V}]^{\alpha \beta i j}$        & $(\bar e_R^\alpha \gamma_\mu e_R^\beta)(\bar d_R^i \gamma^\mu d_R^j)$ & 81 (45)\\
				$[{\mc}_{euRR} ^{V}]^{\alpha \beta i j}$        & $(\bar e_R^\alpha \gamma_\mu e_R^\beta)(\bar u_R^i \gamma^\mu u_R^j)$ & 81 (45)\\
			\hline
			\end{tabularx}
		\end{minipage}
		\begin{minipage}[t]{0.49\textwidth}
			\renewcommand{\arraystretch}{1.5}
			\begin{tabularx}{\textwidth}[t]{|c|Y c|}
				\hline
				\multicolumn{3}{|c|}{Vector operators $ LLRR $} \\
				\hline
				& NC & Count\\
				$[{\mc}_{edLR} ^{V}]^{\alpha \beta i j}$        & $(\bar e_L^\alpha \gamma_\mu e_L^\beta)(\bar d_R^i \gamma^\mu d_R^j)$ & 81 (45)\\
				$[{\mc}_{euLR} ^{V}]^{\alpha \beta i j}$        & $(\bar e_L^\alpha \gamma_\mu e_L^\beta)(\bar u_R^i \gamma^\mu u_R^j)$ & 81 (45)\\
				$[{\mc}_{\nu dLR} ^{V}]^{\alpha \beta i j}$        & $(\bar \nu_L^\alpha \gamma_\mu \nu_L^\beta)(\bar d_R^i \gamma^\mu d_R^j)$ & 81 (45)\\
				$[{\mc}_{\nu uLR} ^{V}]^{\alpha \beta i j}$        & $(\bar \nu_L^\alpha \gamma_\mu \nu_L^\beta)(\bar u_R^i \gamma^\mu u_R^j)$ & 81 (45)\\
				& CC & \\
				$[{\mc}_{LR} ^{V}]^{\alpha \beta i j}$        & $(\bar e_L^\alpha \gamma_\mu \nu_L^\beta)(\bar u_R^i \gamma^\mu d_R^j)$ & 162 (81)\\[3ex]
                \hline
				\multicolumn{3}{|c|}{Vector operators $ RRLL $} \\
				\hline
				& NC & Count\\
				$[{\mc}_{edRL} ^{V}]^{\alpha \beta i j}$        & $(\bar e_R^\alpha \gamma_\mu e_R^\beta)(\bar d_L^i \gamma^\mu d_L^j)$ & 81 (45)\\
				$[{\mc}_{euRL} ^{V}]^{\alpha \beta i j}$        & $(\bar e_R^\alpha \gamma_\mu e_R^\beta)(\bar u_L^i \gamma^\mu u_L^j)$ & 81 (45)\\
			\hline
			\end{tabularx}
		\end{minipage}
		\begin{minipage}[t]{0.46\textwidth}
			\medskip
			\renewcommand{\arraystretch}{1.5}
			\begin{tabularx}{\textwidth}[t]{|c|Y c|}
				\hline
				\multicolumn{3}{|c|}{Scalar operators with $d_R$} \\
				\hline
				& NC & Count\\
				$[{\mc}_{ed, RLLR} ^{S}]^{\alpha \beta i j}$        & $(\bar e_R^\alpha \, e_L^\beta)(\bar d_L^i \, d_R^j)$ & 162 (81)\\
				$[{\mc}_{ed, RLRL} ^{S}]^{\alpha \beta i j}$        & $(\bar e_R^\alpha \, e_L^\beta)(\bar d_R^i \, d_L^j)$ & 162 (81)\\
				& CC &\\
				$[{\mc}_{RLLR} ^{S}]^{\alpha \beta i j}$        & $(\bar e_R^\alpha \, \nu_L^\beta)(\bar u_L^i \, d_R^j)$& 162 (81)\\[3ex]
                \hline
				\multicolumn{3}{|c|}{Scalar operators with $u_R$} \\
				\hline
				& NC & Count\\
				$[{\mc}_{eu, RLLR} ^{S}]^{\alpha \beta i j}$        & $(\bar e_R^\alpha \, e_L^\beta)(\bar u_L^i \, u_R^j)$ & 162 (81)\\
				$[{\mc}_{eu, RLRL} ^{S}]^{\alpha \beta i j}$        & $(\bar e_R^\alpha \, e_L^\beta)(\bar u_R^i \, u_L^j)$ & 162 (81)\\
				& CC &\\
				$[{\mc}_{RLRL} ^{S}]^{\alpha \beta i j}$        & $(\bar e_R^\alpha \, \nu_L^\beta)(\bar u_R^i \, d_L^j)$& 162 (81)\\
                \hline
			\end{tabularx}
		\end{minipage}
		\begin{minipage}[t]{0.52\textwidth}
			\medskip
			\renewcommand{\arraystretch}{1.5}
			\begin{tabularx}{\textwidth}[t]{|c|Y c|}
				\hline
				\multicolumn{3}{|c|}{Tensor operators with $d_R$} \\
				\hline
				& NC & Count\\
				$[{\mc}_{ed, RLLR} ^{T}]^{\alpha \beta i j}$        & $(\bar e_R^\alpha \sigma_{\mu \nu} e_L^\beta)(\bar d_L^i \sigma^{\mu \nu} d_R^j)$ & 162 (81)\\
				$[{\mc}_{ed, RLRL} ^{T}]^{\alpha \beta i j}$        & $(\bar e_R^\alpha \sigma_{\mu \nu} e_L^\beta)(\bar d_R^i \sigma^{\mu \nu} d_L^j)$ & 162 (81)\\
				& CC &\\
				$[{\mc}_{RLLR} ^{T}]^{\alpha \beta i j}$        & $(\bar e_R^\alpha \sigma_{\mu \nu} \nu_L^\beta)(\bar u_L^i \sigma^{\mu \nu} d_R^j)$& 162 (81)\\[3ex]
                \hline
				\multicolumn{3}{|c|}{Tensor operators with $u_R$} \\
				\hline
				& NC & Count\\
				$[{\mc}_{eu, RLLR} ^{T}]^{\alpha \beta i j}$        & $(\bar e_R^\alpha \sigma_{\mu \nu} e_L^\beta)(\bar u_L^i \sigma^{\mu \nu} u_R^j)$ & 162(81)\\
				$[{\mc}_{eu, RLRL} ^{T}]^{\alpha \beta i j}$        & $(\bar e_R^\alpha \sigma_{\mu \nu} e_L^\beta)(\bar u_R^i \sigma^{\mu \nu} u_L^j)$ & 162(81) \\
				& CC &\\
				$[{\mc}_{RLRL} ^{T}]^{\alpha \beta i j}$        & $(\bar e_R^\alpha \sigma_{\mu \nu} \nu_L^\beta)(\bar u_R^i \sigma^{\mu \nu} d_L^j)$& 162(81)\\
                \hline
			\end{tabularx}
		\end{minipage}
		\caption{\label{HEFToplist} Semileptonic operators in HEFT. Here ${\mc}$'s are the WCs for the corresponding operators in the flavor basis. The indices $\alpha, \beta$ denote lepton families and $i, j$ denote quark families. NC and CC correspond to neutral-current and charged-current operators. Count denotes the number of independent operators; the number inside the brackets is the number of independent operators if all WCs were real. Note that for vector CC operators as well as all the scalar and tensor operators we have not explicitly listed their hermitian conjugates but included them in the count. }
	\end{center}
\end{table}

\begin{table}[t]
	\begin{center}
		\small
		\begin{minipage}[t]{0.48\textwidth}
			\centering
			\renewcommand{\arraystretch}{1.5}
			\begin{tabularx}{\textwidth}[t]{|c|Y c|}
                    \hline
                    \multicolumn{3}{|c|}{Vector operators $ LLLL $} \\
                    \hline
                    &Operator & Count\\
                    $[{\mC}_{{\ell} q} ^{(1)}]^{\alpha \beta i j}$        & $(\bar l^\alpha \gamma_\mu l^\beta)(\bar q^i \gamma^\mu q^j)$ & 81 (45)\\
                    $[{\mC}_{{\ell} q} ^{(3)}]^{\alpha \beta i j}$        & $(\bar l^\alpha \gamma_\mu \tau^I l^\beta)(\bar q^i \gamma^\mu \tau^I q^j)$ & 81 (45)\\[3ex]
                    \hline
                    \hline
                    \multicolumn{3}{|c|}{Vector operators $ RRRR $} \\
                    \hline
                    &Operator & Count\\
                    $[{\mC}_{ed}]^{\alpha \beta i j}$        & $(\bar e^\alpha \gamma_\mu e^\beta)(\bar d_R^i \gamma^\mu d_R^j)$ & 81 (45)\\
                    $[{\mC}_{eu}]^{\alpha \beta i j}$        & $(\bar e^\alpha \gamma_\mu e^\beta)(\bar u_R^i \gamma^\mu u_R^j)$ & 81 (45)\\
                    \hline
			\end{tabularx}
		\end{minipage}
		\begin{minipage}[t]{0.48\textwidth}
			\centering
			\renewcommand{\arraystretch}{1.5}
			\begin{tabularx}{\textwidth}[t]{|c|Y c|}
                    \hline
                    \multicolumn{3}{|c|}{Vector operators $ LLRR $} \\
                    \hline
                    &Operator & Count\\
                    $[{\mC}_{{\ell} d} ]^{\alpha \beta i j}$        & $(\bar l^\alpha \gamma_\mu l^\beta)(\bar d_R^i \gamma^\mu d_R^j)$ & 81 (45)\\
                    $[{\mC}_{{\ell} u} ]^{\alpha \beta i j}$        & $(\bar l^\alpha \gamma_\mu l^\beta)(\bar u_R^i \gamma^\mu u_R^j)$ & 81 (45)\\[3ex]
                    \hline
                    \hline
                    \multicolumn{3}{|c|}{Vector operators $ RRLL $} \\
                    \hline
                    &Operator & Count\\
                    $[{\mC}_{eq}]^{\alpha \beta i j}$        & $(\bar e_R^\alpha \gamma_\mu e_R^\beta)(\bar q^i \gamma^\mu q^j)$ & 81 (45)\\
                    &&\\
                    \hline
			\end{tabularx}
		\end{minipage}
		\begin{minipage}[t]{0.48\textwidth}
			\centering
			\medskip
			\renewcommand{\arraystretch}{1.5}
			\begin{tabularx}{\textwidth}[t]{|c|Y c|}
                    \hline
                    \multicolumn{3}{|c|}{Scalar operators with $d_R$} \\
                    \hline
                    &Operator & Count\\
                    $[{\mC}_{{\ell} e dq}]^{\alpha \beta i j}$        & $(\bar l^\alpha_a \, e_R^\beta)(\bar d_R^i \, q^j_a)$ & 162 (81)\\
                    \hline
			\end{tabularx}
		\end{minipage}
		\begin{minipage}[t]{0.48\textwidth}
			\centering
			\medskip
			\renewcommand{\arraystretch}{1.5}
			\begin{tabularx}{\textwidth}[t]{|c|Y c|}
                    \hline
                    \multicolumn{3}{|c|}{Scalar operators with $u_R$} \\
                    \hline
                    &Operator & Count\\
                    $[{\mC}_{{\ell} e qu} ^{(1)}]^{\alpha \beta i j}$        & $(\bar l^\alpha_a \, e_R^\beta) \epsilon_{ab}(\bar q^i_b \, u_R^j)$ & 162 (81)\\
                    \hline
			\end{tabularx}
		\end{minipage}
		\vspace{1cm}
		\begin{minipage}[c]{0.97\textwidth}
			\medskip
			\centering
			\renewcommand{\arraystretch}{1.5}
			\begin{tabularx}{\textwidth}[t]{|c|Y c|}
				\hline
				\multicolumn{3}{|c|}{Tensor operators} \\
				\hline
				& Operator & Count\\
				$[{\mC}_{{\ell} e qu} ^{(3)}]^{\alpha \beta i j}$        & $(\bar l^\alpha_a \sigma_{\mu \nu} e_R^\beta) \epsilon_{ab} (\bar q^i_b \sigma^{\mu \nu} u_R^j)$ & 162 (81)\\
                \hline
			\end{tabularx}
		\end{minipage}
		\caption{\label{SMEFToplist} Semileptonic operators in SMEFT. Here ${\mC}$'s are the WCs for the corresponding operators in the flavor basis. The indices $\alpha, \beta$ denote lepton families and $i, j$ denote quark families. Here $l = (\nu_L, e_L)^T$, $q = (u_L, d_L)^T$, $\tau^I$ are the Pauli matrices and $\epsilon_{ab}$ is the $(2\times 2)$ anti-symmetric matrix with $\epsilon_{12} = 1$. Count denotes the number of independent operators; the number inside the brackets is the number of independent operators if all WCs were real. Note that for   the scalar and tensor operators we have not explicitly listed their hermitian conjugates but included them in the count.}
	\end{center}
\end{table}

\subsection{Predictions for semileptonic operators at high energies}\label{four-fermionic}

We begin our analysis with the 3240 (1674) semileptonic four-fermion  operators\footnote{For non-hermitian operators, we consider the operator and its hermitian conjugate as two distinct operators,  as one can treat $(O+O^\dagger)$ and $i(O-O^\dagger)$ as two separate operators with real Wilson coefficients.} present in HEFT (see Table~\ref{HEFToplist}), where the number within the parenthesis denotes the number of independent operators if the WCs of all these operators were real. Note that each entry in Table~\ref{HEFToplist} represents multiple operators corresponding to different possible values for the family indices. The first entry $[{\mc}_{edLL}^{V}]^{\alpha \beta i j}$, for instance, represents  81 (45) operators, since the indices $\alpha$, $\beta$ denote three lepton families and the indices $i$, $j$ denote three quark families. In Table~\ref{SMEFToplist}, we list the 1053 (558) semileptonic four-fermion operators in SMEFT which would give rise to the above HEFT operators.

The operators in Table~\ref{HEFToplist} and \ref{SMEFToplist} are divided into categories based on their Lorentz structure and the chiralities of the fields involved. In the following, we discuss the mapping between SMEFT and HEFT operators and the resulting SMEFT predictions for each of these categories.  

\paragraph{LLLL vector operators:} 

In this category, there are 486 (261) independent operators in HEFT, as shown in Table~\ref{HEFToplist},  which correspond to the 162 (90) SMEFT operators shown in Table~\ref{SMEFToplist}. The SMEFT operators, when expanded in the unitary gauge, give the following contributions to the HEFT Wilson coefficients:
\begin{align}
	[{\mc}_{\nu uLL}^{V}]^{\alpha \beta i j}&= ([{\mC}_{{\ell} q}^{(1)}]^{\alpha \beta i j}+[{\mC}_{{\ell} q}^{(3)}]^{\alpha \beta i j})~,\quad
	[{\mc}_{euLL}^{V}]^{\alpha \beta i j}=([{\mC}_{{\ell} q}^{(1)}]^{\alpha \beta i j}-[{\mC}_{{\ell} q}^{(3)}]^{\alpha \beta i j}),\label{codef1}\\
	[{\mc}_{\nu d LL}^{V}]^{\alpha \beta i j}&= ([{\mC}_{{\ell} q}^{(1)}]^{\alpha \beta i j}-[{\mC}_{{\ell} q}^{(3)}]^{\alpha\beta i j})~,\quad 
	[{\mc}_{ed LL}^{V}]^{\alpha \beta i j}=([{\mC}_{{\ell} q}^{(1)}]^{\alpha \beta i j}+[{\mC}_{{\ell} q}^{(3)}]^{\alpha\beta i j})~,\\
	[{\mc}_{LL}^{V}]^{\alpha \beta i j} &=  2\, [{\mC}_{{\ell} q}^{(3)}]^{\alpha \beta i j}~,
\end{align}
where we have written both the SMEFT and HEFT WCs in the flavor basis. One can easily read off the  324 (171) SMEFT predictions implied by the above equations:
\begin{align}
    [{\mc}_{euLL}^{V}]^{\alpha \beta i j} &= [{\mc}_{\nu dLL}^{V}]^{\alpha \beta i j}~, \label{corLL1}\\
    [{\mc}_{edLL}^{V}]^{\alpha \beta i j} &= [{\mc}_{\nu uLL}^{V}]^{\alpha \beta i j}~, \label{corLL2}\\
    \,[{\mc}_{LL} ^{V}]^{\alpha \beta i j} &=  [{\mc}_{edLL}^{V}]^{\alpha \beta i j} - [{\mc}_{\nu dLL}^{V}]^{\alpha \beta i j}~.\label{corLL3}
\end{align}
These predictions are in the flavor basis. We would like to have the relations in terms of HEFT operators in the mass basis for later matching with the LEFT operators and with the observables. This can be achieved by the use of unitary matrices $S_{L,R}$ and $K_{L,R}$ for quarks and leptons, respectively. The fields are transformed as  
\begin{align}
	u_L^i &\rightarrow (S_L^{u})^{ij}u_L^{j}~,\quad d_L^i \rightarrow (S_L^{d})^{ij}d_L^{j}~,\quad
	u_R^i \rightarrow (S_R^{u})^{ij}u_R^{j}~,\quad d_R^i \rightarrow (S_R^{d})^{ij}d_L^{j}~,\\
	e_L^\alpha &\rightarrow (K_L^{e})^{\alpha\beta}e_L^{\beta}~,\quad \nu_L^\alpha \to (K_L^{\nu})^{\alpha\beta}\nu_L^{\beta}~,\quad
	e_R^\alpha \to (K_R^{e})^{\alpha\beta}e_R^{\beta}~.
\end{align} 
The relation in eq.\,(\ref{corLL1}) gets transformed in the mass basis as\footnote{ A hat on top of the HEFT WC indicates that it is in the mass basis, otherwise it is in the flavor basis.}
\begin{align}
	(K_L^{e})^{\alpha\rho }\, (S_L^{u})^{ik}\,	[\hat {\mc}_{euLL}^{V}]^{\rho \sigma k l} \, (S_L^{u\dagger})^{{\ell} j} (K_L^{e\dagger})^{\sigma\beta}  &=(K_L^{\nu})^{\alpha\rho }\, (S_L^{d})^{ik}\, [\hat{\mc}_{\nu dLL}^{V}]^{\rho \sigma k l}\, (S_L^{d\dagger})^{{\ell} j} (K_L^{\nu\dagger})^{\sigma\beta}.
\end{align}
Suppressing the lepton and quark family indices, the above equation can be rewritten in a compact form as
\begin{align}
K_L^e\,S_L^u\,\hat {\mc}_{euLL}^V\,S_L^{u\dagger}\,K_L^{e\dagger} &= K_L^\nu\,S_L^d\,\hat {\mc}_{\nu dLL}^V\,S_L^{d\dagger}\,K_L^{\nu\dagger}~,\label{corLL1.2}
\end{align}
where the matrices $S$ and $K$ carry only quark and lepton indices, respectively. This relation may be further expressed as 
\begin{align}
	V^\dagger \, \hat {\mc}_{euLL}^{V} \,V& = U^\dagger \,\hat {\mc}_{\nu dLL}^{V} \, U~,\label{corLL1.3}
\end{align}
using the CKM and PMNS matrices
\begin{align}
	V &\equiv V_{CKM} = S_L^{u\dagger}\,S_L^{d}   ~~\textrm{and}\quad
    U \equiv U_{PMNS}  = K_L^{\nu\dagger}\,K_L^{e} ~.\label{ckm_pmns_def}
\end{align}
Following similar steps, we can rewrite the relations from eqs.\,(\ref{corLL2}) and (\ref{corLL3}) in the mass basis as
\begin{align}
	 V \, \hat {\mc}_{edLL}^{V} \,V^\dagger& = U^\dagger \,\hat {\mc}_{\nu uLL}^{V} \, U~,\label{corLL2.1}\\
	 V^\dagger\,\hat {\mc}_{LL}^{V}\,U&= \hat {\mc}_{edLL}^{V} - U^\dagger\,\hat{\mc}_{\nu dLL}^{V}\,U~.\label{corLL3.1} 
\end{align}
Note that the final SMEFT predictions, i.e. the relations among the HEFT WCs shown in eqs.\,(\ref{corLL1.3}), (\ref{corLL2.1}) and (\ref{corLL3.1}), involve only the physically measurable CKM and PMNS matrices. This makes the relations completely independent of any UV flavor assumption. The relations in eqs.~(\ref{corLL1.3}) and (\ref{corLL2.1}) were derived previously in Ref.~\cite{Bissmann:2020mfi}.

\paragraph{LLRR vector operators: }

Similar to the previous case, in this category there are 486 (261) independent HEFT operators and 162 (90) independent SMEFT operators, as shown in Table~\ref{HEFToplist} and \ref{SMEFToplist}, respectively. The HEFT WCs can be written in terms of the SMEFT ones as follows:
\begin{align}
	[{\mc}_{\nu uLR}^{V}]^{\alpha \beta i j}&= \,[{\mC}_{{\ell} u}]^{\alpha \beta i j},\quad
	[{\mc}_{e uLR}^{V}]^{\alpha \beta i j}= \,[{\mC}_{{\ell} u}]^{\alpha \beta i j},\label{heft-smeft-LR1}\\
	[{\mc}_{\nu dLR}^{V}]^{\alpha \beta i j}&= \,[{\mC}_{{\ell} d}]^{\alpha \beta i j}~,\quad
	[{\mc}_{e dLR}^{V}]^{\alpha \beta i j}= \,[{\mC}_{{\ell} d}]^{\alpha \beta i j}~,\quad [{\mc}_{LR}^{V}]^{\alpha \beta i j} = 0\label{heft-smeft-LR2}
    ~.
\end{align}
Thus, here we get 324 (171) relations among the HEFT coefficients. In the flavor basis and the mass basis, these relations are
\begin{align}
{\mc}_{euLR}^{V} = {\mc}_{\nu uLR}^{V}\quad &\Rightarrow\quad \hat{\mc}_{euLR}^{V} = U^\dagger\,\hat{\mc}_{\nu uLR}^{V}\,U~,\label{corLR1}\\
{\mc}_{edLR}^{V} = {\mc}_{\nu dLR}^{V}\quad &\Rightarrow\quad \hat{\mc}_{edLR}^{V} = U^\dagger\,\hat{\mc}_{\nu dLR}^{V}\,U~,\label{corLR2}\\
{\mc}_{LR}^{V} = 0\quad &\Rightarrow\quad \hat{\mc}_{LR}^{V} = 0~.\label{corLR3}
\end{align}
Note that in this category, only right-handed quark fields appear and the rotation matrices for the right-handed quarks cancel out in the relations when translated to the mass basis. As a result, there is no CKM matrix in these relations and only the PMNS matrix $U$ appears for the leptons. The relations above show that the charged-current HEFT WCs vanish for this category of operators. This is because in SMEFT, as noted in \cite{Burgess:2021ylu, Cata:2015lta}, right-handed quarks cannot participate in charged-current semileptonic processes at dimension 6 level due to hypercharge conservation.

\begin{table}[h!]
	\begin{center}
		\small
		\begin{minipage}[t]{0.90\textwidth}
			\centering
			\renewcommand{\arraystretch}{1.6}
			\begin{tabular}[t]{|c| c | c|}
				\hline
				Category & Analytic relations & Count\\
				\hline
				\multirow{3}{*}{\rotatebox{0}{$ LLLL $}} & 
				$ V^\dagger_{ik} \, [\hat {\mc}_{euLL}^{V}]^{\alpha \beta k l} \,V_{{\ell} j} = U^\dagger_{\alpha \rho} \,[\hat {\mc}_{\nu dLL}^{V}]^{\rho \sigma i j} \, U_{\sigma \beta}$& 81 (45)\\
				&
				$ V_{ik} \, [\hat {\mc}_{edLL}^{V}]^{\alpha \beta k l} \,V^{\dagger}_{{\ell} j} = U^\dagger_{\alpha \rho} \,[\hat {\mc}_{\nu uLL}^{V}]^{\rho \sigma i j} \, U_{\sigma \beta}$ & 81 (45)\\
				& $ V^\dagger_{ik} \, [\hat{\mc}_{LL} ^{V}]^{\alpha \beta k j} =  [\hat {\mc}_{edLL}^{V}]^{\alpha \rho i j}\,U^\dagger_{\rho \beta} - U^\dagger_{\alpha \sigma} \, [{\mc}_{\nu dLL}^{V}]^{\sigma \beta i j}$ & 162 (81)\\
				\hline\hline
				\multirow{1}{*}{\rotatebox{0}{$ RRRR $}} &No relations &\\
				\hline\hline
				\multirow{3}{*}{\rotatebox{0}{$ LLRR $}} & $[\hat {\mc}_{edLR}^{V}]^{\alpha \beta i j} = U^\dagger_{\alpha \rho} \, [\hat {\mc}_{\nu dLR}^{V}]^{\rho \sigma i j}\,U_{\rho \beta}$ & 81 (45)\\
				& $[\hat {\mc}_{euLR}^{V}]^{\alpha \beta i j} = U^\dagger_{\alpha \rho} \, [\hat {\mc}_{\nu u LR}^{V}]^{\rho \sigma i j}\,U_{\rho \beta}$  & 81 (45)\\
                 & $[\hat{\mc}_{LR}^V]^{\alpha\beta i j} = 0$ & 162 (81)\\
				\hline \hline
				\multirow{2}{*}{\rotatebox{0}{$ RRLL $}} & \multirow{2}{*}{$[\hat {\mc}_{edRL}^{V}]^{\alpha \beta i j} = V^\dagger_{ i k} \, [\hat {\mc}_{e u RL}^{V}]^{\rho \sigma k l}\,V_{ lj}$} & \multirow{2}{*}{81 (45)}\\
				& &\\
				\hline\hline
				\multirow{2}{*}{\rotatebox{0}{Scalar ($d_R$)}} & $ V_{ik}\, [\hat {\mc}_{ed, RLLR}^{S}]^{\alpha \beta k j} =  [\hat {\mc}_{RLLR}^{S}]^{\alpha \rho i j}\,U_{\rho \beta}$& 162 (81)\\
				& $[\hat {\mc}_{e d,  RLRL}^{S}]^{\alpha \beta i j} = 0$ & 162 (81)\\
				\hline\hline
				\multirow{2}{*}{\rotatebox{0}{Scalar ($u_R$)}} & $[\hat {\mc}_{eu, RLRL}^{S}]^{\alpha \beta i k} \, V_{kj}=  -[\hat{\mc}_{RLRL}^{S}]^{\alpha \rho i j} \, U_{\rho \beta}$ & 162 (81)\\
				& $[\hat{\mc}_{e u,  RLLR}^{S}]^{\alpha \beta i j} = 0$ & 162 (81)\\
				\hline\hline
				\multirow{2}{*}{\rotatebox{0}{Tensor ($d_R$)}} & $[\hat {\mc}_{ed,\, \textrm{all}} ^{T}]^{\alpha \beta i j} = 0$ &  324 (162)\\
				& $[\hat {\mc}_{RLLR} ^{T}]^{\alpha \beta i j} = 0$ &162 (81) \\
				\hline\hline
				\multirow{2}{*}{\rotatebox{0}{Tensor ($u_R$)}} & $[\hat {\mc}_{eu, RLRL}^{T}]^{\alpha \beta i k} \, V_{kj}=  -[\hat {\mc}_{RLRL}^{T}]^{\alpha \rho i j} \, U_{\rho \beta}$ & 162 (81)\\
				& $[\hat{\mc}_{e u,  RLLR}^{T}]^{\alpha \beta i j} = 0$ & 162 (81)\\
				\hline\hline
				\multirow{2}{*}{\rotatebox{0}{$Z$ and $W^{\pm}$}} & $[\hat{\mc}_{ud_LW}]^{ij} = \frac{1}{\sqrt{2}} \cos\theta_w\, ([\hat{\mc}_{u_LZ}]^{ik}\,V_{kj} - V_{ik}\,[\hat{\mc}_{d_LZ}]^{kj})$& 18 (9)\\
				& $[\hat{\mc}_{e\nu_LW}]^{\alpha\rho}\,U_{\rho \beta} =  \frac{1}{\sqrt{2}}\cos\theta_w \, ([\hat{\mc}_{e_LZ}]^{\alpha\beta} - U^\dagger_{\alpha\rho}\,[\hat{\mc}_{\nu_LZ}]^{\rho\sigma}\,U_{\sigma \beta})$& 18 (9)\\
				\hline
			\end{tabular}
		\end{minipage}
		\caption{\label{CorrTable} Linear relations among the HEFT semileptonic WCs in the mass basis predicted by the SMEFT. Summation over repeated indices is implicit. Count denotes the number of independent operators; the number inside the brackets is the number of independent operators if all WCs are real.}
	\end{center}
\end{table}

\paragraph{\textbf{RRRR vector operators: }}

Right-handed fermions are not charged under $SU(2)_L$. Thus even in SMEFT, the up-type and down-type right-handed fields can appear independently in neutral-current semileptonic operators, as in HEFT.  This makes the number of neutral-current operators of RRRR type in HEFT and SMEFT to be the same as shown in Tables~\ref{HEFToplist} and \ref{SMEFToplist}, respectively. Furthermore, in the absence of right-handed neutrinos, there are no charged-current operators either in HEFT or in SMEFT in this category. 
As a result, in this category, there are no relations among the HEFT coefficients.

\paragraph{\textbf{RRLL vector operators: }}
	
In the case of vector operators involving right-handed leptons and left-handed quarks, there are 162 (90) independent operators in HEFT and 81 (45) in SMEFT, respectively. This results in 81 (45) relations among the HEFT WCs. The mapping between HEFT and SMEFT WCs in the flavor basis and the resulting relations in the mass basis for this category are
\begin{align}
	{\mc}_{ed}^V = \,{\mC}_{eq}~,&\quad {\mc}_{eu}^V =  \,{\mC}_{eq}~\quad	\Rightarrow \quad \hat{\mc}_{edRL}^{V} = V^\dagger\,\hat{\mc}_{euRL}^{V}\,V~.
\end{align}  
Note that the PMNS matrix does not appear in these relations as only right-handed electrons are involved and the corresponding flavor rotations cancel out. Furthermore, there are no charged-current operators in this category as there is no right-handed neutrino in SM.

\paragraph{\textbf{Scalar operators: }}

There are 486 (243) scalar semileptonic operators with right-handed down-type quarks and 486 (243) operators with right-handed up-type quarks in HEFT. In SMEFT, there are 162 (90) operators for each of these scenarios. We find 324 (153) relations among the HEFT coefficients for each scenario. Mapping of these operators between HEFT and SMEFT in the flavor basis gives 
\begin{align}
	[{\mc}_{ed,RLLR}^{S}]^{\alpha \beta i j} & = [{\mC}_{{\ell} e dq}]^{\beta \alpha j i *}~, &[{\mc}_{eu,RLLR}^{S}]^{\alpha \beta i j} & = 0~, \\
	[{\mc}_{ed,RLRL}^{S}]^{\alpha \beta i j} & = 0~, &[{\mc}_{eu,RLRL}^{S}]^{\alpha \beta i j} & = -[{\mC}_{{\ell} e qu}]^{\beta \alpha j i *}~,\\
	[{\mc}_{RLLR}^{S}]^{\alpha \beta i j} &= [{\mC}_{{\ell} e dq}]^{\beta \alpha j i\,*}~, & [{\mc}_{RLRL}^{S}]^{\alpha \beta i j} &= [{\mC}_{{\ell} e qu}]^{\beta \alpha j i\,*}~.
\end{align}
From the above equations, we get the following relations among the HEFT WCs in the mass basis: 
\begin{align}
	V\,\hat{\mc}_{ed,RLLR}^{S} &= \hat{\mc}_{RLLR}^{S}\,U~, & \hat{\mc}_{eu,RLRL}^{S}\,V&= -\hat{\mc}_{RLRL}^{S}\,U~,\\
	\hat{\mc}_{ed,RLRL}^{S} & = 0~, &\hat{\mc}_{eu,RLLR}^{S}& = 0~.
\end{align}
Note that both the above relations in eq.\,(2.27) represent relations among neutral-current scalar operators (on the LHS) and charged-current scalar operators (on the RHS).  The WCs in eq.\,(2.28) vanish\footnote{ Note that the vanishing of these HEFT WCs correspond to the relations $C_S =- C_P$ and $C_S^\prime = C_P^\prime$ in the conventional LEFT for the UV4f models as noted in \cite{Alonso:2014csa, Cata:2015lta}.} since the corresponding SMEFT operators would not satisfy $U(1)_Y$. 

\paragraph{\textbf{Tensor operators: }}

There is no tensor operator with right-handed down-type quarks in SMEFT as these operators cannot conserve $U(1)_Y$ hypercharge. Thus, all the tensor operators with right-handed down-type quarks in HEFT get zero contribution from SMEFT. As a result, SMEFT predicts 486 (243) constraints on such HEFT WCs:
\begin{align}
    \hat{\mc}_{ed,RLLR}^{T} &= 0~,\quad \hat{\mc}_{ed,RLRL}^{T}  = 0~,\quad \hat{\mc}_{RLLR}^{T} =0~.
\end{align}
For the case of tensor operators with right-handed up-type quarks, the mapping and relations are exactly the same as the scalar operators:
\begin{align}
     \hat{\mc}_{eu,RLRL}^{T}\,V&=  -\hat{\mc}_{RLRL}^{T}\,U~, \quad \hat{\mc}_{eu,RLLR}^{T} = 0~.
\end{align}
The reason for the vanishing of the WCs in the last equation is again that the corresponding SMEFT operators would not preserve the $U(1)_Y$ hypercharge symmetry. See also references \cite{Alonso:2014csa, Cata:2015lta}.

In Table~\ref{CorrTable}, we present all the relations among the HEFT WCs corresponding to four-fermion semileptonic operators, which would be predicted by SMEFT. We express these relations in the mass basis and explicitly put the indices for the quark and the lepton families.  

\subsection{Predictions for the couplings of $Z$, $W^\pm$ and $h$ to fermions}\label{sec:ZWH}

\begin{table}[t]
	\begin{minipage}{0.48\textwidth}
		\centering
		\renewcommand{\arraystretch}{1.3}
		\begin{tabular}{|c|cc|}
            \hline
            \multicolumn{3}{|c|}{HEFT}\\
            \hline
            \hline
            \multicolumn{3}{|c|}{$LL$ quarks}\\
            \hline
            & Operator & Count\\
            $[{\mc}_{u_LZ}]^{ij}$ & $(\bar u_L^i \gamma^\mu u_L^j)\,Z_\mu$& 9(6)\\
            $[{\mc}_{d_LZ}]^{ij}$ & $(\bar d_L^i \gamma^\mu d_L^j)\, Z_\mu$&9(6)\\
            $[{\mc}_{ud_LW}]^{ij}$ & $(\bar u_L^i \gamma^\mu d_L^j)\, W_\mu^+$& 18(9)\\
            \hline
            \hline
            \multicolumn{3}{|c|}{$RR$ quarks}\\
            \hline
            & Operator & Count\\
            $[{\mc}_{u_RZ}]^{ij}$ & $(\bar u_R^i \gamma^\mu u_R^j)\,Z_\mu$& 9(6)\\
            $[{\mc}_{d_RZ}]^{ij}$ & $(\bar d_R^i \gamma^\mu d_R^j)\, Z_\mu$&9(6)\\
            $[{\mc}_{ud_RW}]^{ij}$ & $(\bar u_R^i \gamma^\mu d_R^j)\, W_\mu^+$&18(9)\\
            \hline
            \hline
            \multicolumn{3}{|c|}{$LL$ leptons}\\
            \hline
            & Operator & Count\\
            $[{\mc}_{\nu_LZ}]^{\alpha\beta}$ & $(\bar \nu_L^\alpha \gamma^\mu \nu_L^\beta)\, Z_\mu$&9(6)\\
            $[{\mc}_{e_LZ}]^{\alpha\beta}$ & $(\bar e_L^\alpha \gamma^\mu e_L^\beta)\,Z_\mu$&9(6)\\
            $[{\mc}_{e \nu_LW}]^{\alpha\beta}$ & $(\bar e_L^\alpha \gamma^\mu \nu_L^\beta)\, W_\mu^+$&18(9)\\
            \hline
            \hline
            \multicolumn{3}{|c|}{$RR$ leptons}\\
            \hline
            & Operator & Count\\
            $[{\mc}_{e_RZ}]^{\alpha\beta}$ & $(\bar e_R^\alpha \gamma^\mu e_R^\beta)\,Z_\mu$&9(6)\\
            \hline
            \hline
            \multicolumn{3}{|c|}{Scalar operators}\\
            \hline
            & Operator & Count\\
            $[{\mc}_{eh}]^{\alpha\beta}$ & $(\bar e_L^\alpha \, e_R^\beta)\,h$ & 9(6)\\
            $[{\mc}_{dh}]^{ij}$ & $(\bar d_L^i \, d_R^j)\,h$ & 9(6)\\
            $[{\mc}_{uh}]^{ij}$ & $(\bar u_L^i \, e_R^j)\,h$ & 9(6)\\
            \hline			
		\end{tabular}
	\end{minipage}
    \hspace{-0.5cm}
	\begin{minipage}{0.48\textwidth}
		\centering
		\renewcommand{\arraystretch}{1.3}
		\begin{tabular}{|c|cc|}
            \hline
            \multicolumn{3}{|c|}{SMEFT}\\
            \hline
            \hline
            \multicolumn{3}{|c|}{$LL$ quarks}\\
            \hline
            & Operator & Count\\
            $[{\mC}_{Hq}^{(1)}]^{ij}$ & $(H^\dagger \overleftrightarrow{D}_\mu H)(\bar q^i \gamma^\mu q^j)$&9(6)\\
            $[{\mC}_{Hq}^{(3)}]^{ij}$ & $(H^\dagger \overleftrightarrow{D}_\mu \,\tau^IH)(\bar q^i \gamma^\mu \,\tau^I q^j)$&9(6)\\
            & &\\
            \hline
            \hline
            \multicolumn{3}{|c|}{$RR$ quarks}\\
            \hline
            & Operator & Count\\
            $[{\mC}_{Hu}]^{ij}$ & $(H^\dagger \overleftrightarrow{D}_\mu H)(\bar u_R^i \gamma^\mu u_R^j)$&9(6)\\
            $[{\mC}_{Hd}]^{ij}$ & $(H^\dagger \overleftrightarrow{D}_\mu \,\tau^IH)(\bar d_R^i \gamma^\mu \,\tau^I d_R^j)$&9(6)\\
            $[{\mC}_{Hud}]^{ij}$& $(\widetilde{H}^\dagger \overleftrightarrow{D}_\mu \,H)(\bar u_R^i \gamma^\mu \,d_R^j)$& 18(9)\\
            \hline
            \hline
            \multicolumn{3}{|c|}{$LL$ leptons}\\
            \hline
            & Operator & Count\\
            $[{\mC}_{Hl}^{(1)}]^{\alpha\beta}$ & $(H^\dagger \overleftrightarrow{D}_\mu H)(\bar l^\alpha \gamma^\mu l^\beta)$&9(6)\\
            $[{\mC}_{Hl}^{(3)}]^{\alpha\beta}$ & $(H^\dagger \overleftrightarrow{D}_\mu \,\tau^IH)(\bar l^\alpha \gamma^\mu \,\tau^I l^\beta)$&9(6)\\
            & &\\
            \hline
            \hline
            \multicolumn{3}{|c|}{$RR$ leptons}\\
            \hline
            & Operator & Count\\
            $[{\mC}_{He}]^{\alpha\beta}$ & $(H^\dagger \overleftrightarrow{D}_\mu H)(\bar e_R^\alpha \gamma^\mu e_R^\beta)$&9(6)\\
            \hline
            \hline
            \multicolumn{3}{|c|}{Scalar quarks}\\
            \hline
            & Operator & Count\\
            $[{\mC}_{eH}]^{\alpha\beta}$ & $(H^\dagger\,H)\,(\bar l^{\alpha}\,e_R^\beta H)$ & 9(6)\\
            $[{\mC}_{dH}]^{ij}$ & $(H^\dagger\,H)\,(\bar q^{i}\,d_R^j H)$ & 9(6)\\
            $[{\mC}_{uH}]^{ij}$ & $(H^\dagger\,H)\,(\bar q^{i}\,u_R^j \widetilde{H})$ & 9(6)\\
            \hline	
		\end{tabular}
	\end{minipage}
	\caption{\label{heftZW} Left column: HEFT operators representing the couplings of  $Z$,  $W^{\pm}$ and $h$ with fermions. Right column: SMEFT operators contributing to the corresponding HEFT operators (following notations of \cite{Grzadkowski:2010es}). Count denotes the number of independent operators; the number inside the brackets is the number of independent operators if all WCs were real. The SMEFT basis in which these operators are written is defined in Appendix~\ref{app: SMEFTbasis}.}
\end{table}

In addition to quarks and leptons, HEFT also involves $Z, W^\pm$ and $h$ bosons as degrees of freedom. The BSM couplings of these bosons to the fermions appear as HEFT WCs, as shown in eq.\,(\ref{heftlag}). These WCs contribute to low-energy semileptonic processes via $Z, W^\pm$ and $h$ exchange diagrams. They can also be probed independently by studying decays of  $Z, W^\pm$, and $h$. However, when the BSM couplings of these bosons to fermions are parameterized in terms of SMEFT WCs, the number of independent WCs are less than the total number of relevant HEFT WCs. Thus, SMEFT predicts relations among the corresponding HEFT WCs, as earlier. In this section, we derive these relations. 

In Table~\ref{heftZW}, we show the 144 (87) HEFT operators and 108 (69) SMEFT operators that give rise to $Z, W^\pm$ and $h$ couplings to fermions. Once again, the HEFT operators have been presented in the unitary gauge as $U(1)_{em}$ invariant terms. These operators can be rewritten in an $SU(2)_L \times U(1)_Y$ invariant form where this symmetry is non-linearly realized, as shown in Appendix~\ref{app:HEFTbasis}. 
Note that the SMEFT basis we have used is not the commonly used Warsaw basis. The details of our basis and our rationale for it have been presented in Appendix~\ref{app: SMEFTbasis}.

While the number of dimension-6 SMEFT and HEFT operators are the same for the coupling with $h$, the number of HEFT operators contributing to $Z$ and $W^\pm$ coupling deviations to left-handed quarks or leptons is 36 (21) and it exceeds the number of contributing  SMEFT operators, 18 (12). This implies 18 (12) relations among the HEFT WCs. The expressions for the HEFT WCs for these operators in terms of the SMEFT ones can be written in the flavor basis as 
\begin{align}
\label{ZW1}
    [{\mc}_{u_LZ}]^{ij} &= \eta_{LZ} \, ([{\mC}_{Hq}^{(1)}]^{ij} - [{\mC}_{Hq}^{(3)}]^{ij}), &[{\mc}_{\nu_LZ}]^{\alpha\beta} &= \eta_{LZ} \, ([{\mC}_{Hl}^{(1)}]^{\alpha\beta} - [{\mC}_{Hl}^{(3)}]^{\alpha\beta})~,\\
    [{\mc}_{d_LZ}]^{ij} &= \eta_{LZ} \, ([{\mC}_{Hq}^{(1)}]^{ij} + [{\mC}_{Hq}^{(3)}]^{ij})~, &[{\mc}_{e_LZ}]^{\alpha\beta} &= \eta_{LZ} \, ([{\mC}_{Hl}^{(1)}]^{\alpha\beta} + [{\mC}_{Hl}^{(3)}]^{\alpha\beta})~,\label{ZW2}\\
    [{\mc}_{ud_LW}]^{ij} &= \eta_{LW} \, [{\mC}_{Hq}^{(3)}]^{ij}~, &[{\mc}_{e\nu_LW}]^{\alpha\beta} &= \eta_{LW} \, [{\mC}_{Hq}^{(3)}]^{\alpha\beta}~.\label{ZW3}
\end{align}
Here $\eta_{LZ} = -g/(2\,\cos\theta_W)$, where $\theta_W$ is the Weinberg angle, and $\eta_{LW} = g/(\sqrt{2})$. These expressions can then be used to derive  the relations among the HEFT WCs:
\begin{align}
    [\hat{\mc}_{ud_LW}]^{ij} &= \frac{1}{\sqrt{2}} \cos\theta_w\, ([\hat{\mc}_{u_LZ}]^{ik}\,V_{kj} - V_{ik}\,[\hat{\mc}_{d_LZ}]^{kj})~,\label{ZWcorr1}\\
    [\hat{\mc}_{e\nu_LW}]^{\alpha\rho}\,U_{\rho \beta} &=  \frac{1}{\sqrt{2}}\cos\theta_w \, ([\hat{\mc}_{e_LZ}]^{\alpha\beta} - U^\dagger_{\alpha\rho}\,[\hat{\mc}_{\nu_LZ}]^{\rho\sigma}\,U_{\sigma \beta})~.\label{ZWcorr2}
\end{align}
These relations are also shown in the last two rows of Table~\ref{CorrTable}.
Once again,  the relations in the mass basis contain only the physically measurable CKM and PMNS matrices.

The relations in eq.\,(\ref{ZWcorr1}) and (\ref{ZWcorr2}) should be independent of the choice of the SMEFT basis. In the Warsaw basis, the additional operators  $ {\cal O}_T = (H^\dagger \, \overleftrightarrow{D} H)^2$ and ${\cal O}_{WB} = g g' (H^\dagger \tau^I\,H)\,W_{\mu\nu}^I\,B^{\mu\nu}$ would contribute to the couplings of gauge bosons to the fermions by affecting their mass and kinetic terms. However, their contributions in the above two relations cancel out. This can be more transparently seen in the SMEFT basis that we use, where these two operators are traded for two other operators which do not affect the gauge boson couplings (see Appendix~\ref{app: SMEFTbasis}).

\subsection{Predictions for semileptonic operators at low energies}\label{sec: leftcor}
In the previous two subsections, we discussed the predictions of SMEFT at high energies, i.e., relations among HEFT WCs above the EW scale. Now we consider the SMEFT predictions for the low-energy observables where the relevant effective field theory is LEFT.
The forms of LEFT operators are the same as in Table~\ref{HEFToplist} apart from the fact that the operators involving the top quark would not be included in the LEFT lagrangian.
We will now rewrite the relations derived in the previous two subsections in terms of the LEFT WCs. 
In order to carry this out, we need the matching relations between the WCs of HEFT and LEFT operators. For instance, for operators of the $LLLL$ category, the matching relations in the flavor basis\footnote{In our notation, $\tilde{\tempC}$ corresponds to LEFT coefficient in the flavor basis and ${\tempC}$ in the mass basis.} are
\begin{align}
	[\tilde{\tempC}_{{\ell} q LL}^{V}]^{\alpha \beta i j}&=\omega \,[{\mc}_{{\ell} q LL}^{V}]^{\alpha \beta i j} + k_{{\ell}_L} \, [{\mc}_{q_L Z}]^{ij} \, \delta_{\alpha\beta} + k_{q_L}\, [{\mc}_{ l_L Z}]^{\alpha\beta}\,\delta_{ij}~,\label{left-heftLL1}\\
	[\tilde{\tempC}_{LL}^{V}]^{\alpha \beta i j}&=\omega \,[{\mc}_{LL}^{V}]^{\alpha \beta i j} + k_{e\nu W} \, [{\mc}_{ud_LW}]^{ij} \, \delta_{\alpha\beta} + k_{ud W}\, [{\mc}_{e\nu_LW}]^{\alpha\beta}\,\delta_{ij}~,\label{left-heftLLCC}
\end{align}
where $l\in \{\nu,e\}$, $q\in \{u,d\}$, $\omega = v^2/(2\,\Lambda^2)$, and the $k$ coefficients are 
\begin{align}
    k_{f_L} &= \frac{2\cos\theta_w}{g}(T^3_f-Q_f \sin^2\theta_{w})~,\quad \textrm{and}~~k_{e\nu W} = k_{udW} = \frac{\sqrt{2}}{g}~,
\end{align}
with $f_L\in \{\nu_L, e_L, u_L, d_L\}$.

In this work, we have not considered effects of the renormalization group (RG) running of the LEFT coefficients from the weak scale to the scale of the relevant experiments. These would need to be included using the RG  equations of Ref.~\cite{Jenkins:2013zja} for a more precise phenomenological treatment.

Using matching relations like eq.~(\ref{left-heftLL1}) and eq.~(\ref{left-heftLLCC}), we can now rewrite the relations of Table~\ref{CorrTable}, which were written in terms of the HEFT WCs, the LEFT WCs, and the BSM couplings of $Z$ and $W^\pm$. For the $LLLL$ operators, for example, these relations become
\begin{align}
    &V_{ik}\left[[{\tempC}_{e dLL}^{V}]^{\alpha \beta k l}- \left( k_{e_L} \, [\hat{\mc}_{d_LZ}]^{kl} \, \delta_{\alpha\beta} + k_{d_L}\, [\hat{\mc}_{e_LZ}]^{\alpha\beta}\,\delta_{kl}\right)\right]V^\dagger_{{\ell} j}\nonumber\\
    &~~= U^\dagger_{\alpha \rho}\left[[{\tempC}_{\nu uLL}^{V}]^{\rho \sigma i j}-\chi ~ \left(k_{\nu_L} \, [\hat{\mc}_{u_LZ}]^{ij} \, \delta_{\rho\sigma} + k_{u_L}\, [\hat{\mc}_{\nu_LZ}]^{\rho \sigma}\,\delta_{ij}\right)\right] U_{\sigma\beta},\label{leftheftcor1}\\
    &V^\dagger_{ik}\left[[{\tempC}_{e uLL}^{V}]^{\alpha \beta k l}-\chi~ \left( k_{e_L} \, [\hat{\mc}_{u_LZ}]^{kl} \, \delta_{\alpha\beta} + k_{u_L}\, [\hat{\mc}_{e_LZ}]^{\alpha\beta}\,\delta_{kl}\right)\right]V_{{\ell} j} \nonumber\\
    &~~= U^\dagger_{\alpha \rho}\left[[{\tempC}_{\nu dLL}^{V}]^{\rho \sigma i j}- \left(k_{\nu_L} \, [\hat{\mc}_{d_LZ}]^{ij} \, \delta_{\rho\sigma} + k_{d_L}\, [\hat{\mc}_{\nu_LZ}]^{\rho \sigma}\,\delta_{ij}\right)\right] U_{\sigma\beta}~,\label{leftheftcor2}\\
    & V^\dagger_{ik} \, \left[[{\tempC}_{LL}^{V}]^{\alpha \beta k j} - \chi  \left( k_{e\nu W} \, [\hat{\mc}_{ud_LW}]^{kj} \, \delta_{\alpha\beta} + [k_{ud W}]^{kj}\, [\hat{\mc}_{e\nu_LW}]^{\alpha\beta}\right)\right] \nonumber\\
    &~~=  \left[[{\tempC}_{e d LL}^{V}]^{\alpha \rho i j}-  \left( k_{e_L} \, [\hat{\mc}_{d_LZ}]^{ij} \, \delta_{\alpha\rho} + k_{d_L}\, [\hat{\mc}_{e_LZ}]^{\alpha\rho}\,\delta_{ij}\right)\right]\,U^\dagger_{\rho \beta}\nonumber\\
    & ~~~- U^\dagger_{\alpha \sigma} \, \left[[{\tempC}_{\nu d LL}^{V}]^{\sigma \beta i j}-  \left( k_{\nu_L} \, [\hat{\mc}_{d_LZ}]^{ij} \, \delta_{\sigma\beta} + k_{d_L}\, [\hat{\mc}_{\nu_LZ}]^{\sigma\beta}\,\delta_{ij}\right)\right]~, \label{leftheftcor3}
\end{align}
where, for WCs involving top quark, we have defined\footnote{This convention is used for the purpose of giving eqs.\,(\ref{leftheftcor1}) to (\ref{leftheftcor3}) a unified form for all quarks. This is an exception to our normal convention, where we use `$C$' only for LEFT WCs.} $[{\tempC}_{{\ell} q LL}^{V}]^{\alpha \beta i j}\equiv\omega [\hat{\mc}_{{\ell} q LL}^{V}]^{\alpha \beta i j}$ and $[{\tempC}_{LL}^{V}]^{\alpha \beta i j}\equiv \omega [\hat{\mc}_{LL}^{V}]^{\alpha \beta i j}$. In eq.\,(\ref{leftheftcor3}), $\chi=0$ ($\chi=1$) if the respective four-fermion operator contains (does not contain) the top quark. The introduction of $\chi$ ensures that the HEFT WCs are replaced by LEFT ones for all the four-fermion operators not containing the top quark.
The relations for the WCs in the other categories can be similarly derived and have been presented in Appendix~\ref{app:left-heft}. 

We now mention two important scenarios where the SMEFT predictions derived in this section can be simplified. First,  note that apart from neutrino physics experiments, it is impossible to distinguish the different flavors of neutrinos in observables. These observables thus depend on combinations of WCs with neutrino flavor indices summed over and are independent of the basis used for neutrinos. In particular, we can choose to work in a basis aligned to the charged-lepton flavor basis. This amounts to substituting $U=1$ in all the SMEFT  predictions, whether it is for HEFT WCs in Table~\ref{tab: implications} or for LEFT WCs such as those in eqs.\,(\ref{leftheftcor1}-\ref{leftheftcor3}) or in Appendix~\ref{app:left-heft}. Secondly, in the    UV4f scenario where there are no modifications to $Z, W^\pm$ and $h$ couplings with respect to SM, the matching equations in eq.\,(\ref{left-heftLLCC})  get simplified and we can obtain SMEFT predictions involving LEFT WCs simply by substituting $[\hat{\mc}]^{\alpha \beta i j}$ by $[{\tempC}]^{\alpha \beta i j}$  in Table~\ref{tab: implications}. This scenario becomes more relevant in the phenomenological applications of the SMEFT predictions that we present in the following section.

\section{SMEFT-predicted constraints on new physics}\label{sec: indirect_bounds}
In this section, we will show how the SMEFT predictions derived in Sec.~\ref{sec: SMEFTtoHEFT} can be used to obtain bounds on the LEFT Wilson coefficients $[ {\tempC}]^{\alpha\beta i j}$. We utilize the fact that the SMEFT predictions give analytic equations that can connect strongly constrained WCs to poorly constrained ones, thus allowing us to extract stronger bounds on the latter.  In this section, we restrict ourselves to  UV4f models, where the UV physics generates only four-fermionic operators in SMEFT, so that the operators discussed in Sec.~\ref{sec:ZWH} are absent. While a more general analysis using the constraints on  $Z$ and $W^\pm$ couplings (see Ref.~\cite{Efrati:2015eaa})  is possible, our primary aim here is to illustrate the power of the SMEFT predictions and thus we focus on the very well-motivated UV4f scenario.   As discussed in the previous section, in this scenario we can use the relations in  Table~\ref{CorrTable} by simply replacing $[\hat{\mc}]^{\alpha \beta i j}$ by $[{\tempC}]^{\alpha \beta i j}$. Furthermore, as explained at the end of the previous section, the observables in this section will be insensitive to the flavor of neutrinos, and hence we can take $U \to 1$ in the SMEFT predictions. 

We further restrict ourselves to the operators involving only left-handed quarks and leptons (i.e. $LLLL$ discussed in Sec.~\ref{sec: leftcor}) as these provide leading corrections with respect to SM.\footnote{ While low-energy flavor observables get interference level corrections from both $RRLL$ and $LLLL$  operators, as far as high $p_T$ observables are concerned, only $LLLL$ operator contributions can interfere with SM contribution if fermion masses are neglected.}  The relations amongst $LLLL$  operators in  UV4f models are given by
\begin{align}
	[{\tempC}_{euLL}^{V}]^{\alpha \beta i j} &= V_{ik}\,[{\tempC}_{\nu d LL}^V]^{\alpha \beta k l} V^\dagger_{{\ell} j}~,\label{corLFV1}\\
	[{\tempC}_{ed LL}^{V}]^{\alpha \beta i j} &= V^\dagger_{ik}\,[{\tempC}_{\nu u LL}^V]^{\alpha \beta k l} V_{{\ell} j}~,\label{corLFV2}\\
	[{\tempC}_{LL}^{V}]^{\alpha \beta i j} &= V_{ik}\,([{\tempC}_{ed LL}^{V}]^{\alpha \beta k j} - [{\tempC}_{\nu d LL}^{V}]^{\alpha \beta k j})~,\label{corLFV3}
\end{align}
as can be obtained from eqs.\,(\ref{leftheftcor1} --\ref{leftheftcor3}).   Recall that RG effects have been ignored in deriving the above relations,  and  as discussed in Sec.~\ref{sec: leftcor},  the WCs in the above equations that involve the top quark have been defined as $[{\tempC}]^{\alpha \beta i j}\equiv \omega\,[\hat{\mc}]^{\alpha \beta i j}$. These WCs can be constrained using data from top production and decays. All other WCs in the above equation are the standard LEFT WCs of eqs.\,(\ref{LEFT1}) and (\ref{LEFT2}).

Note that eqs.\,(\ref{corLFV1}-\ref{corLFV3}) involve 486 (261) WCs $[{\tempC}]^{\alpha \beta i j}$,   which arise from 162 (90) SMEFT coefficients. These three equations therefore correspond to 324 (171) relations among the  WCs. 
Note that in several earlier analyses  (e.g. \cite{Fuentes-Martin:2020lea, Bause:2020auq}) the  WCs have been assumed to be real. This is of course valid for the WCs of Hermitian operators, i.e where $\alpha=\beta$ and $i=j$. However, as eqs.\,(\ref{corLFV1}--\ref{corLFV3}) show, all the WCs are related linearly with complex coefficients (i.e, combinations of CKM matrix elements)  which makes it inconsistent for all of them to be real. Note that even if all the WCs of SMEFT in the UV scale are real in the flavor basis,  phases will appear in $[{\tempC}]^{\alpha \beta i j}$ through CKM elements while matching. We in our analysis consider complex values for all the WCs of non-Hermitian operators.

In the rest of this section, we focus on deriving bounds on the WCs from semileptonic processes. To start with, in Sec.~\ref{sec: indNC1} to Sec.~\ref{sec: indCC} we consider only processes involving muon and muon neutrinos, i.e $\alpha=\beta=2$. This is because many of the direct bounds from the muon channel are quite stringent compared to those from the electron or tau channel.
The terms in eqs.\,(\ref{corLFV1}) and (\ref{corLFV2}) contain only neutral current WCs. On the other hand, in eq.\,(\ref{corLFV3}) charged current WCs are expressed in terms of neutral current WCs. Based on these relations, in Sec.~\ref{sec: indNC1} and \ref{sec: indNC2} we obtain indirect bounds on neutral current WCs appearing in eqs.\,(\ref{corLFV1}) and (\ref{corLFV2}) respectively. In Sec.~\ref{sec: indCC} we discuss about the indirect bounds for charged current WCs. In Sec.~\ref{sec: LFV}, we further indicate how these relations may be used in conjunction with constraints on lepton flavor violating decays to constrain Wilson coefficients involving other lepton families. 

\subsection{Bounds on neutral-current WCs involving $(\nu d)$ and  $(e u)$}\label{sec: indNC1}

There are   6 complex and 6 real neutral-current WCs in eq.\,(\ref{corLFV1}) with $\alpha =\beta = 2$. These WCs correspond to operators either with neutrinos and down-type quarks ($\nu d LL$), or with charged leptons and up-type quarks ($euLL$). 
We first discuss direct bounds on these WCs. We consider both low-energy observables, such as rare decays,  as well as high-energy observables, such as the high-$p_T$ Drell-Yan process, top decays, etc. While the former can directly bound the LEFT WCs  $[{\tempC}]^{\alpha \beta i j}$, the latter can directly bound only the high energy HEFT WCs, $[\hat{\mc}]^{\alpha \beta i j}$. As we are considering UV4f models here, however,  the bounds on $[\hat{\mc}]^{\alpha \beta i j}$ can be converted to bounds on $[{\tempC}]^{\alpha \beta i j}$ in a straightforward way by keeping only the first term in the matching relations, eq.\,(\ref{left-heftLL1}) and eq.\,(\ref{left-heftLLCC}).

\begin{figure}[t]
    \centering
    \includegraphics[width=\textwidth]{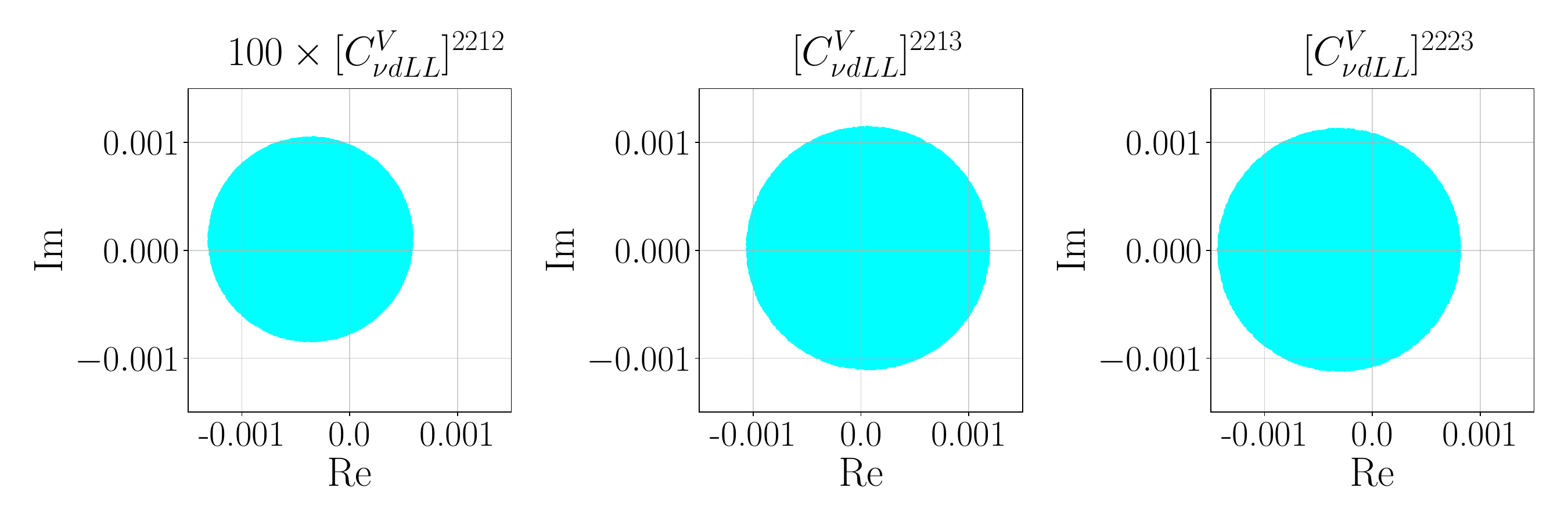}
    \includegraphics[width=\textwidth]{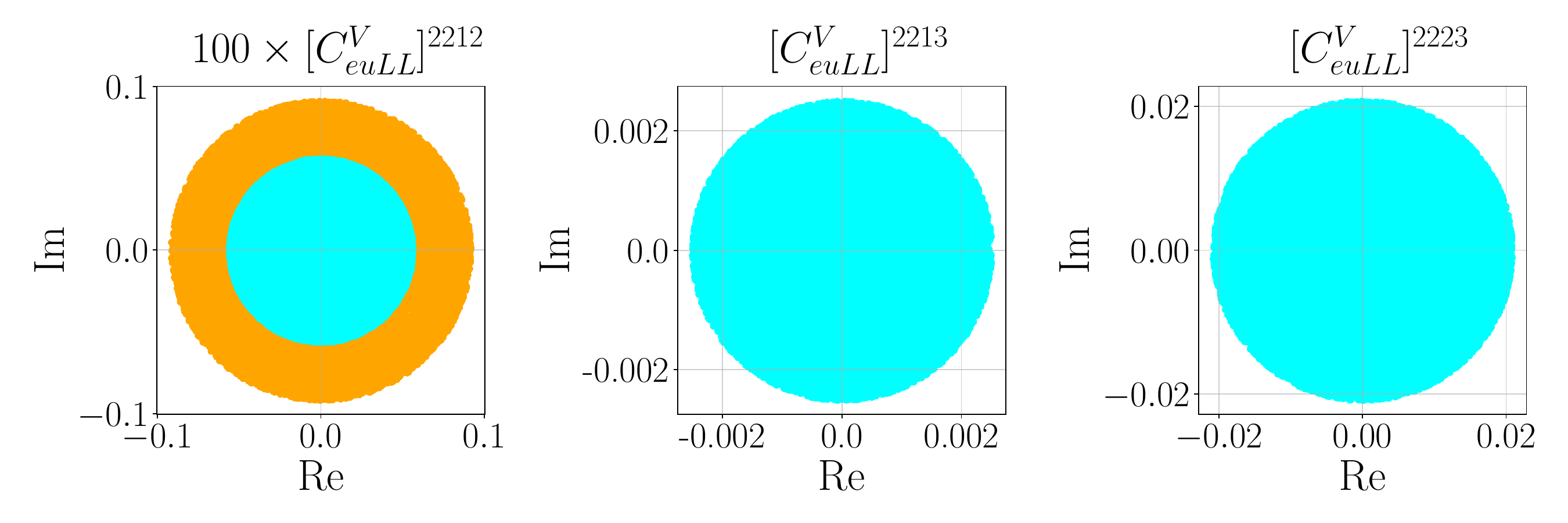}
    \hspace*{0.7cm}\includegraphics[width=0.97\textwidth]{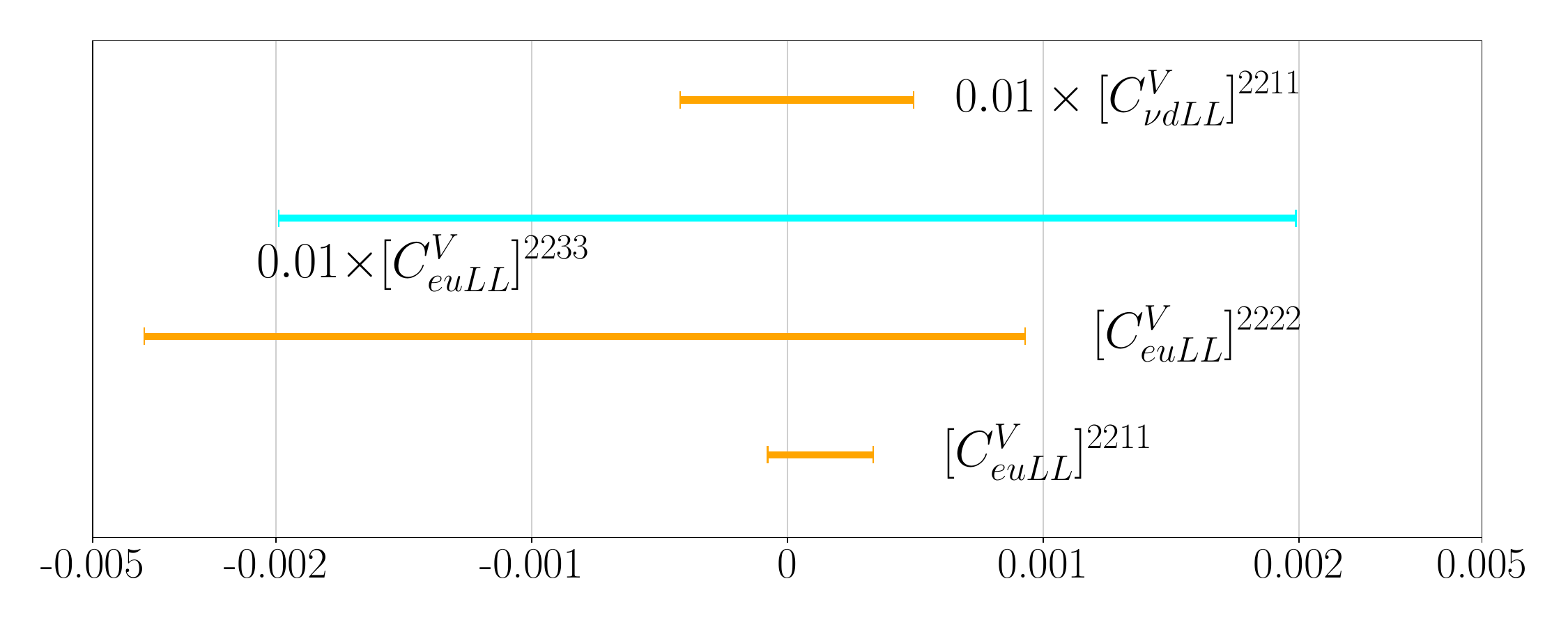}
    \caption{Direct bounds on the complex WCs ${\tempC}_{\nu dLL}^V$ (top panels) and ${\tempC}_{e u LL}^V$ (middle panels). The cyan color represents bounds from rare meson decays, orange represents bounds from high-$pT$ dimuon searches while purple represents bounds from top productions and decays. The WCs shown in the bottom panels are real due to the hermiticity of the corresponding operators. Note that the bottom panel uses the symmetric log scale. See Appendix~\ref{app: numbers} for numerical values of the bounds.} 
    \label{set1direct}
\end{figure}

Direct bounds on the WCs $[{\tempC}_{\nu d LL}^{V}]^{2212}$, $[{\tempC}_{\nu d LL}^{V}]^{2213}$ and $[{\tempC}_{\nu d LL}^{V}]^{2223}$ are obtained from rare decays of $K$ and $B$ mesons. For $[{\tempC}_{\nu d LL}^{V}]^{2212}$, we have used the recent measurement of the branching ratio of  $K^+ \to \pi^+ \nu \nu$ in the NA62 experiment~\cite{NA62:2022qes}. For $[{\tempC}_{\nu d LL}^{V}]^{2213}$ we take the $90\%$ upper bounds on the branching ratios of the decay modes $B^+\to \rho^+\,\nu\,\nu$ and $B^+\to \pi^+\,\nu\,\nu$~\cite{ParticleDataGroup:2022pth}. For $[{\tempC}_{\nu d LL}^{V}]^{2223}$, we include the recent measurement of $B^+\to K^+ \nu\,\nu$ branching ratio in \cite{Belle-II:2023esi} along with the $90\%$ upper bound on the branching ratio of $B^+\to K^{*+}\,\nu\,\nu$~\cite{ParticleDataGroup:2022pth}. The theoretical values for the discussed mesonic decay modes are calculated using the package `\textit{flavio}'~\cite{Straub:2018kue}. 
The bound on $[{\tempC}_{\nu d LL}^{V}]^{2211}$  is obtained from constraints\footnote{The bounds presented in \cite{Farzan:2017xzy} are for the vector and axial vector WCs. We convert these to bounds on operators in our basis by adding the $1\sigma$ ranges in quadrature.} on non-standard interactions of neutrinos in atmospheric and accelerator neutrino experiments~\cite{Farzan:2017xzy, Escrihuela:2011cf}. These bounds are shown in the top panels of Fig.~\ref{set1direct}.
For the WCs $[{\tempC}_{\nu d LL}^{V}]^{2222}$ and $[{\tempC}_{\nu d LL}^{V}]^{2233}$, there are no direct bounds available.

The direct bounds for WCs containing up-type quarks and charged leptons are obtained from rare decays, high-$p_T$ dilepton searches as well as top production and decays. The WC $[{\tempC}_{e u LL}^{V}]^{2212}$ gets constraints from rare decays of $D$ meson~\cite{Fuentes-Martin:2020lea}. For $[{\tempC}_{e u LL}^{V}]^{2211}$, $[{\tempC}_{e u LL}^{V}]^{2212}$ and $[{\tempC}_{e u LL}^{V}]^{2222}$, strong bounds are obtained from high-$p_T$ dimuon searches at the LHC. In the UV4f scenario and with the approximation of negligible RG effects, these bounds can be taken to be bounds on the LEFT WCs.
We use CMS data for the dimuon mode~\cite{CMS:2021ctt} and the package `\textit{HighPT}'~\cite{Allwicher:2022mcg, Allwicher:2022gkm} which provides bounds on SMEFT WCs. In order to convert these into bounds on isolated LEFT WCs, we turn on those linear combinations of SMEFT WCs which make that particular LEFT WC nonzero, and leave other dimuon modes unaffected. 
Bounds on WCs involving top quark (e.g. $[{\tempC}_{e u LL}^{V}]^{2213}$, $[{\tempC}_{e u LL}^{V}]^{2213}$  and  $[{\tempC}_{e u LL}^{V}]^{2233}$ ) are obtained from data on top production and decays~\cite{Afik:2021jjh}. These direct bounds are shown in Fig.~\ref{set1direct}.

Note that, in order to obtain the direct bounds in Fig.~\ref{set1direct}, we have only bounded the individual contribution of the relevant $LLLL$ operator with $\alpha=\beta=2$ and ignored possible contributions from other operators. Under some very reasonable assumptions, however, including these contributions would not significantly alter the bounds we have obtained. First of all, as far as the dineutrino decay modes are concerned, the experiments cannot distinguish between different neutrino flavors. To extract bounds on the $\bar{\nu}_\mu \nu_\mu$ mode, we assume that there are no large cancellations between the interference contributions of the different neutrino flavor modes.  Also, for low energy observables a linear combination of $LLLL$ WCs and WCs of other vector operators in  Table~\ref{HEFToplist}  enter the interference term in EFT corrections. In the cases where measurements are sensitive to the interference term, there can in principle be flat directions where the bounds obtained here get weakened, but this would again require a fine-tuned cancellation between the interference terms of the $LLLL$ and other vector operators;  we assume such cancellations are absent. Finally, there are operators in Table~\ref{HEFToplist}, such as the scalar and tensor operators, that give contributions proportional to the square of their WCs but the inclusion of such positive definite terms would only strengthen our bounds. Thus, under these assumptions, the direct bounds discussed here hold also in the presence of other operator contributions. 

\begin{figure}[t]
    \centering
    \begin{minipage}{0.8\textwidth}
    \hspace*{-0.5cm}\includegraphics[width=0.49\textwidth]{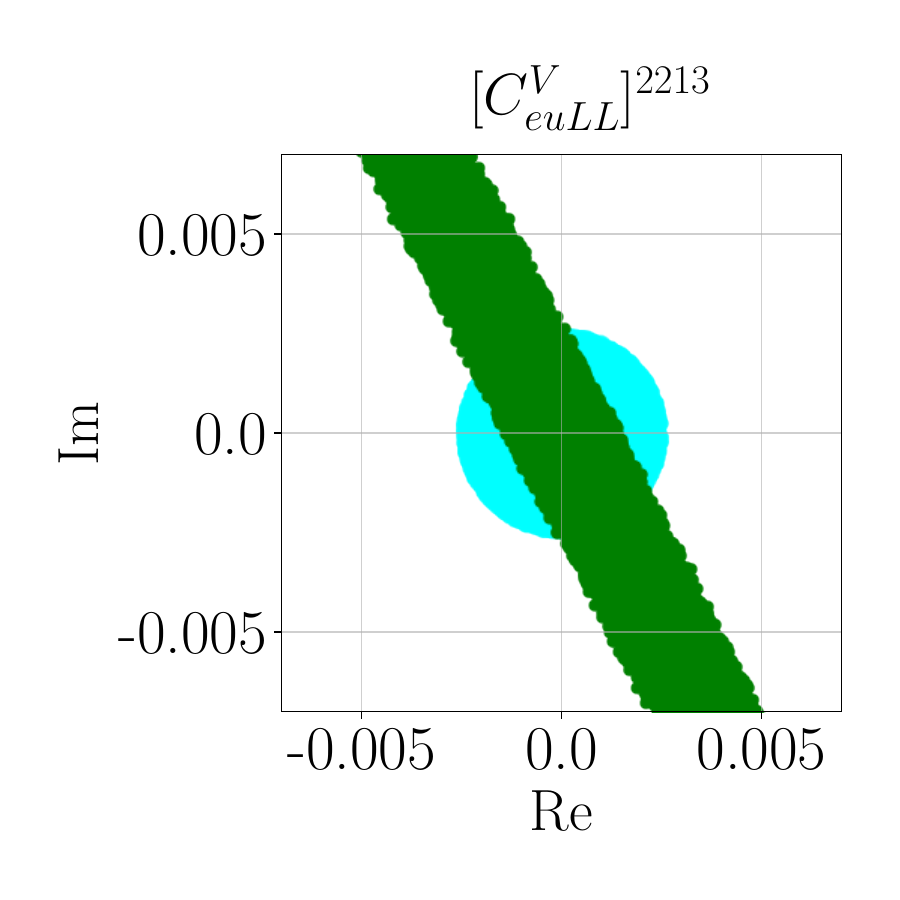}
    \includegraphics[width=0.49\textwidth]{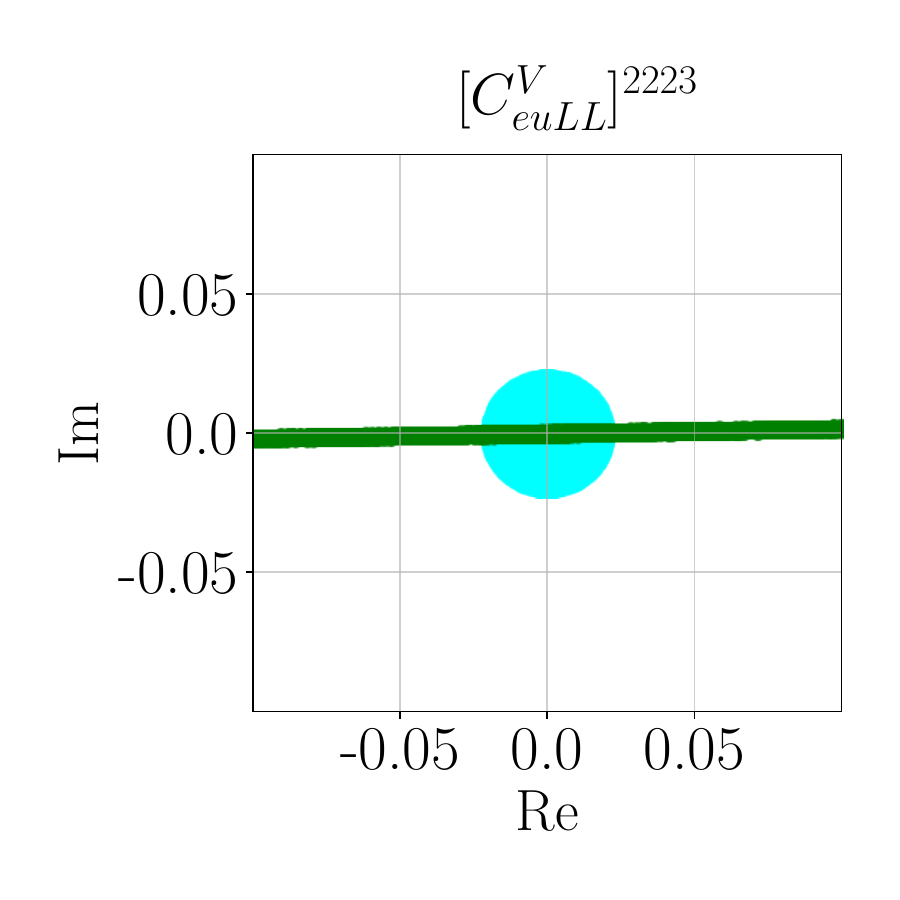}
    \end{minipage}
    \begin{minipage}{0.8\textwidth}
    \includegraphics[width=\textwidth]{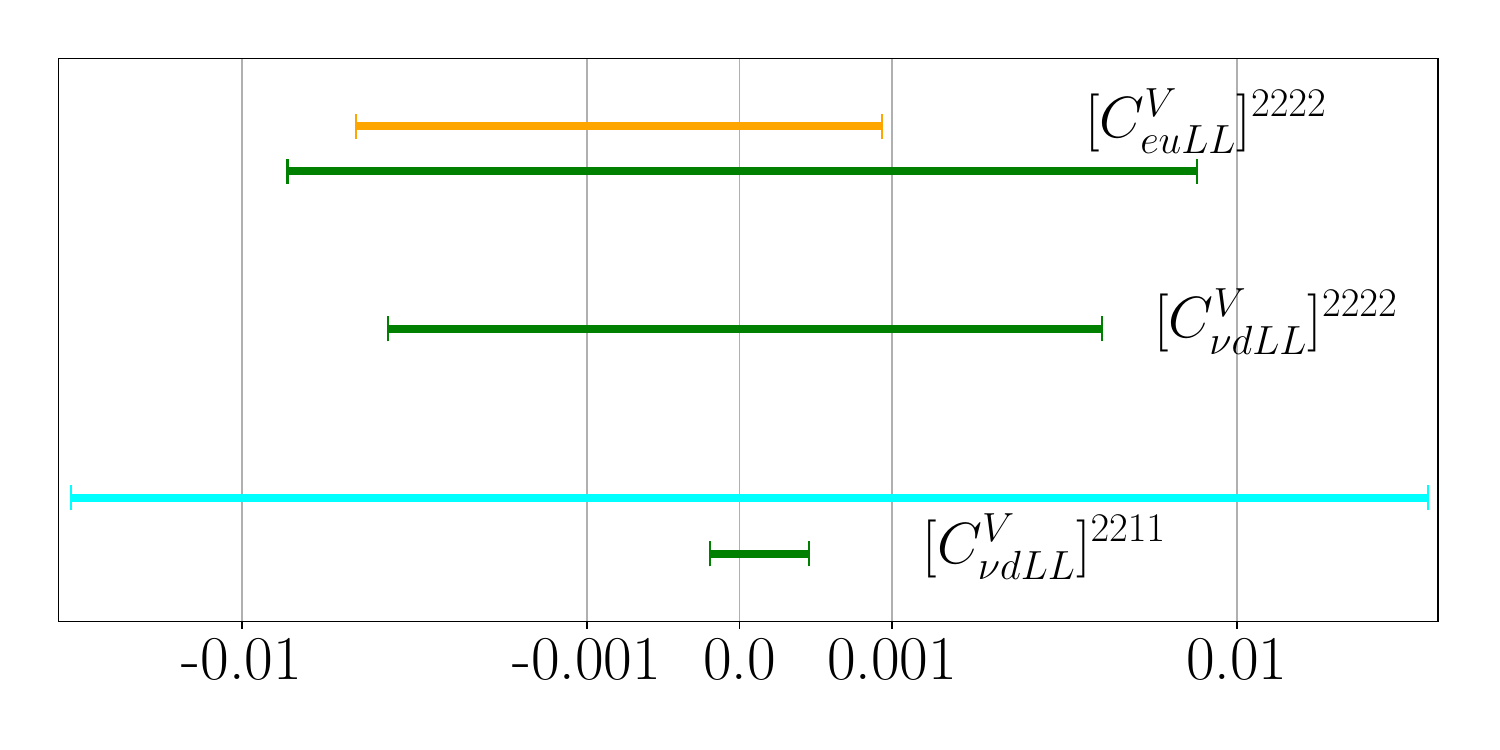}
    \end{minipage}
    \caption{Direct bounds from low-energy (cyan) and high-$p_T$ (orange) processes, along with the indirect (green) bounds on the complex WCs $[{\tempC}_{e u LL}^{V}]^{2 2 13}$ and $[{\tempC}_{e u LL}^{V}]^{2 2 2 3}$ and on the real WCs $[{\tempC}_{\nu d LL}^{V}]^{2 2 11}$, $[{\tempC}_{\nu d LL}^{V}]^{2 2 22}$ and $[{\tempC}_{e u LL}^{V}]^{2 2 22}$. The input parameters used are the four complex WCs $[{\tempC}_{\nu d LL}^{V}]^{2 2 12}$, $[{\tempC}_{\nu d LL}^{V}]^{2 2 13}$, $[{\tempC}_{\nu d LL}^{V}]^{2 2 23}$ and $[{\tempC}_{e u LL}^{V}]^{2 2 12}$ and one real WC $[{\tempC}_{e u LL}^{V}]^{2 2 11}$. Note that the bottom panel uses the symmetric log scale. See Appendix~\ref{app: numbers} for numerical values of the bounds.}
    \label{set1indirect}
\end{figure}

Now we turn to the indirect bounds obtained by using the SMEFT predictions. Counting the real and imaginary parts of the WCs separately, eq.\,(\ref{corLFV1}) involves a total of 18 parameters, connected by 9 linear relations. Our goal is to find indirect bounds on WCs that are weakly bound or have no direct bound, with the help of these relations. To this end, we first choose the 9 parameters which have the most stringent bounds:
\begin{align}
 &\textrm{Re}\left([{\tempC}_{\nu d LL}^V]^{2212}\right), ~\textrm{Im}\left([{\tempC}_{\nu d LL}^V]^{2212}\right),~  \textrm{Re}\left([{\tempC}_{\nu d LL}^V]^{2213}\right),~ \textrm{Im}\left([{\tempC}_{\nu d LL}^V]^{2213}\right),\nonumber\\
 &\textrm{Re}\left([{\tempC}_{\nu d LL}^V]^{2223}\right),
 \textrm{Im}\left([{\tempC}_{\nu d LL}^V]^{2223}\right),~ \textrm{Re}\left([{\tempC}_{e u LL}^V]^{2212}\right), ~\textrm{Im}\left([{\tempC}_{e u LL}^V]^{2212}\right)~, 
 \end{align}
 and the real WC $[{\tempC}_{e u LL}^V]^{2211}$. The remaining 9 parameters can then be written in terms of these using eq.\,(\ref{corLFV1}), and indirect bounds on them may be obtained.  In Fig.~\ref{set1indirect} we show the resultant indirect bounds on these parameters. For the complex WCs $[{\tempC}_{e u LL}^V]^{2213}$ and $[{\tempC}_{e u LL}^V]^{2223}$, the region of intersection between the indirect and the direct bounds can put a tighter constraint on the preferred values. These WCs correspond to single top production along with two leptons or top decays via $t\to c \ell \ell$ and $t\to u \ell \ell$ channels. It may be noticed that the constraints on the imaginary part of $[{\tempC}_{e u LL}^V]^{2223}$ are strong, making $[{\tempC}_{e u LL}^V]^{2223}$ appear almost as a real WC. This feature may be understood as follows. Eq\,(\ref{corLFV1}) implies
  \begin{align}
 [{\tempC}_{\nu d LL}^V]^{2233} = |V_{tb}|^2 [{\tempC}_{e u LL}^V]^{2233} +   V_{cb}^*\,V_{tb} \,[{\tempC}_{e u LL}^V]^{2223} + \mathcal{O}(\lambda^3)~.
 \end{align}
Here $\lambda = \sin(\theta_c)$ where $\theta_c\sim 0.227$ is the Cabbibo angle. Since $[{\tempC}_{e u LL}^V]^{2233}$ and $[{\tempC}_{\nu d LL}^V]^{2233}$ are real and $V_{cb}^*V_{tb}$ is real up to $\mathcal{O}(\lambda^3)$, the only imaginary quantity appearing in this equation is $\textrm{Im}([{\tempC}_{e u LL}^V]^{2223})$; hence it is strongly constrained. 
 
As far as the real WCs are concerned, for $[{\tempC}_{\nu d LL}^V]^{2211}$ we get a better constraint than the available direct bound which may be tested in experiments studying matter effects on neutrino oscillations. At the same time, $[{\tempC}_{\nu d LL}^V]^{2222}$, which has no direct bound, now gets bounded. For $[{\tempC}_{e u LL}^V]^{2222}$, the indirect bound is slightly worse than the direct bound.
For the other two, viz. $[{\tempC}_{\nu dLL}^V]^{2233}$ and $[{\tempC}_{\nu d LL}^V]^{2233}$, the indirect bounds are much worse than the direct bounds. 

Similar relations have been explored in literature in order to put indirect bounds on various EFT coefficients, albeit for a smaller subset of WCs, with some UV flavor assumptions, or by neglecting CKM elements. In \cite{Bause:2020auq, Bause:2021cna, Bause:2021ihn}, similar bounds have been calculated assuming the WCs to be real and neglecting terms in eqs.\,(\ref{corLFV1}-\ref{corLFV3}) having CKM elements that are higher order in $\lambda$. The indirect bounds obtained on the real WCs in \cite{Bause:2020auq}  become weaker when all the CKM matrix elements are inserted.

Note that our choice of the 9 input parameters need not have been the best one for finding the best indirect bounds on any parameter. A different set of 9 input parameters could be optimum. Indeed the best bounds may be obtained by using all the available direct bounds in a combined fit. Since the primary aim of this paper is to illustrate the utility of the linear relations in obtaining indirect bounds, we leave the detailed analysis for future work. 

\begin{figure}[t]
    \centering
    \begin{minipage}{0.45\textwidth}
        \includegraphics[width=\textwidth]{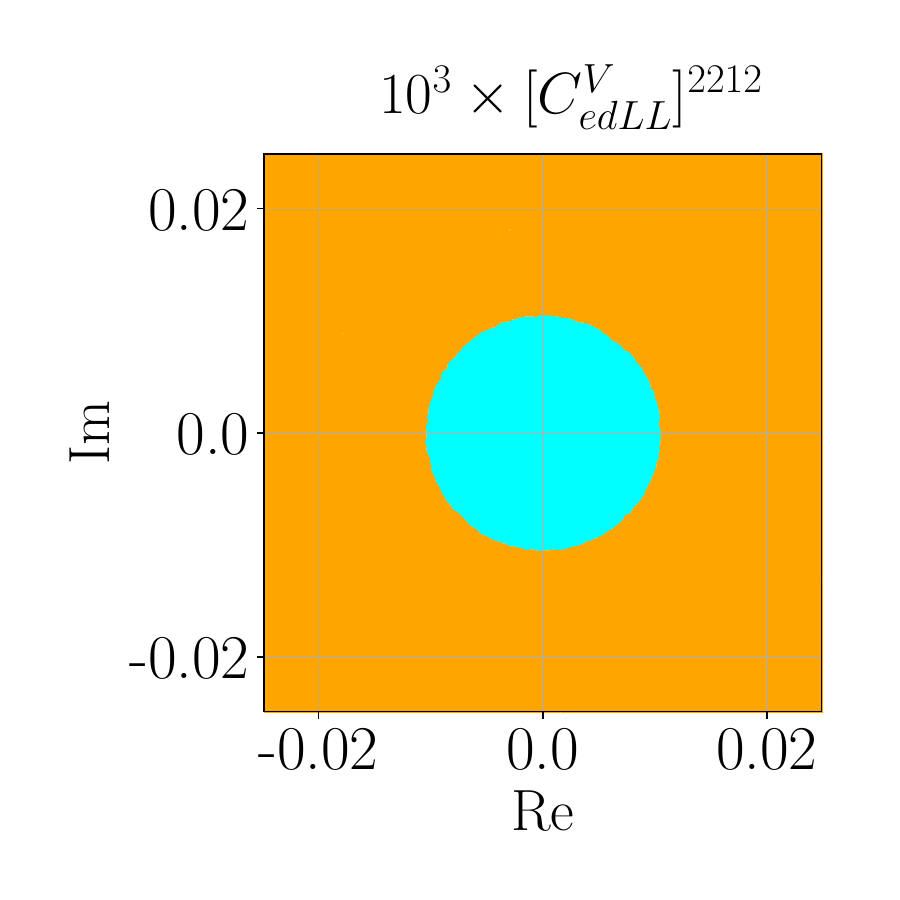}
    \end{minipage}
    \begin{minipage}{0.45\textwidth}
        \includegraphics[width=\textwidth]{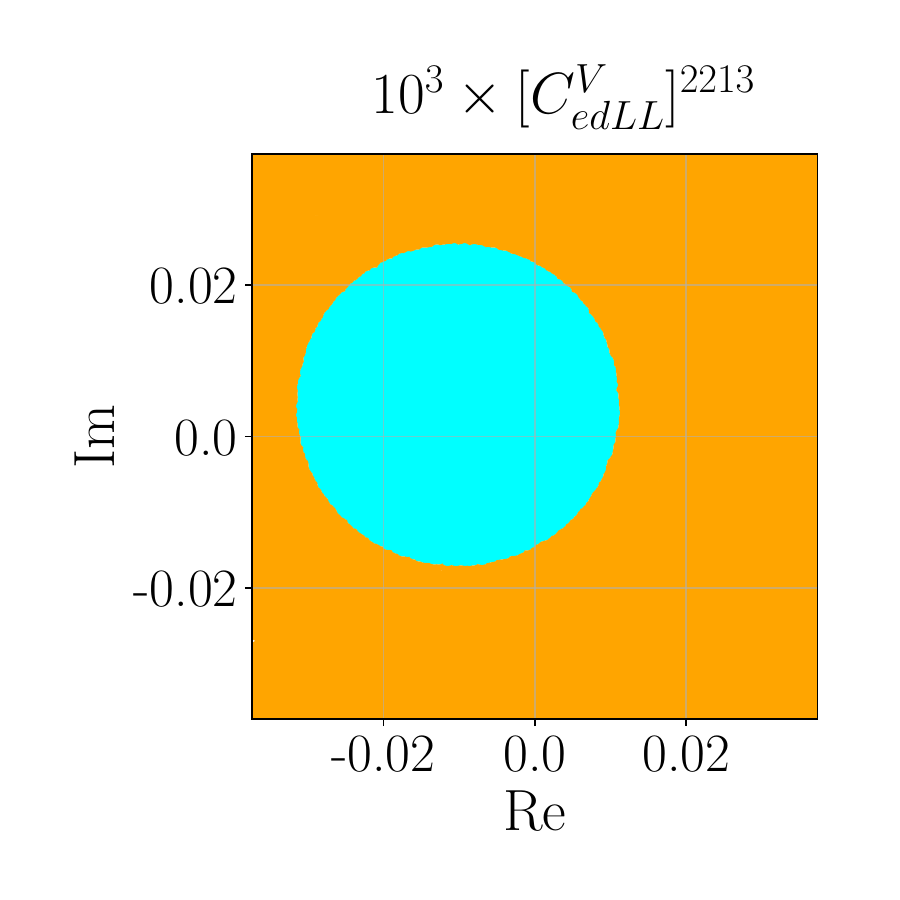}
    \end{minipage}
    \begin{minipage}{0.45\textwidth}
        \includegraphics[width=\textwidth]{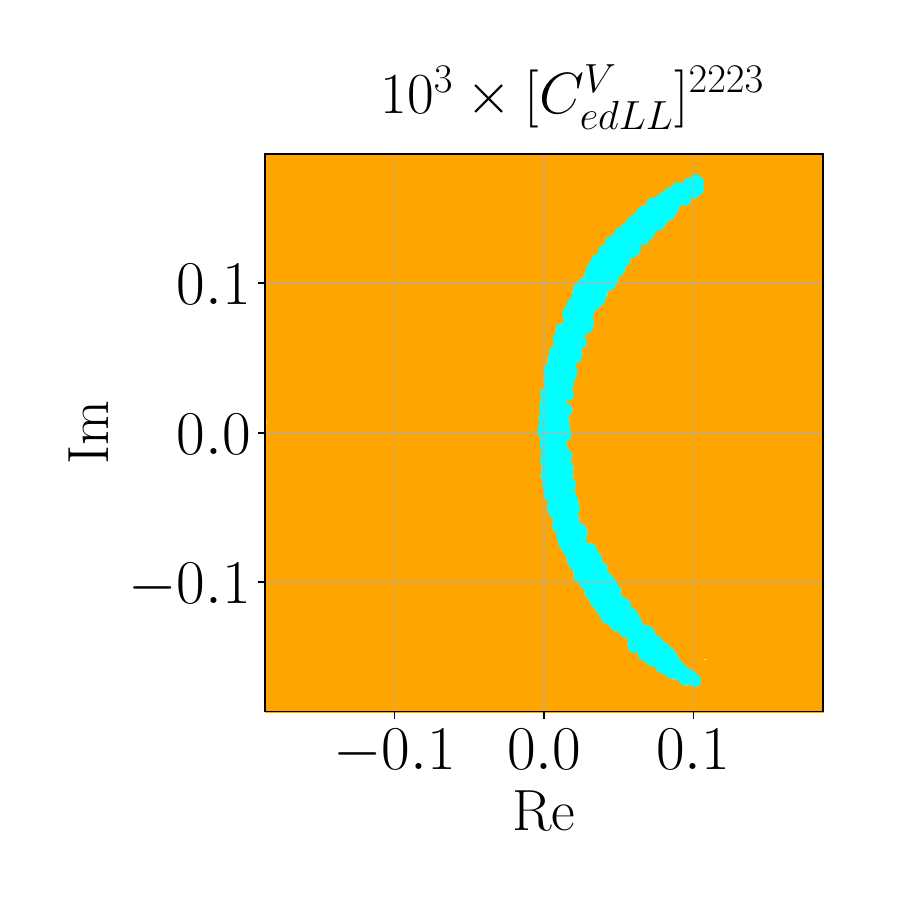}
    \end{minipage}
    \begin{minipage}{0.45\textwidth}
        \hspace{1.5cm}\includegraphics[width=.78\textwidth]{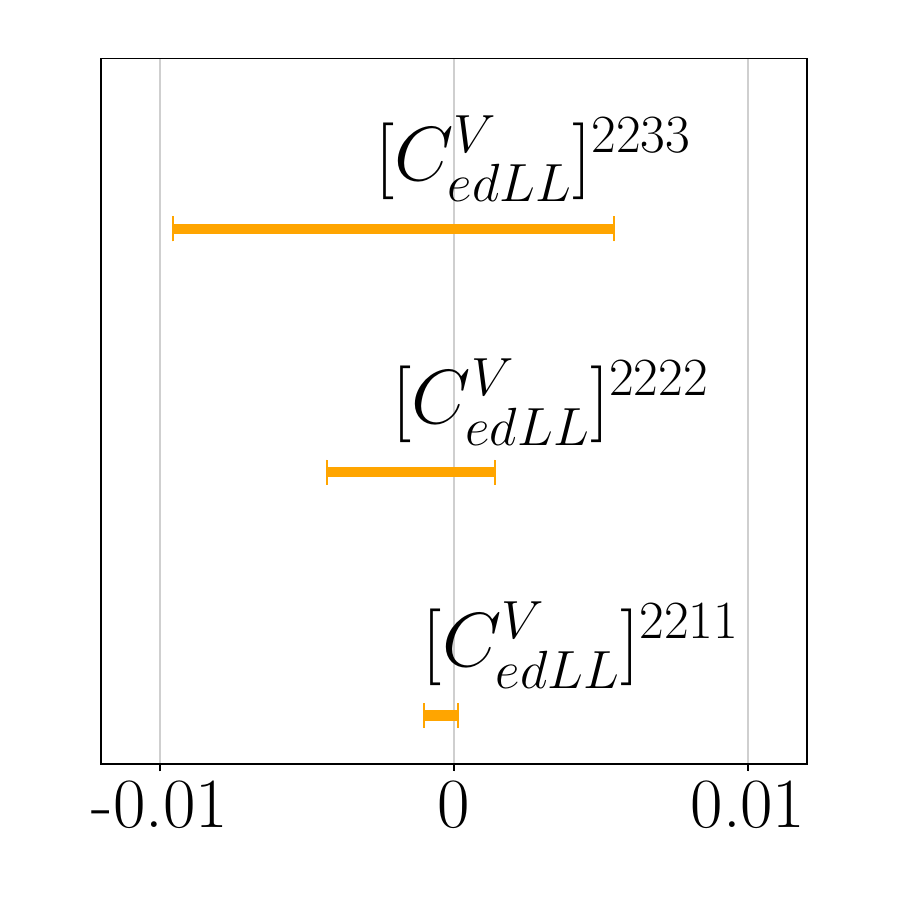}
    \end{minipage}
    \caption{The top panels and the bottom left panel show direct bounds from meson decays (cyan) for the complex WCs $[{\tempC}_{e d LL}^{V}]^{2212}$, $[{\tempC}_{e d LL}^{V}]^{2213}$, $[{\tempC}_{e d LL}^{V}]^{2223}$.  The orange background in these three panels indicates that the parameter space of these complex WCs displayed in this figure is allowed by the high-$p_T$ dimuon searches, and only constrained by meson decays. The bottom right panel shows the constraints from high-$p_T$ dimuon searches (orange) on the real WCs $[{\tempC}_{e d LL}^{V}]^{2211}$, $[{\tempC}_{e d LL}^{V}]^{2222}$ and $[{\tempC}_{e d LL}^{V}]^{2233}$. Note that the bottom-right panel uses the symmetric log scale. See Appendix~\ref{app: numbers} for numerical values of the bounds.}
    \label{set2direct}
\end{figure}

\subsection{Bounds on neutral-current WCs involving $(ed)$ and $(\nu u)$}\label{sec: indNC2}
In this section, we perform a similar analysis as in Sec.~\ref{sec: indNC1} for neutral-current WCs involving the muon family, using the relation in eq.\,(\ref{corLFV2}). The WCs involved correspond to the operators containing either charged leptons and down-type quarks $(edLL)$, or neutrinos and up-type quarks $(\nu uLL)$. 

The bounds on $(edLL)$ WCs are typically stronger since they involve charged muon. 
The WCs $[{\tempC}_{ed LL}^{V}]^{2 2 1 2}$, $[{\tempC}_{ed LL}^{V}]^{2 2 1 3}$ and $[{\tempC}_{ed LL}^{V}]^{2 2 2 3}$ get direct bounds from rare decays of $K$ and $B$ mesons.  Bound on the absolute value of $[{\tempC}_{ed LL}^{V}]^{2 2 1 2}$ is provided in \cite{Bause:2020auq}; we convert this to bounds on the real and the imaginary parts of this WC by taking into account all possible values for its phase. For $[{\tempC}_{ed LL}^{V}]^{2 2 1 3}$, we obtain the bound from the branching ratio measurement of $B^0\to \mu^+\,\mu^-$~\cite{ParticleDataGroup:2022pth}. For the real and the imaginary parts of $[{\tempC}_{ed LL}^{V}]^{2 2 2 3}$, we use a combined fit to the observables $\mathcal{B}(B^{(+,0)}\to K^{(+, 0)}\,\mu^+\,\mu^-)$, $\mathcal{B}(B^{(+,0)}\to K^{*\,(+, 0)}\,\mu^+\,\mu^-)$, $R_{K^{(*)}}$, $\mathcal{B}(B_s\to\mu^+\,\mu^-)$, as well as the angular observables $P_5^\prime$ and $F_L$ in $B^0\to K^{*0}\,\mu^+\,\mu^-$. The high-$p_T$ dimuon searches give bounds  on the three real WCs $[{\tempC}_{ed LL}^{V}]^{2 2 1 1}$, $[{\tempC}_{ed LL}^{V}]^{2 2 2 2}$ and $[{\tempC}_{ed LL}^{V}]^{2 2 3 3}$. We show these bounds in Fig.~\ref{set2direct}. 
Among the $(\nu uLL)$ WCs, only a weak bound is available on $[{\tempC}_{\nu u LL}^{V}]^{2211}$ from constraints on non-standard interactions of neutrinos in atmospheric neutrino experiments~\cite{Farzan:2017xzy}. Once again, while these bounds are on the individual contributions of the respective operators, inclusion of other operators would not significantly alter them given the assumptions stated in Sec.~\ref{sec: indNC1}.

\begin{figure}[t]
    \centering
    \begin{minipage}{0.48\textwidth}
        \includegraphics[width=\textwidth]{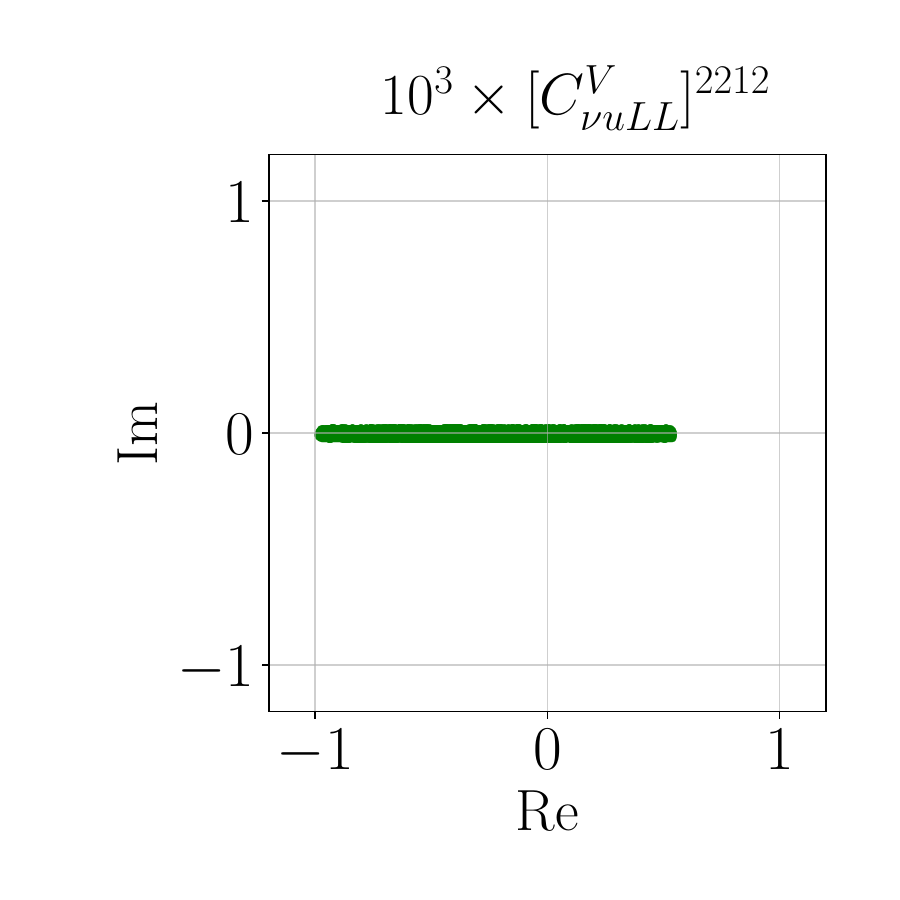}
    \end{minipage}
    \begin{minipage}{0.48\textwidth}
        \includegraphics[width=\textwidth]{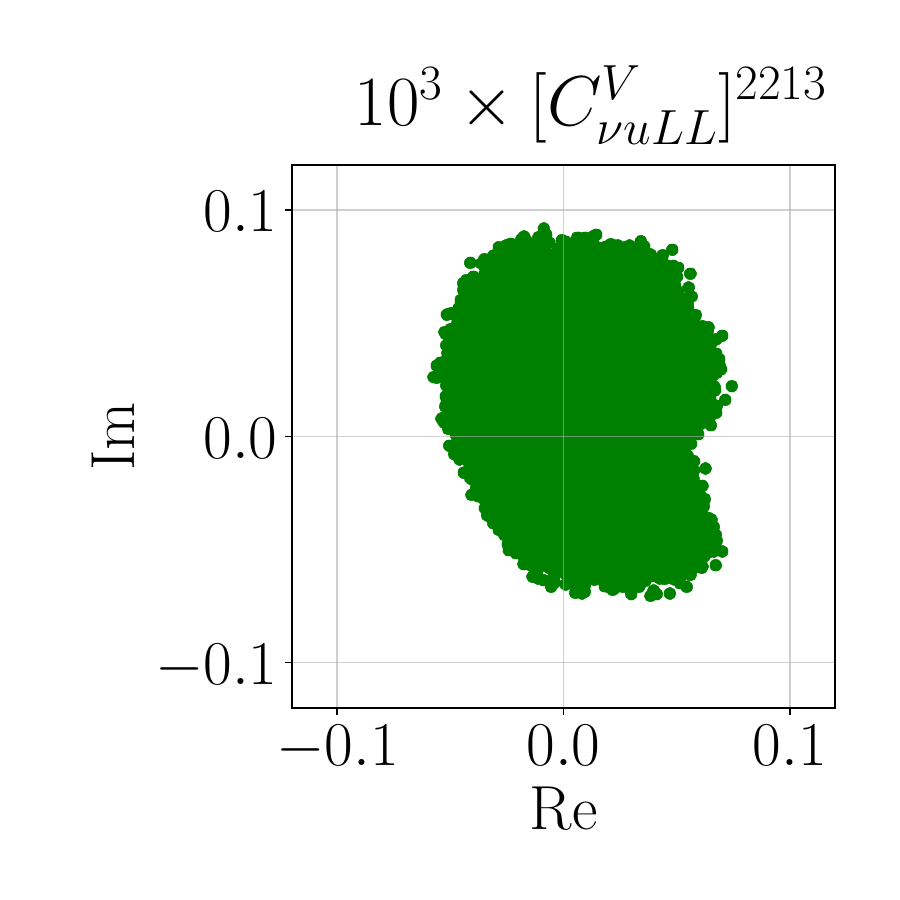}
    \end{minipage}
    \begin{minipage}{0.48\textwidth}
        \includegraphics[width=\textwidth]{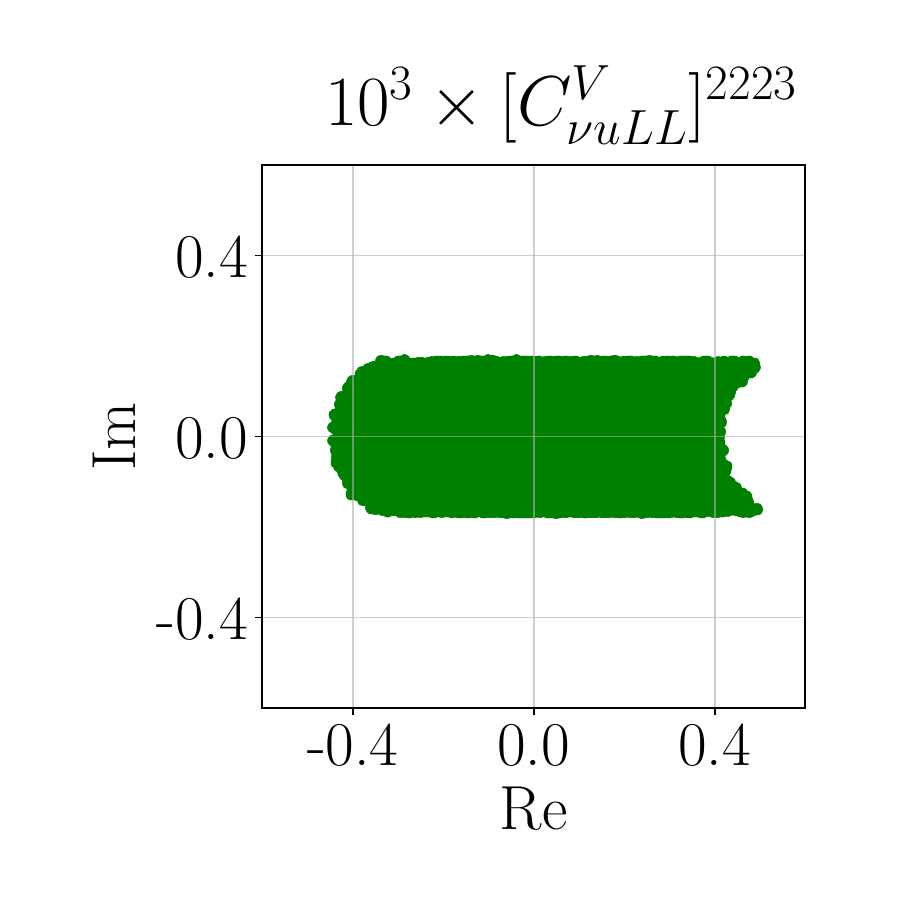}
    \end{minipage}
    \begin{minipage}{0.48\textwidth}
        \hspace{1.4cm}\includegraphics[width=.8\textwidth]{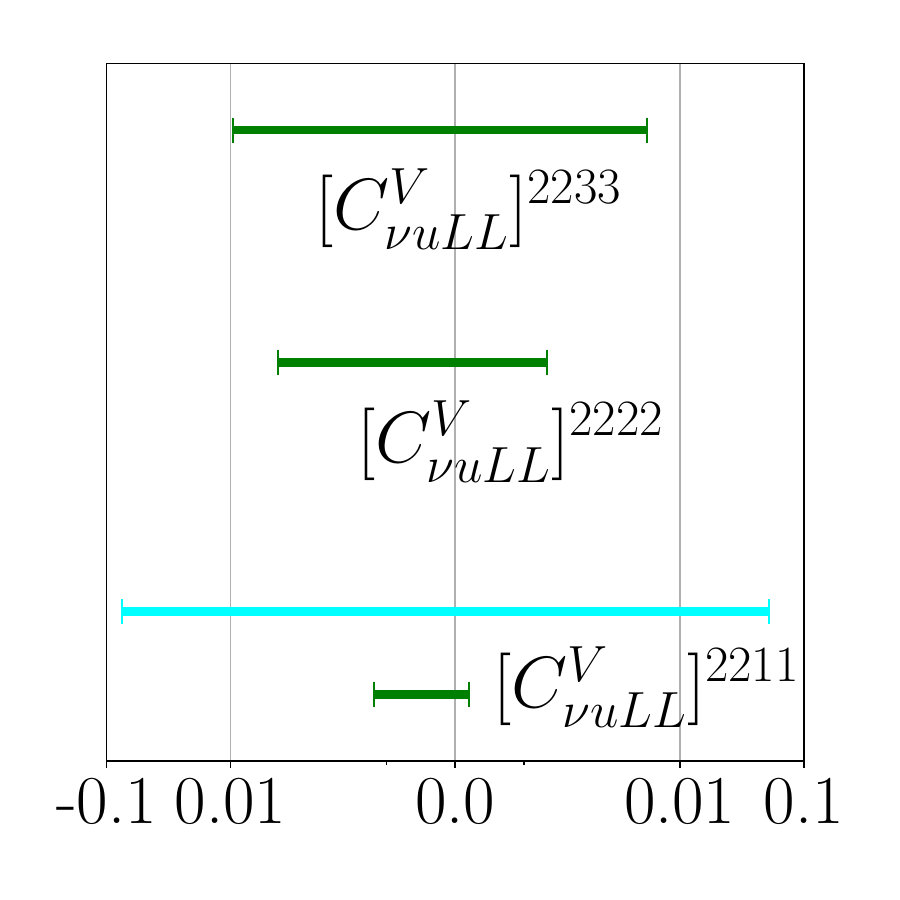}
    \end{minipage}
    \caption{Indirect bounds (green) on the complex WCs $[{\tempC}_{\nu u LL}^{V}]^{2212}$, $[{\tempC}_{\nu u LL}^{V}]^{2213}$, $[{\tempC}_{\nu u LL}^{V}]^{2223}$, and the real WCs $[{\tempC}_{\nu u LL}^{V}]^{2211}$, $[{\tempC}_{\nu u LL}^{V}]^{2222}$, $[{\tempC}_{\nu u LL}^{V}]^{2233}$. The direct bound is available only for the real WC $[{\tempC}_{\nu u LL}^{V}]^{2211}$ (shown in cyan). Note that the bottom-right panel uses the symmetric log scale. See Appendix~\ref{app: numbers} for numerical values of the bounds.}
    \label{set2indirect}
\end{figure}

Counting the real and imaginary parts of the WCs separately, eq.\,(\ref{corLFV1}) involves a total of 18 parameters, connected by 9 linear relations.
In order to get stronger bounds on the $(\nu uLL)$ WCs, we take the 9 parameters corresponding to the $(edLL)$ WCs as inputs and derive the indirect bounds using this relation. These bounds have been shown in Fig.~\ref{set2indirect}. 
It can be seen that the complex WCs $[{\tempC}_{\nu u LL}^V]^{2212}$, $[{\tempC}_{\nu u LL}^V]^{2213}$, $[{\tempC}_{\nu u LL}^V]^{2223}$ and the real WCs $[{\tempC}_{\nu u LL}^V]^{2222}$ and $[{\tempC}_{\nu u LL}^V]^{2233}$, which do not have any direct bounds, get indirect constraints. The first among these WCs would contribute to the invisible decay widths of $D$ mesons while the next two would contribute to the semileptonic top decays,  $t\to u \nu\nu$ and $t \to c \nu\nu$. 
The indirect bound also improves the constraints on $[{\tempC}_{\nu u LL}^{V}]^{2211}$ significantly. This indirect bound would be important for constraining models with neutrino non-standard interactions (NSI)~\cite{Farzan:2017xzy} and can be tested in precision neutrino oscillation experiments.
 
Note that the indirect constraints suggest that $[{\tempC}_{\nu u LL}^V]^{2212}$ is almost real. This can be understood by looking at the  leading-order contributions to $[{\tempC}_{\nu u LL}^V]^{2212}$ in eq.\,(\ref{corLFV2}):
\begin{align}
    [{\tempC}_{\nu u LL}^V]^{2212} &= V_{ud}\,V_{cs}^* [{\tempC}_{ed LL}^V]^{2212} + V_{ud}\,V_{cd}^* [{\tempC}_{ed LL}^V]^{2211} + V_{us}\,V_{cs}^*\,[{\tempC}_{ed LL}^V]^{2222}\nonumber\\
    &~+  V_{us}\,V_{cd}^*\,[{\tempC}_{ed LL}^{V\,*}]^{2212}+\mathcal{O}(\lambda^3)~.
\end{align}
In the above equation, all the CKM coefficients are real up to ${\cal O}(\lambda^3)$. The WCs $[{\tempC}_{e d LL}^V]^{2211}$ and $[{\tempC}_{e d LL}^V]^{2222}$ are real, while $\textrm{Im}([{\tempC}_{\nu u LL}^V]^{2212})$ has strong constraints of $\mathcal{O}(0.02)$. Therefore, the  imaginary part of the left-hand side, i.e. $\textrm{Im}\left([{\tempC}_{\nu u LL}^V]^{2212}\right)$, is strongly constrained. 

\subsection{Bounds on charged-current WCs}  \label{sec: indCC}

Eq.\,(\ref{corLFV3}) allows us to express charged-current WCs as combinations of neutral-current WCs. Restricting to the muon family of lepton, i.e $\alpha=\beta=2$, there are 9 charged-current WCs on the left-hand side of eq.\,(\ref{corLFV3}); all of them can be complex in general. All these charged-current WCs would get indirectly constrained due to the bounds on the neutral-current WCs. In this section, we first show the direct bounds for the 9 charged-current WCs from mesonic decays and from high-$p_T$ monolepton searches. Later, we compare these bounds with the ones derived indirectly using eq.\,(\ref{corLFV3}).

\begin{figure}[t!]
    \centering
    \includegraphics[width=\textwidth]{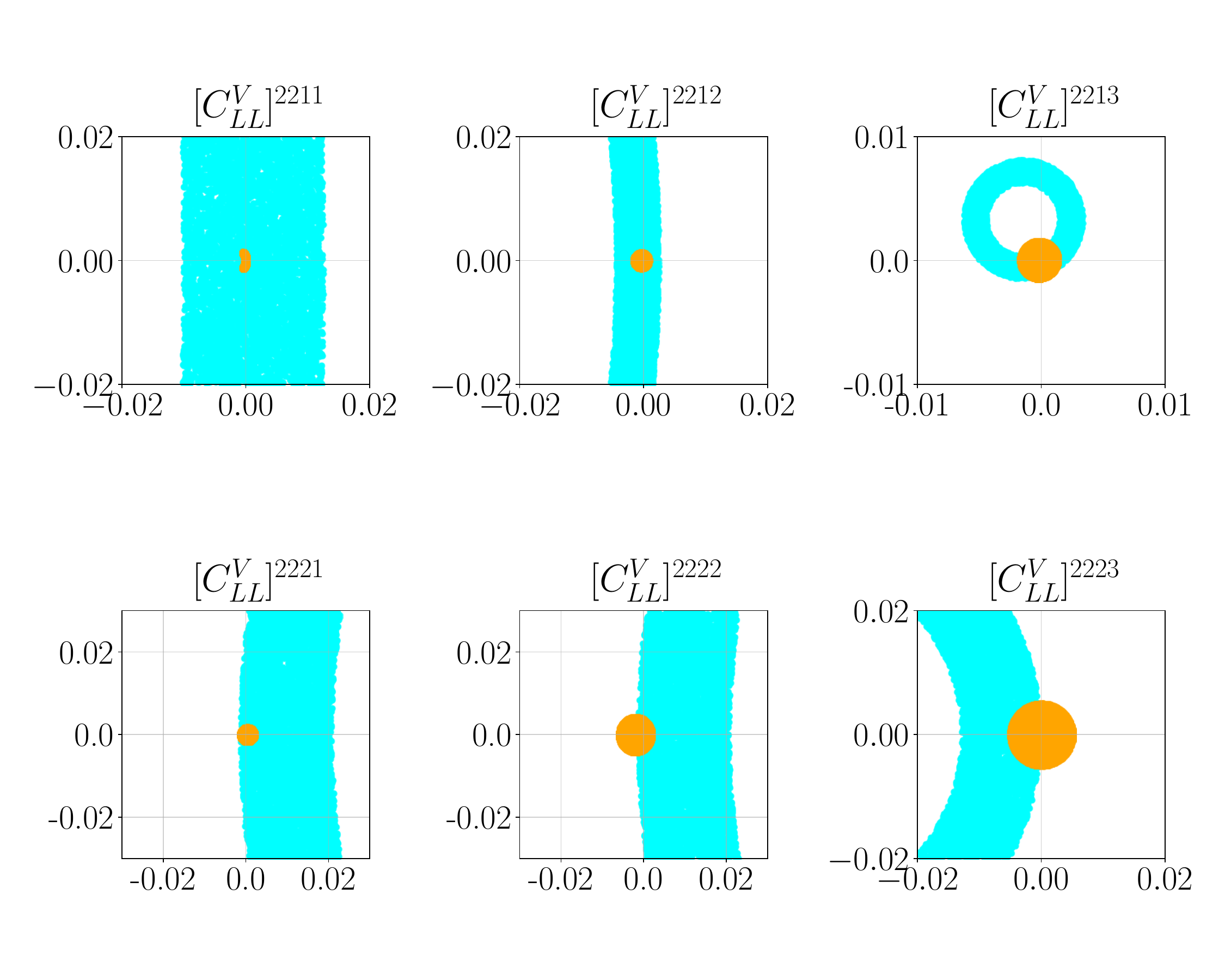}
    \caption{Direct bounds on the charged-current WCs from meson decays (cyan) and from high-$p_T$ mono-muon searches (orange). Note that there are no direct constraints on the WCs associated with charged current decays of top quark. See Appendix~\ref{app: numbers} for numerical values of the bounds.}
    \label{set3direct}
\end{figure}

For the WCs $[{\tempC}_{LL}^V]^{2211}$, $[{\tempC}_{LL}^V]^{2212}$, $[{\tempC}_{LL}^V]^{2213}$, $[{\tempC}_{LL}^V]^{2221}$, $[{\tempC}_{LL}^V]^{2222}$ and $[{\tempC}_{LL}^V]^{2223}$, we obtain direct bounds using 
the branching ratios~\cite{ParticleDataGroup:2022pth} of the decay modes $\pi^+\to \mu^+\nu$, $K^+\to\pi\mu^+\nu$, $B^+\to \pi^0l\nu$
$D^+\to \mu^+\nu$, $D_s\to \mu\nu$ and $B^+\to D l\nu$, respectively.  However, stronger bounds can be obtained for these WCs from high-$p_T$ monolepton searches. In order to do this\footnote{In \cite{Allwicher:2022gkm}, bounds on SMEFT coefficients are provided using high-$p_T$ single lepton and dilepton searches. 
However, no combination of SMEFT coefficients can map to a single charged-current LEFT coefficient without generating other LEFT coefficient that can contribute to the same single charged-lepton final state mode. Therefore we calculate these bounds independently}, we generate bin-wise events in MadGraph~\cite{Alwall:2014hca}. Note that the charged-current NP would not change the shape of the $q^2$ dependence from the SM prediction, since the relevant charged-current operators in SM and NP are identical.  We use the results from the ATLAS analysis in Ref.~\cite{ATLAS:2019lsy}, and incorporate the effect of their cuts by using a re-scaling factor on our generated events such that they reproduce the ATLAS data for SM. We then perform a $\chi^2$ fit for the isolated charged-current WC to obtain bound on the NP WC. These direct bounds obtained from the meson decays (cyan) and from the high-$p_T$ mono-muon searches (orange) are shown in Fig.~\ref{set3direct}. 

\begin{figure}[t]
    \centering
    \includegraphics[width=\textwidth]{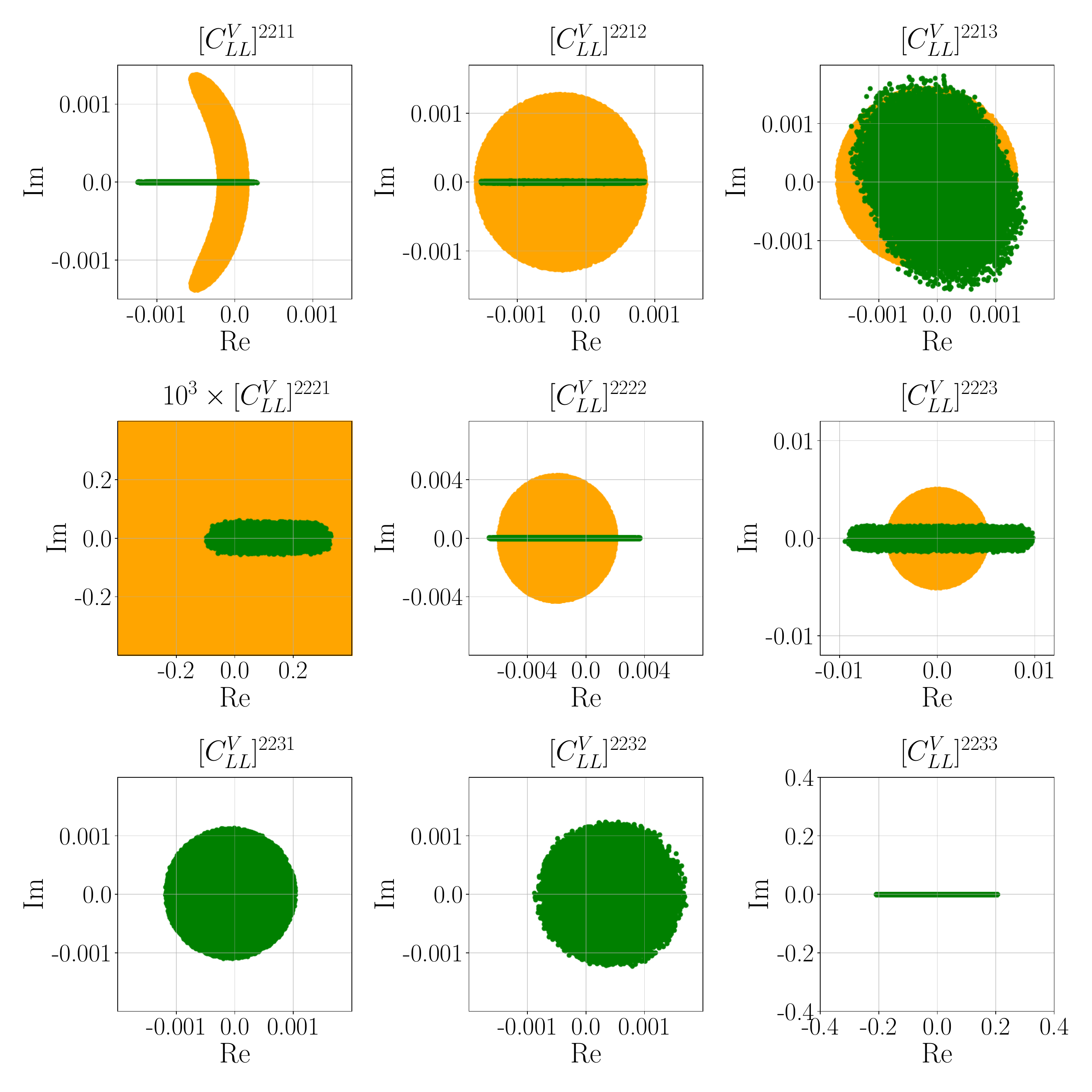}
    \caption{Direct bounds (orange) on the charged-current WCs from high-$p_T$ mono-muon searches along with the indirect bounds (green) obtained using eq.\,(\ref{corLFV3}). Note that the quantities in the bottom panels have no direct bounds. See Appendix~\ref{app: numbers} for numerical values of the bounds.}
    \label{set3indirect}
\end{figure}

In order to obtain indirect bounds, we use the best available bounds (direct or indirect) for the neutral-current WCs appearing on the right-hand side of eq.\,(\ref{corLFV3}). These indirect bounds (green) along with the best available direct bounds (orange) are shown in Fig.~\ref{set3indirect}. The figure shows that
this method provides constraints on $[{\tempC}_{LL}^V]^{2231}$, $[{\tempC}_{LL}^V]^{2232}$ and $[{\tempC}_{LL}^V]^{2233}$, where no direct bounds were available.  For $[{\tempC}_{LL}^V]^{2221}$, the indirect constraints are significantly stronger than the direct bounds.  These WCs would contribute to branching ratios of semileptonic decays of top quark and D meson decays, viz.  $D\to \pi \mu \nu$, etc.

In addition, the imaginary parts of $[{\tempC}_{LL}^V]^{2211}$, $[{\tempC}_{LL}^V]^{2212}$, $[{\tempC}_{LL}^V]^{2222}$ and $[{\tempC}_{LL}^V]^{2223}$ are constrained more strongly. These WCs would contribute to branching ratios of meson decays, viz. $K\to \pi \mu\nu$, $B\to D \mu \nu$, etc. 
The reason for strong indirect constraints on the imaginary parts of these four WCs may be understood using eq.\,(\ref{corLFV3}). For example, $[{\tempC}_{LL}^V]^{2211}$ may be written using eq.\,(\ref{corLFV3}) as
\begin{align}
    [{\tempC}_{LL}^V]^{2211} &= V_{ud}\, ([{\tempC}_{ed LL}^V]^{2211}-[{\tempC}_{\nu d LL}^V]^{2211}) + V_{us}\,([{\tempC}_{ed LL}^V]^{2212}-[{\tempC}_{\nu d LL}^V]^{2212}) + \mathcal{O}(\lambda^3)~.
\end{align}
The CKM coefficients appearing on the right-hand side of the above equation are real up to $\mathcal{O}(\lambda^3)$. The WCs $[{\tempC}_{ed LL}^V]^{2211}$ and $[{\tempC}_{\nu dLL}^V]^{2211}$ are real. Furthermore, the bounds on the imaginary parts of $[{\tempC}_{ed LL}^V]^{2212}$ and $[{\tempC}_{\nu d LL}^V]^{2212}$ are of $\mathcal{O}(0.02)$ and these WCs appear with a CKM coefficient of $\mathcal{O}(\lambda)$. Thus, the imaginary part of the  WC on the left-hand side, i.e. $[{\tempC}_{LL}^V]^{2211}$, is strongly constrained. Using similar arguments, we can show that the WCs $[{\tempC}_{LL}^V]^{2212}$, $[{\tempC}_{LL}^V]^{2222}$ and $[{\tempC}_{LL}^V]^{2233}$ are expected to be dominantly real.

\subsection{Predictions for lepton flavor violating observables}\label{sec: LFV}

So far we have considered the relations only among WCs involving one lepton family i.e. muon. In this subsection, we expand our discussion to SMEFT predictions that include all lepton families, while remaining in the UV4f scenario. These relations will relate diverse reaction channels like rare decays of $B$, $D$ and $K$ mesons as well as lepton flavor violating (LFV) processes such as $\tau\rightarrow {\ell} \,q_i \,q_j$ and ${\ell}\, N \to{\ell}^\prime \, N$. 
Focusing once again on UV4f models, we shall indicate the methodology by one example and present a set of relevant processes in Table~\ref{tab: implications}.

\begin{table}[t]
    \centering
    \renewcommand{\cellalign}{cc}
    \renewcommand{\theadalign}{cc}
    \begin{tabular}{|c|c|p{2cm}|p{2cm}|c|}
        \hline
        Eq. $\downarrow$  & LHS WC & RHS WCs  & Transitions & Processes\\
        \hline
        \multirow{4}{*}{\rotatebox{0}{(\ref{corLFV1})}} & $[{\tempC}_{euLL}^{V}]^{{\ell} 3 1 1}$ & \makecell*{$[{\tempC}_{\nu d LL}^{V}]^{{\ell} 3 1 1}$ \\$[{\tempC}_{\nu d LL}^{V}]^{{\ell} 3 1 2}$} & \makecell[cc]{$\tau \to u\,u\,{\ell}$ \\ $s\to d \,\nu \,\nu$} & \makecell{LFV $\tau$ decay,\\ $K\to \pi \nu \nu$}\\
        \cline{2-5}
        & $[{\tempC}_{euLL}^{V}]^{{\ell}{\ell}^\prime 1 1}$ & \makecell*{$[{\tempC}_{\nu d LL}^{V}]^{{\ell}{\ell}^\prime 1 1}$ \\$[{\tempC}_{\nu d LL}^{V}]^{{\ell}{\ell}^\prime 1 2}$} & \makecell*{${\ell} \, u \to{\ell}^\prime \, u$ \\ $s\to d \,\nu \,\nu$} & \makecell*{LFV ${\ell}\,N\to{\ell}^\prime\, N$ \\$K\to \pi \nu \nu$}\\
        \cline{2-5}
        & $[{\tempC}_{euLL}^{V}]^{{\ell} 3 1 2}$ & $[{\tempC}_{\nu d LL}^{V}]^{{\ell} 3 1 2}$ & \makecell*{$c\, u \to \tau\,{\ell}$ \\ $s\to d \,\nu \,\nu$} & \makecell*{LFV $D$ decays, \\ $K\to \pi \nu \nu$~\cite{Bause:2020auq}}\\
        \cline{2-5}
        & $[{\tempC}_{euLL}^{V}]^{{\ell}{\ell}^\prime i 3}$ & \makecell*{$[{\tempC}_{\nu d LL}^{V}]^{{\ell}{\ell}^\prime 1 3}$ \\ $[{\tempC}_{\nu d LL}^{V}]^{{\ell}{\ell}^\prime 2 3}$} & \makecell*{$t\to u_{i}\, {\ell}\,{\ell}^\prime$\\ $b\to d \,\nu \,\nu$ \\$b\to s \,\nu \,\nu$} & \makecell*{LFV top decay, \\$B$ decays to dineutrinos~\cite{Bause:2020auq}}\\
        \hline
        \multirow{3}{*}{\rotatebox{0}{(\ref{corLFV2})}} & $[{\tempC}_{e d LL}^{V}]^{{\ell} 3 i j}$ & $[{\tempC}_{\nu u LL}^{V}]^{{\ell} 3 1 2}$ & \makecell*{$\tau \to d\,d\,{\ell}$\\ $c\to u\,\nu\,\nu$} & \makecell*{LFV $\tau$ decay, \\$D$ decay to dineutrinos}\\
        \cline{2-5}
        & $[{\tempC}_{e d LL}^{V}]^{{\ell}{\ell}^\prime i j}$ & $[{\tempC}_{\nu u LL}^{V}]^{{\ell}{\ell}^\prime 1 2}$ & \makecell*{ ${\ell}\,d \to{\ell}^\prime\, d$ \\ $c\to u\,\nu\,\nu$} &\makecell*{ LFV ${\ell}\, N\to{\ell}^\prime \,N$ \\$D$ decay to dineutrinos}\\
        \cline{2-5}
        & $[{\tempC}_{edLL}^{V}]^{{\ell}{\ell}^\prime i 3}$ & \makecell*{$[{\tempC}_{\nu u LL}^{V}]^{{\ell}{\ell}^\prime 1 3}$ \\ $[{\tempC}_{\nu u LL}^{V}]^{{\ell}{\ell}^\prime 2 3}$} & \makecell*{$b\to d_{i}\, {\ell}\,{\ell}^\prime$\\ $t\to u \,\nu \,\nu$ \\$t\to c \,\nu \,\nu$} & \makecell*{LFV B decay, \\ top decays to dineutrinos}\\
        \hline
        \multirow{3}{*}{\rotatebox{0}{(\ref{corLFV3})}} & $[{\tempC}_{LL}^{V}]^{{\ell} 3 1 1}$ & \makecell*{$[{\tempC}_{e d LL}^{V}]^{{\ell} 3 1 1}$ \\$[{\tempC}_{e d LL}^{V}]^{{\ell} 3 1 2}$\\ $[{\tempC}_{\nu d LL}^{V}]^{{\ell} 3 1 1}$} & \makecell*{$\tau \to u\,d\,\nu$ \\$\tau \to d\,d\,{\ell}$\\ $\tau \to d\,s\,{\ell}$ \\ $ s\to d \nu\,\nu$} & \makecell*{CC decay of $\tau$ \\ LFV $\tau$ decay, \\ $K\to \pi\, \nu\,\nu$}\\
        \cline{2-5}
        & $[{\tempC}_{LL}^{V}]^{{\ell}{\ell}^\prime 1 1}$ & \makecell*{$[{\tempC}_{e d LL}^{V}]^{{\ell}{\ell}^\prime 1 1}$ \\ $[{\tempC}_{e d LL}^{V}]^{{\ell}{\ell}^\prime 1 2}$\\ $[{\tempC}_{\nu d LL}^{V}]^{{\ell}{\ell}^\prime 1 1}$} & \makecell*{${\ell} \to u\,d\,\nu$ \\${\ell} \, d \to{\ell}^\prime\,d$\\ $s \to d\,{\ell}\,{\ell}^\prime$ \\  $ s\to d \nu\,\nu$} & \makecell*{ LFV ${\ell}\, N\to{\ell}^\prime \,N$ \\ $K\to \pi \, {\ell}\,{\ell}^\prime$ \\ $K\to \pi\, \nu\,\nu$}\\
        \cline{2-5}
        & $[{\tempC}_{LL}^{V}]^{{\ell}{\ell}^\prime i 3}$ &  \makecell*{$[{\tempC}_{e d LL}^{V}]^{{\ell}{\ell}^\prime 1 3}$\\ $[{\tempC}_{e d LL}^{V}]^{{\ell}{\ell}^\prime 2 3}$\\ $[{\tempC}_{\nu d LL}^{V}]^{{\ell}{\ell}^\prime 1 3}$ \\$[{\tempC}_{\nu d LL}^{V}]^{{\ell}{\ell}^\prime 2 3}$} & \makecell*{$b\to u_i \,{\ell}\, \nu$ \\$b\to d_i\,{\ell}\,{\ell}^\prime$ \\$b\to d_i\,\nu\,\nu$} &\makecell*{ CC decay of $B$ meson,\\ LFV $B$ decays, \\$B$ decays to dineutrinos~\cite{Bause:2021ihn}}\\
        \hline
    \end{tabular}
    \caption{Correlations among different WCs involving all lepton families, derived from eqs.\,(\ref{corLFV1}-\ref{corLFV3}). The second column shows the WC appearing on the left hand side of these equations, whereas the third column contains the WCs appearing on the right-hand side of those equations with large CKM coefficients, with values   $\mathcal{O}(\lambda)$ or more. }
    \label{tab: implications}
\end{table}

From eq.\,(\ref{corLFV1}), we get the following relation among the LEFT WCs:
\begin{align}
    [{\tempC}_{euLL}^{V}]^{{\ell} 3 1 1} &= |V_{ud}|^2 \,[{\tempC}_{\nu d LL}^{V}]^{{\ell} 3 1 1} + (V_{ud}^*\,V_{cd} \,[{\tempC}_{\nu d LL}^{V}]^{{\ell} 3 1 2} + \mbox{c.c.}) + (V_{ud}^*\,V_{td} \,[{\tempC}_{\nu d LL}^{V}]^{{\ell} 3 1 3}+ \mbox{c.c.})\nonumber\\
    &~ + |V_{cd}|^2 \,[{\tempC}_{\nu d LL}^{V}]^{{\ell} 3 2 2} + (V_{cd}^*\,V_{td} \,[{\tempC}_{\nu d LL}^{V}]^{{\ell} 3 2 3}+ \mbox{c.c.}) + |V_{td}|^2 \,[{\tempC}_{\nu d LL}^{V}]^{{\ell} 3 3 3}~, \label{corLFV1.1}
\end{align}
where $\ell=1~{\rm or}~2$. Among the CKM coefficients in this equation, the leading ones are  $|V_{ud}|^2\sim \mathcal{O}(1)$ and $|V_{ud}^* V_{cd}|\sim \mathcal{O}(\lambda)$. All the other coefficients are $\mathcal{O}(\lambda^2)$ or smaller. Therefore, at the leading order, this equation connects the three WCs $[{\tempC}_{euLL}^{V}]^{{\ell} 3 1 1}$, $[{\tempC}_{\nu d LL}^{V}]^{{\ell} 3 1 1}$ and $[{\tempC}_{\nu d LL}^{V}]^{{\ell} 3 1 2}$. Hence the new physics WCs contributing to LFV  tau decays and $K\to \pi \, \nu\,\nu$ are related to each other.

Further relations involving other lepton and quark families are given in Table~\ref{tab: implications}. Some similar relations have been presented in \cite{Fuentes-Martin:2020lea}.  Note that such discussions in earlier literature often assume some flavor structure for the quark sector. We emphasize again that in our discussion, the implications presented in this section are independent of any NP flavor structure assumption for the quarks.

So far, we have discussed observables that are insensitive to the flavor of neutrinos. Neutrino experiments that are sensitive to neutrino flavor can probe the neutrino non-standard interactions (NSI) generated by the operators in Table~\ref{HEFToplist} containing neutrinos. The predictions in eq.\,(\ref{corLFV1})-eq.\,(\ref{corLFV3}) in UV4f models (or the more general predictions in Table~\ref{CorrTable}) would then imply constraints on NSI from charged LFV. We discuss this in more detail in Sec.~\ref{implibroken}.

\section{SMEFT-predicted evidence for new physics}  \label{sec: phenoevidence}

In this section, we discuss how to use the SMEFT predictions derived in Sec.~\ref{sec: SMEFTtoHEFT} in the event that measurements provide evidence for certain new physics WCs to be nonzero. We will show that, given the SMEFT predictions derived in this work, it is in general not consistent to assume a single non-zero WC to explain an excess in a certain channel.\footnote{For some operators, such as the $RRRR$ vector operators, there are no constraints implied by SMEFT. For this category of operators, therefore, we can have a single non-zero WC.}    In fact, we will show that for certain operators a   non-zero WC  must be accompanied by multiple other WCs that are non-vanishing. This would imply that the observed excess must be accompanied by correlated excesses in many other channels.
This is because the  SMEFT  predictions in Table~\ref{tab: implications} are linear equations involving multiple WCs, implying that it is not possible for only one of these coefficients to be nonzero. 

For example, consider the situation where an observed deviation from SM in a particular channel indicates that one of the $LLLL$ LEFT WCs  is non-vanishing.  In SMEFT, this LEFT coefficient might arise either from a four-fermion operator or an operator inducing an off-diagonal $W$ or $Z$ coupling to fermions.

The former situation is realized in the UV4f models, where
we can use SMEFT predictions in eq.\,(\ref{corLFV1}).
These are 6 linear equations
involving 12 (possibly) complex WCs when $\alpha =\beta$.
If one of these WCs, (say $C_1$) is found to be nonzero, we can write these equations in a form where $C_{7}$ to $C_{12}$ are expressed as linear combinations of $C_1$ to $C_6$. Then, as long as the coefficient of $C_1$ is nonzero in all these equations (as is generically observed to be the case), all the 6 coefficients $C_7$ to $C_{12}$ also have to be nonzero. For one of them to be vanishing, we will need one of the other coefficients, $C_2$ to $C_6$, to be nonzero in order to cancel the $C_1$ contribution. Thus, the nonvanishing nature of $C_1$ necessarily implies that overall at least 7 WCs are nonvanishing in principle. Of course, depending on the CKM coefficients, the magnitudes of these coefficients may be small or large.

When $\alpha\neq \beta$, eq.\,(\ref{corLFV1}) gives 9 linear equations, therefore one nonzero WC among these will imply at least 10 of the WCs of the type ($\nu d$) or ($eu$) nonvanishing. As eq.\,(\ref{corLFV2}) is completely decoupled from eq.\,(\ref{corLFV1}), it is of course still consistent for all the WCs appearing in it to vanish. The charged current WCs in eq.\,(\ref{corLFV3}), however, cannot all vanish and one can use similar arguments to conclude that at least 3 of them must be nonzero whether or not  $\alpha$ equals  $\beta$.

Similarly, from eq\,(\ref{corLFV2}), for $\alpha =\beta$ $(\alpha\neq\beta)$ we have 6 (9) linear relations. These imply that, if one of the WCs of the kind  ($ed$) or ($\nu u$) is found to be nonzero, then a total of at least 7 (10) WCs of these kinds should be nonzero in principle. Again, by eq.\,(\ref{corLFV3})  a non-zero neutral-current WC  will lead to at least 3 non-vanishing charged-current WCs. The CKM coefficients will guide us regarding which of these WCs are likely to have larger magnitudes. Thus, these relations direct us toward specific decay channels where deviation from SM is expected to be present. 

In the latter situation, i.e. when the LEFT operators arise from modifications of $W/Z$ couplings, the low-energy pattern of deviations is very different. For example, if one of the  $Z$ coupling to down quarks gets BSM corrections,  the penultimate row of Table~\ref{CorrTable} would imply at least three $W$-coupling modifications. Alternatively, if all
the $W$ couplings are to be at their SM value, this would 
imply modifications of at least 10 of the 18 $Z$ couplings to up and down-type quarks.   Once the  $W$ and $Z$ are integrated out, each   $W$-coupling modification will induce  3 non-vanishing semileptonic  LEFT WCs,  and each   $Z$-coupling modification will induce 6 non-vanishing semileptonic LEFT WCs.  Studying the pattern of BSM deviations can, therefore,  help pinpoint the underlying UV  physics. 
We shall not consider this scenario further in this section.

\subsection{Implications of the measured excess in $B\to K \nu \nu$}

\begin{figure}[t]
    \centering
    \includegraphics[width=\textwidth]{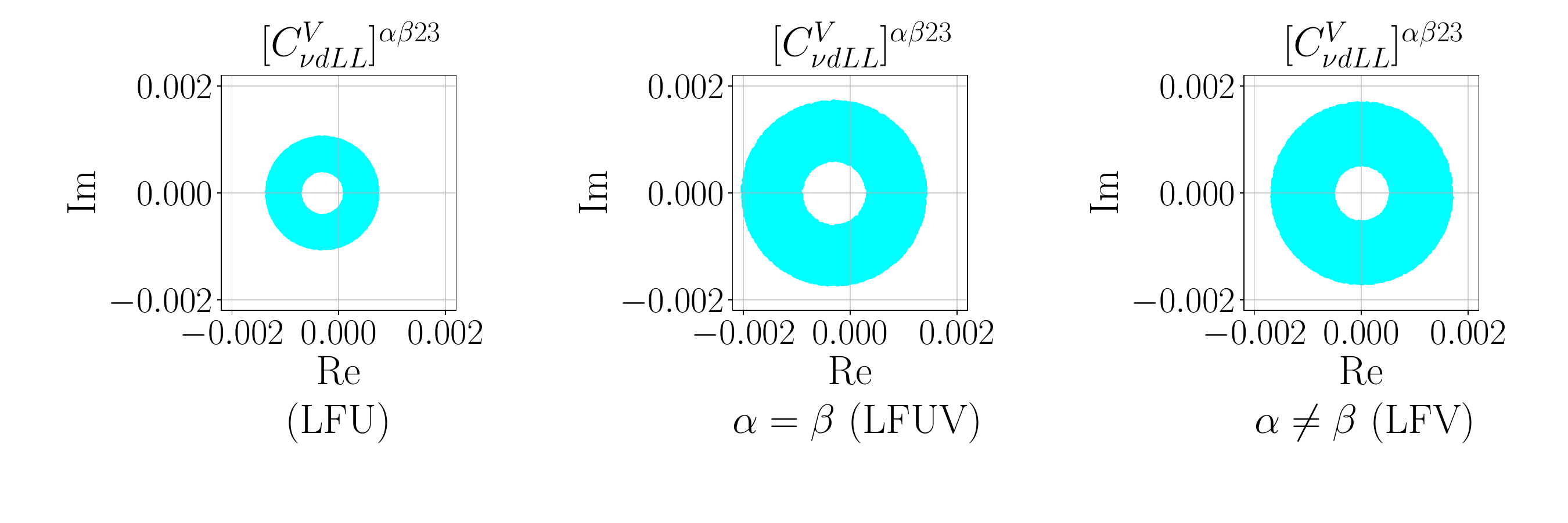}
    \caption{Preferred parameter region at $90\%$ C.L. for $[{\tempC}_{\nu d LL}^{V}]^{\alpha\beta 23}$ in order to explain the observed excess in $B\to K \nu \nu$ branching ratio. The left panel shows lepton flavor universal (LFU) scenario, where $[{\tempC}_{\nu d LL}^{V}]^{\alpha\beta 23}$ is nonzero and equal for all $\alpha = \beta \in \{e\,,\mu\,,\tau\}$. The middle panel shows lepton flavor nonuniversal (LFUV) scenario where $[{\tempC}_{\nu d LL}^{V}]^{\alpha\beta 23}$ is nonzero only for one value of $\alpha =\beta$. The right panel depicts the LFV scenario with $\alpha \neq \beta$ and only one $[{\tempC}_{\nu d LL}^{V}]^{\alpha\beta 23}$ nonzero.}
    \label{bsnunuplots}
\end{figure}

In the recent measurement of $B\to K \nu \nu$ at Belle II~\cite{Belle-II:2023esi}, the observed branching ratio has $3.5\sigma$ excess over the SM value. If this excess were to be explained in terms of the LEFT coefficients $[{\tempC}_{\nu d LL}^{V}]^{\alpha \beta 2 3}$, the required values of these WCs in various scenarios are shown in Fig.~\ref{bsnunuplots}. In the first scenario, we assume that new physics turns on a lepton flavor universal (LFU) combination of WCs whereas in the second (i.e. LFUV)  and third (i.e. LFV) scenarios, we assume that a single WC is turned on with $\alpha =\beta$ and $\alpha \neq \beta$, respectively.

From this figure, it is clear that the coefficient $[{\tempC}_{\nu d LL}^{V}]^{\alpha \beta 2 3}$ is non-vanishing at $90\%$ C.L. for all scenarios considered. As discussed earlier, a nonzero $[{\tempC}_{\nu d LL}^{V}]^{\alpha \beta 2 3}$ will indicate at least seven (ten) non-vanishing WCs appearing in eq.\,(\ref{corLFV1}) for $\alpha =\beta$ $(\alpha\neq \beta)$.

For example,  in the LFUV  (LFV) scenarios, eq.\,(\ref{corLFV1}) corresponds to 27 (54) equations of the form
\begin{align}
    [{\tempC}_{euLL}^{V}]^{\alpha \beta i j} &= V_{i2}\,[{\tempC}_{\nu d LL}^V]^{\alpha \beta 2 3} V^\dagger_{3j} + ... 
\end{align}
in UV4f models. Since the CKM coefficients $V_{cs}V_{tb}^*$ and $V_{us}V_{tb}^*$, which are $\mathcal{O}(1)$ and $\mathcal{O}(\lambda)$ respectively, are significant, it is expected that in the absence of any cancellation coming from other $[{\tempC}_{\nu dLL}^{V}]^{\alpha \beta i j}$ elements, the WCs $[{\tempC}_{euLL}^{V}]^{\alpha \beta 1 3}$ and $[{\tempC}_{euLL}^{V}]^{\alpha \beta 2 3}$ will have significant nonzero values. Thus the modes $t\to c\, e^\alpha\,e^\beta$ and $t \to u e^\alpha \, e^\beta$ will be the ones where there can be potential new physics. Currently the bounds on these coefficients are $|[{\tempC}_{euLL}^{V}]^{\alpha \beta 1 3}| < 0.003$ and $|[{\tempC}_{euLL}^{V}]^{\alpha \beta 2 3}| < 0.02$,  respectively. Exploration of these modes further may lead to discovery of further anomalies in these two channels. These processes will also test the solution of $B\to K \nu\nu$ anomaly in terms of $[{\tempC}_{\nu d LL}^{V}]^{\alpha \beta 2 3}$. This demonstrates that the semileptonic neutral-current top decays will be strong probes of the origin of the $B\to K \nu\nu$ anomaly in the context of SMEFT.

Eq.\,(\ref{corLFV3}), in this LFUV (LFV) scenario, gives the 9 (18) equations of the form
\begin{align}
     [{\tempC}_{LL}^{V}]^{\alpha \beta i 3} &= V_{i2}\,([{\tempC}_{ed LL}^{V}]^{\alpha \beta 2 3} - [{\tempC}_{\nu d LL}^{V}]^{\alpha \beta 2 3})~.
\end{align}
%
Since the CKM coefficients $V_{cs}$ and $V_{us}$, which are $\mathcal{O}(1)$ and $\mathcal{O}(\lambda)$, respectively, are significant, it is expected that in the absence of any cancellation coming from other $[{\tempC}_{\nu d LL}^{V}]^{\alpha \beta i j}$ or $[{\tempC}_{e d LL}^{V}]^{\alpha \beta i j}$ elements, the WCs $[{\tempC}_{LL}^{V}]^{\alpha \beta 2 3}$ and $[{\tempC}_{LL}^{V}]^{\alpha \beta 1 3}$ will have significant nonzero values. Thus, charged-current semileptonic $B$ meson decays would also be sensitive probes of the origin of $B\to K \nu\nu$  anomaly. 

Similar discussions can be found in \cite{Bause:2023mfe, Chen:2024jlj}. In \cite{Bause:2023mfe}, relations among the WCs in the $LLRR$ category, as shown in Table~\ref{CorrTable}, have been used to relate $b_R\to s_R\tau \tau$ and $b_R\to s_R \nu\nu$. These relations, as discussed in \cite{Bause:2023mfe}, predict excess branching fractions for the modes $B\to K^{(*)}\tau\tau$, $B_s\to \tau\tau$, etc. In \cite{Chen:2024jlj},  matching relations have been derived among the SMEFT and LEFT WCs,   assuming up-alignment. These have been then used to obtain the effects of the observed excess in $B\to K\nu\nu$ on other processes, namely, $B\to D^{(*)}\ell\nu_\ell$, $B\to K^{(*)}\ell^+\ell^-$, $B_q\to \tau\nu_\tau$, $B_s\to \tau^+\tau^-$, $D_s\to \tau^+\nu_\tau$, etc.

\subsection{Implications of the $R(D^{(*)})$ anomalies}  

\begin{figure}[h]
    \centering
    \includegraphics[width=0.8\textwidth]{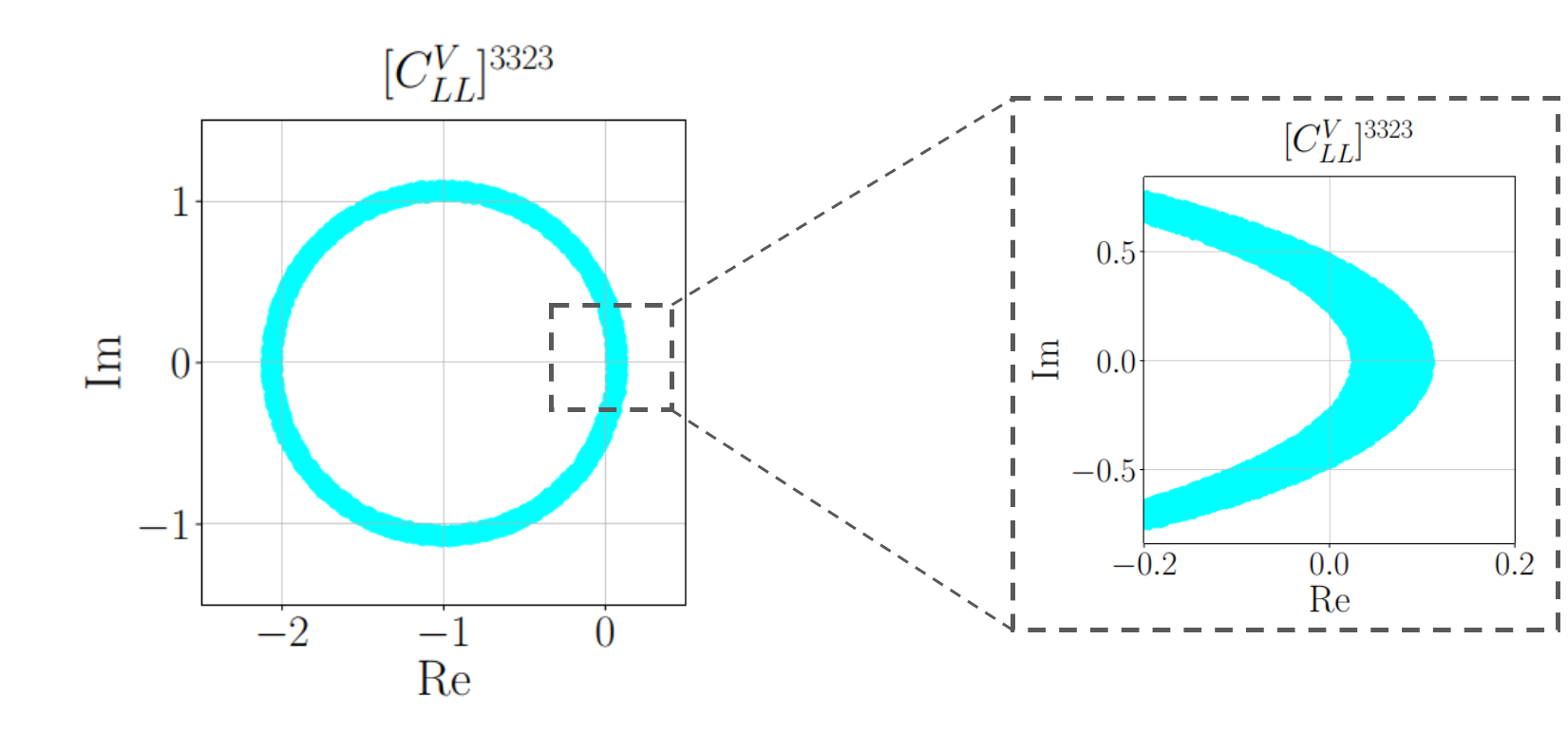}
    \caption{Preferred parameter region at $90\%$ C.L. for the WC $[{\tempC}_{LL}^{V}]^{3323}$.}
    \label{bctaunuplot}  
\end{figure}

One possible explanation of  multiple anomalies observed in the $b\to c \tau \bar \nu$ channels, such as $R(D)$, $R(D^{*})$ and $R(J/\psi)$, is to have nonzero values for the LEFT WC $[{\tempC}_{LL}^{V}]^{3 3 2 3}$. We show the preferred range of this WC at $90\%$ C.L. in Fig.~\ref{bctaunuplot}. Note that this preferred range does not include the point $[{\tempC}_{LL}^{V}]^{3 3 2 3} = 0$. 

From eq.\,(\ref{corLFV3}), we can write $[{\tempC}_{LL}^{V}]^{3 3 2 3}$ in terms of the neutral-current WCs as
\begin{align}
    [{\tempC}_{LL} ^{V}]^{3 3 2 3} &=  V_{cd}\left[[\hat {\tempC}_{edLL}^{V}]^{3 3 1 3} - [{\tempC}_{\nu dLL}^{V}]^{3 3 1 3}\right] + V_{cs}\left[[\hat {\tempC}_{edLL}^{V}]^{3 3 2 3} - [{\tempC}_{\nu dLL}^{V}]^{3 3 2 3}\right]\nonumber\\
    &~+V_{cb}\left[[\hat {\tempC}_{edLL}^{V}]^{3 3 3 3} - [{\tempC}_{\nu dLL}^{V}]^{3 3 3 3}\right]~.\label{LL3323rel}
\end{align}
Since $[{\tempC}_{LL}^{V}]^{3 3 2 3}\neq 0$, it suggests that 
at least one WC appearing on the right-hand side of eq.\,(\ref{LL3323rel}) has to be nonzero. Relevant interesting modes could be of the type $b\to d \tau \tau$, $b \to s \tau \tau$, $b \to d \nu \nu$ and $b \to s \nu \nu$ which suggests that the NP can manifest in observables related to processes such as $B \to \tau \tau$, $B_s \to \tau \tau$, $B\to K^{(*)} \tau \tau$, $B\to K^{(*)}\nu\nu$, etc. 

\subsection{Implications of the violation of SMEFT predictions} \label{implibroken}

In this subsection, we consider a scenario where many anomalies have been observed and multiple LEFT coefficients must have non-zero values to explain them. According to our results, these LEFT WCs must obey the SMEFT predictions of Table~\ref{tab: implications}. We now discuss what an observation of a violation of these predictions would imply. 

First of all, if low-energy measurements indicate a violation of the UV4f predictions in eq.\,(\ref{corLFV1}--\ref{corLFV3}), it may only mean that the UV model is not in the UV4f category, but still maps to SMEFT when heavier degrees of freedom are integrated out. It would only indicate that we are outside the UV4f region of Fig.~\ref{vendiag}, and not necessarily outside the SMEFT region. We must then check whether or not the more general predictions  Table~\ref{CorrTable} are obeyed. This would require looking for deviations in $W$ and $Z$ decays and/or high-$p_T$ Drell-Yan data.\footnote{If we use only the high-$p_T$ data to test our predictions, we can directly use Table~\ref{CorrTable} and thus test the validity of our predictions without taking into account $Z$ and  $W^\pm$ decays.} 

If the violation of the predictions persists at the level of Table~\ref{CorrTable} (or the equations in Appendix.~\ref{app:left-heft}), it would imply that one of the assumptions used in deriving these predictions is incorrect. Note, first of all,  that we have only included  dim-6 operators in our analysis. Inclusion of dimension-8 (dim-8) operators will result in a violation of these predictions at ${\cal O}(v^4 /\Lambda^4)$. For instance, the dim-8 operator 
\begin{equation}
  \left[ {\cal O}_{{\ell} q3}\right]^{\alpha \beta i j}=(\bar l^\alpha \gamma_\mu  l^\beta)(\bar q^i \gamma^\mu \tau^I q^j)(H^\dagger \tau^I H)
\end{equation}
will break the equality in the first row of Table~\ref{CorrTable}, as follows: 
\begin{equation}
    V^\dagger_{ik} \, [\hat {\mc}_{euLL}^{V}]^{\alpha \beta k l} \,V_{{\ell} j} - U^\dagger_{\alpha \rho} \,[\hat {\mc}_{\nu dLL}^{V}]^{\rho \sigma i j} \, U_{\sigma \beta} \sim v^4 /\Lambda^4\left[{\cal C}_{{\ell} q3}\right]  ~.
\end{equation}
Similarly, other operators at dim-8 or higher order will introduce a breaking of the other predictions in Table~\ref{CorrTable}. Such effects are, however, higher order in the SMEFT expansion parameter $v^2/\Lambda^2$, and are thus expected to be small.

If larger, ${\cal O}(1)$ violations of the predictions are observed, it would indicate something more radical, namely, that one of the assumptions of SMEFT itself is violated and we lie outside the SMEFT region of Fig.~\ref{vendiag}. This would be the case if (i) the scale of new physics is below the weak scale, (ii) there is heavy new physics that does not decouple because it gets a large fraction of its mass from the electroweak vacuum expectation value~\cite{Banta:2021dek, Cohen:2020xca}, or (iii) the observed 125 GeV scalar, $h$, is not a part of the SU(2) doublet that breaks the electroweak symmetry ~\cite{Falkowski:2019tft, Cohen:2020xca, Banta:2021dek, Cata:2015lta}.

As an example, consider the case of neutrino NSI that are induced by operators containing neutrinos in Table~\ref{HEFToplist}. As mentioned in Sec.~\ref{sec: LFV}, for a given choice of the quark flavor indices, eqs.~(\ref{corLFV1}--\ref{corLFV3}) (or the equations in Table~\ref{CorrTable}), imply relations between the NSI and the stringently constrained lepton flavor violating operators~\cite{Proceedings:2019qno}.  These predictions can, however,  be evaded in new physics scenarios where dim-8 operators become important. For instance, if the leading contribution to the NSI  is  from dim-8 (and not dim-6) operators like
\begin{equation}
  \left[ {\cal O}_{l3q}\right]^{\alpha \beta i j}=(\tilde{H}^\dagger \tau^I \tilde{H})(\bar l^\alpha \gamma_\mu \tau^I  l^\beta)(\bar q^i \gamma^\mu  q^j)~,
\end{equation}
new physics affects only the neutrino and not the charged-lepton sector. Even in this case, however,  dim-6 charged-lepton flavor-violating effects will be generated at loop level~\cite{Ardu:2022pzk}. A  more natural way of decoupling these two sectors is if the new physics scale is below the electroweak scale (see, e.g. Ref.~\cite{Farzan:2019xor}).

\section{Concluding remarks}\label{sec: conclusion}

In this work, we have systematically derived the consequences of the  $SU(2)_L\times U(1)_Y$ invariance of SMEFT on semileptonic flavor observables.  These consequences arise from the fact that a complete parametrization of BSM deviations in flavor physics observables can be only achieved by writing a lagrangian that respects   $U(1)_{em}$ and not the full symmetry of SMEFT. For instance, while the left-handed up and down type fermions form $SU(2)_L$ doublets and always appear together in  SMEFT operators, as far as flavor observables are concerned, searches in the up and down sectors are completely independent. Therefore,  BSM deviations in these channels must be parameterized by independent operators.

To be more precise, while the most general  $U(1)_{em}$ invariant lagrangian has 3240 independent semileptonic four-fermion operators  (see Table~\ref{HEFToplist})  and another set of 144 operators that contribute to semileptonic processes via $Z, W^\pm$ and $h$ exchange (see Table~\ref{heftZW}), the number of dim-6 SMEFT operators in these categories are 1053 (see Table~\ref{SMEFToplist}) and 108 (see Table~\ref{heftZW}), respectively. This then results in 2223 constraints in the space of  WCs of the $U(1)_{em}$ invariant lagrangian that can be thought of as predictions of SMEFT at the dim-6 level. One of the main results of this work is the derivation of these 2223 constraints. We present these constraints as linear relations among the WCs of the $U(1)_{em}$ invariant lagrangian, in Table~\ref{CorrTable}. These relations are a succinct expression of the consequences of the    $SU(2)_L\times U(1)_Y$ invariance of SMEFT for semileptonic operators. They are completely independent of UV flavor assumptions as we find that the elements of the rotation matrices of the left-handed and right-handed up-type and down-type fermions do not individually appear in them but only in combinations that form CKM and PMNS elements.   We then show how these relations can be written in terms of LEFT  WCs by integrating out the $Z,W^\pm$ and $h$ bosons. We refer the reader to Fig.~\ref{vendiag} where this scenario has been pictorially represented.

The  $U(1)_{em}$ invariant lagrangian we have considered is in fact equivalent, in the unitary gauge,   to the HEFT lagrangian which is generally written in an $SU(2)_L\times U(1)_Y$ invariant form but with the gauge symmetry being realized non-linearly.   We show this explicitly in Appendix~\ref{app:HEFTbasis} where we present a one-to-one mapping between the invariant HEFT operators and the list of   $U(1)_{em}$  invariant operators in Table~\ref{HEFToplist} and Table~\ref{heftZW}. In the process, we find some  HEFT operators that were missed in earlier literature and others that were considered but are actually redundant.

The SMEFT predictions we have derived have powerful phenomenological consequences as they connect observables in different sectors, such as rare decays in the kaon, B-Meson and charm sectors; decays of the top, $Z, W^\pm$ and $\tau$; lepton flavor violating observables and even neutrino NSI. On the one hand, they can be used to express poorly constrained WCs in terms of strongly constrained ones, thus allowing us to put new stronger indirect bounds on the former. On the other hand, if evidence for new physics is seen, they in general imply that BSM effects cannot appear in a single isolated channel because these linear relations imply that if one WC is non-zero, multiple others also must be non-vanishing. 

To illustrate the usefulness of these relations in phenomenology, we focus on the well-motivated UV4f scenario, where the UV physics only involves four-fermion operators, and HEFT WCs corresponding to BSM couplings of the $Z$, $W^{\pm}$ and Higgs to fermions are absent.   We further restrict ourselves to the operators with only left-handed fermions, i.e the $LLLL$ class of operators. In this scenario, there are three sets of relations among the LEFT WCs. The first set relates  the WCs of the neutral-current operators $(\bar \nu_L\gamma_\mu \nu_L)\,(\bar d_L \gamma^\mu d_L)$ and  $(\bar e_L\gamma_\mu e_L)\,(\bar u_L \gamma^\mu u_L)$. The second set consists of relations among the WCs of the neutral-current operators $(\bar e_L\gamma_\mu e_L)\,(\bar d_L \gamma^\mu d_L)$ and  $(\bar \nu_L\gamma_\mu \nu_L)\,(\bar u_L \gamma^\mu u_L)$. In the third set, the charged-current WCs are related to the above neutral-current coefficients.

The main phenomenological results of this work are as follows: 
\begin{enumerate}
     \item \textbf{Indirect bounds from SMEFT predictions: } In   Sec.~\ref{sec: indNC1}-~\ref{sec: indCC}, we consider  $LLLL$ operators in UV4f models. Using bounds from meson decays and high-$p_T$ Drell-Yan searches and applying the SMEFT predictions, we obtain novel bounds on WCs related to $d\bar d\to\nu\bar \nu$, $u_i\to u_j\nu \bar \nu$ and top decays, that are much stronger than the direct bounds. Our main results are summarised in Fig.~\ref{set1indirect},~\ref{set2indirect} and~\ref{set3indirect} .
     \item \textbf{Connecting quark and lepton flavor violation:} In Sec.~\ref{sec: LFV}, we show how the SMEFT predictions derived by us connect flavor violation in the quark and lepton sectors. In Table~\ref{tab: implications},  we present a list of processes spanning diverse observation channels (e.g. LFV tau decays, LFV ${\ell} N\to {\ell}^\prime N$ transitions, rare semileptonic $B$, $D$ and $K$ decays, top production and decays, etc.) that are connected via our analytic relations among the WCs.
     \item \textbf{Evidence for new physics from SMEFT predictions:} In Sec.~\ref{sec: phenoevidence}, we show that the relations among the WCs of the type $LLLL$ imply that a single  nonzero WC requires that there are at least 9 other nonzero WCs. We then discuss the specific cases of the observed excess in the $B\to K \nu\nu$ branching fraction and $R(D^{(*)})$ anomalies, and list other search channels that should see a correlated signal if these anomalies survive in the future.
\end{enumerate}

In future studies, we aim to extend the approach developed here and apply it to other flavor physics observables.  In this work, we have considered only a subset of operators appearing in LEFT, HEFT and SMEFT, namely the set of semileptonic operators. In future work, we will extend our analysis by including all operators up to dimension-6 in order to find SMEFT-predicted relations among the corresponding LEFT and HEFT WCs.  
These predictions  will allow us to interconnect many other important flavor observables.
For instance, predictions can be obtained for dipole operators connected to observables such as the $b \to s \gamma$  process, for four quark operators that are associated to the $\Delta F=2$ meson-mixing processes and nonleptonic meson decays, for four-lepton operators associated to LFV processes such as $\mu \to 3 e$, etc.
We, thus,   hope that this work will initiate a rich program in quark and lepton flavor phenomenology that uncovers many more SMEFT-predicted links between observables.

\acknowledgments
We would like to thank Tuhin S. Roy, Ketan M. Patel, Abhishek M. Iyer, Arnab Roy, Samadrita Mukherjee, Dibya S. Chattopadhyay and Radhika Vinze for useful discussions. This work is supported by the Department of Atomic Energy, Government of India, under Project Identification Number RTI 4002.  We acknowledge the use of computational facilities of the Department of Theoretical Physics at Tata Institute of Fundamental Research, Mumbai. We would also like to thank Ajay Salve and  Kapil Ghadiali for technical assistance. 

\appendix

\section{Semileptonic HEFT operators in $SU(2)_L \times U(1)_Y$ invariant form} 
 \label{app:HEFTbasis}

In Table~\ref{HEFToplist} and Table~\ref{heftZW}, we have presented all possible $U(1)_{em}$-invariant semileptonic operators relevant to this work. In this Appendix, we show that these operators can be rewritten as $SU(2)_L \times U(1)_Y$ invariant operators of HEFT  with the symmetry realized non-linearly.  Following  the notation and approach used in Ref.~\cite{Buchalla:2012qq}, we introduce the Goldstone matrix $U = \exp(2i\varphi^aT^a/v)$,
where $\varphi_a$ are the Goldstones of the breaking of $SU(2)_L \times SU(2)_R\to SU(2)_V$. Under $SU(2)_L \times SU(2)_R$, the matrix $U$ transforms as $U\to g_L U  g_R^\dagger$, where $g_L$ and $g_R$ are the respective group elements.  We also introduce the $SU(2)_R$ quark and lepton doublets denoted by $r\equiv (u_R, d_R)^T$ and  $\eta \equiv (0, e_R)^T$, respectively.

As the correct symmetry-breaking pattern in SM is $SU(2)_L \times U(1)_Y\to U(1)_{em}$, and not $SU(2)_L \times SU(2)_R\to SU(2)_V$,  one must include explicit sources of  $SU(2)_R$ breaking (see for eg. Ref.~\cite{murayama}). For bosonic operators, this is usually done by introducing the two spurions $L_\mu=U D_\mu U^\dagger$ and $\tau_L= U T_3 U^\dagger$. For fermionic operators, we need other sources of $SU(2)_R$ breaking. As shown in Ref.~\cite{Buchalla:2012qq}, this can be achieved by including factors of $U P_i$ in the operators where the projection matrices $P_i$  are defined as
\begin{align}
    P_{\pm} \equiv \frac{1}{2} \pm T_3, \quad P_{12} \equiv T_1 + iT_2, \quad P_{21} \equiv T_1 - iT_2~.
\end{align}
In the above equation, $T_i$ are the $SU(2)_L$ generators.  One can keep track of the hypercharge invariance of the operators by keeping in mind that, while  $UP_+$ and $UP_{12}$  extract the $Y=-1$ components of $U$, the projections $UP_-$ and $UP_{21}$ extract the $Y=1$ components of $U$.

\begin{table}[t]
    \centering
    \renewcommand{\arraystretch}{1.2}
    \begin{tabular}{|c|c|}
        \hline
        \multicolumn{1}{|c|}{$LLLL$} & \multicolumn{1}{c|}{$LLRR$} \\
        \hline
        ${\cataO}_{LL3}=({\bar l\gamma_\mu l}) \,({\bar q\gamma^\mu q })$ & ${\cataO}_{LR5}=({\bar l\gamma_\mu l})\, ({\bar u\gamma^\mu u}) $ \\
        ${\cataO}_{LL4}= ({\bar l\gamma_\mu \tau^a l})\, ({\bar q\gamma^\mu \tau^a q })$ & ${\cataO}_{LR6}=({\bar l\gamma_\mu l})\, ({\bar d\gamma^\mu d}) $\\
        ${\cataO}_{LL10}= ({\bar l\gamma_\mu U\tau^3 U^\dagger l})\, ({\bar q\gamma^\mu U\tau^3 U^\dagger q })$ & ${\cataO}_{FY11}=({\bar{\ell}UP_- r})\, ({\bar r P_+ U^\dagger l})$ \\
        ${\cataO}_{LL11}= ({\bar l\gamma_\mu l})\, ({\bar q\gamma^\mu U\tau^3 U^\dagger q})$ & ${\cataO}_{LR14}=({\bar l\gamma_\mu U\tau^3 U^\dagger l})\, ({\bar u\gamma^\mu u})$ \\
        ${\cataO}_{LL12}= ({\bar l\gamma_\mu U\tau^3 U^\dagger l})\, ({\bar q\gamma^\mu q})$ & ${\cataO}_{LR15}= ({\bar l\gamma_\mu U\tau^3 U^\dagger l})\, ({\bar d\gamma^\mu d})$ \\
        %
        %
        ${\cataO}_{LL14}= ({\bar l\gamma_\mu q})\, ({\bar q\gamma^\mu U\tau^3 U^\dagger l})$ & \\
        \hline
        \hline
        \multicolumn{1}{|c|}{$RRLL$} & \multicolumn{1}{c|}{$RRRR$} \\
        \hline
        ${\cataO}_{LR7}= ({\bar e\gamma_\mu e})\, ({\bar q\gamma^\mu q})$ & ${\cataO}_{RR5}= ({\bar e\gamma_\mu e})\, ({\bar u\gamma^\mu u})$ \\
        ${\cataO}_{LR16}= ({\bar e\gamma_\mu e})\, ({\bar q\gamma^\mu U\tau^3 U^\dagger q})$& ${\cataO}_{RR6}= ({\bar e\gamma_\mu e})\, ({\bar d\gamma^\mu d})$ \\
        \hline
        \hline
        \multicolumn{1}{|c|}{Scalar with $d_R$} & \multicolumn{1}{c|}{Tensor with $d_R$} \\
        \hline
        ${\cataO}_{FY7}= ({\bar q UP_- r})\, ({\bar{\ell}UP_- \eta})$ & ${\cataO}_{FY8}= ({\bar q\sigma^{\mu\nu} UP_- r})\, ({\bar l\sigma_{\mu\nu} UP_- \eta})$ \\
        ${\cataO}_{LR9}= ({\bar q\gamma^\mu l})\, ({\bar e\gamma_\mu d})$ & { $\bullet$ ${\cataO}_{ST13}=({\bar r P_-\sigma^{\mu\nu} U q})\, ({\bar l\sigma_{\mu\nu} UP_- \eta})$}\\
        ${\cataO}_{LR18}= ({\bar q\gamma^\mu U\tau^3 U^\dagger l})\, ({\bar e\gamma_\mu d})$ & { $\bullet$ ${\cataO}_{ST14}=({\bar q\sigma^{\mu\nu} U P_{12} r})\, ({\bar \eta \sigma_{\mu\nu} U P_{21} l})$}\\
        \hline
        \hline
        \multicolumn{1}{|c|}{Scalar with $u_R$} & \multicolumn{1}{c|}{Tensor with $u_R$} \\
        \hline
        %
        ${\cataO}_{ST9}= ({\bar q UP_+ r})\, ({\bar{\ell}UP_- \eta})$ & ${\cataO}_{ST11}= ({\bar q\sigma^{\mu\nu} UP_+ r})\, ({\bar l\sigma_{\mu\nu} UP_- \eta})$\\
        ${\cataO}_{FY9}= ({\bar{\ell}UP_- \eta})\, ({\bar r P_+ U^\dagger q})$ & { $\bullet$ ${\cataO}_{ST16}= ({\bar r P_+\sigma^{\mu\nu} U q})\, ({\bar l\sigma_{\mu\nu} UP_- \eta})$}\\
        ${\cataO}_{ST10}= ({\bar q UP_{21} r})\, ({\bar{\ell}UP_{12} \eta}) $ & ${\cataO}_{ST12}= ({\bar q\sigma^{\mu\nu} UP_{21} r})\, ({\bar l\sigma_{\mu\nu} UP_{12} \eta})$\\
        \hline
    \end{tabular}
    \caption{\label{heftlq} List of semileptonic $SU(2)_L \times U(1)_Y$ invariant HEFT operators. Note that this list is somewhat different from the list presented in \cite{Buchalla:2012qq} (see the text for more details). Some redundant operators present in \cite{Buchalla:2012qq} are omitted from this list. On the other hand, some operators (preceded by a bullet) which were absent in \cite{Buchalla:2012qq} have been added and have been named using a similar nomenclature. Note that $\tau^a=T^a/2$ are the Pauli matrices.}
\end{table}

We first consider four-fermion operators.  In Table~\ref{heftlq}, we present all possible  $SU(2)_L \times U(1)_Y$ invariant HEFT operators with two quarks and two leptons,  up to dimension 6. Note that this list has some differences from the list of operators presented in Ref.~\cite{Buchalla:2012qq} that we will discuss in detail in the following. Working in the unitary gauge, i.e. taking  $U \to 1$, we now write each of the operators in Table~\ref{heftlq} in terms of the operators in Table~\ref{HEFToplist}. This would confirm that there is a one-to-one mapping between these two sets of operators in the unitary gauge. 

\noindent \textbf{For $LLLL$ vector operators:}
%
\begin{align}
	{\cataO}_{LL3} & = {\tempO}_{\nu u LL}^V + {\tempO}_{e u LL}^V + {\tempO}_{\nu d LL}^V + {\tempO}_{e d LL}^V ~, \label{CataLL1}\\
	 {\cataO}_{LL4} & = {\tempO}_{\nu u LL}^V - {\tempO}_{e u LL}^V - {\tempO}_{\nu d LL}^V + {\tempO}_{e d LL}^V + 2\,{\tempO}_{LL}^V + 2\,{\tempO}_{LL}^{\prime V}~, \\
    {\cataO}_{LL10} & = {\tempO}_{\nu u LL}^V - {\tempO}_{e u LL}^V - {\tempO}_{\nu d LL}^V + {\tempO}_{e d LL}^V ~, \\
	{\cataO}_{LL11} & = {\tempO}_{\nu u LL}^V + {\tempO}_{e u LL}^V - {\tempO}_{\nu d LL}^V - {\tempO}_{e d LL}^V ~, \\
	{\cataO}_{LL12} & = {\tempO}_{\nu u LL}^V - {\tempO}_{e u LL}^V + {\tempO}_{\nu d LL}^V - {\tempO}_{e d LL}^V ~, \\
     {\cataO}_{LL14} & = {\tempO}_{\nu u LL}^V + {\tempO}_{LL}^V - {\tempO}_{LL}^{\prime V} - {\tempO}_{e d LL}^V ~,
\end{align}
where we have suppressed the quark and lepton flavor indices. Here  $[{\tempO}_{LL}^{\prime V}]^{\alpha\beta ij} = ([{\tempO}_{LL}^{V}]^{\beta\alpha j i})^\dagger$ and $[{\tempO}_{LL}^{V}]^{\alpha\beta ij}$ are two independent operators. The  6 operators listed in Table~\ref{heftlq}, therefore, receive contributions from  6 independent operators of this category in Table~\ref{HEFToplist}, providing a one-to-one mapping between these two lists.  In this category, there is one more operator in \cite{Buchalla:2012qq} i.e ${\cataO}_{LL13}=(\bar q\gamma^\mu U\tau^3  U^\dagger l)(\bar l\gamma_\mu U\tau^3  U^\dagger q)$. But this operator is not independent of the 6 operators appearing on the `$LLLL$' block of Table~\ref{heftlq}. Indeed, it can be written as
\begin{align}
    {\cataO}_{LL13} & = {\tempO}_{\nu u LL}^V - {\tempO}_{ LL}^V - ({\tempO}_{LL}^{\prime V}) + {\tempO}_{e d LL}^V~,
\end{align}
which is equivalent to the relation,
\begin{align}
	 {\cataO}_{LL13}&= \frac{1}{2}({\cataO}_{LL3} + 2\,{\cataO}_{LL10} - {\cataO}_{LL4} )~.
\end{align}
This operator has therefore been omitted in our list.

\noindent \textbf{For $LLRR$ vector operators:}
%
\begin{align}
	{\cataO}_{LR5} &= {\tempO}_{\nu u LR}^{V} + {\tempO}_{eu LR}^{V}~, &{\cataO}_{LR6} &= {\tempO}_{\nu d LR}^{V} + {\tempO}_{e d LR}^{V}~,\label{CataLR5}\\
	{\cataO}_{FY{11}} &=-\frac{1}{2}\,{\tempO}_{LR}^{V}~, &{\cataO}_{LR14} &= {\tempO}_{\nu u LR}^{V} - {\tempO}_{eu LR}^{V}~,\\
	{\cataO}_{LR15} &= {\tempO}_{\nu d LR}^{V} - {\tempO}_{e d LR}^{V}~,
	%
	%
\end{align}
Note that the operator ${\cataO}_{FY11}$ as defined in Table~\ref{heftlq} consists of two scalar currents. However, this operator maps to a vector operator after the Fierz transformation and hence it is included in the category $LLRR$.

\noindent \textbf{For $RRLL$ vector operators:}
%
\begin{align}
	{\cataO}_{LR7} &= {\tempO}_{eu RL}^{V} + {\tempO}_{ed RL}^{V}~,\quad {\cataO}_{LR16} = {\tempO}_{eu RL}^{V} - {\tempO}_{ed RL}^{V}~.\label{CataLR1}
\end{align}

\noindent \textbf{For $RRRR$ vector operators:}
%
\begin{align}
	{\cataO}_{RR5} &=  {\tempO}_{e u RR}^{V}~,\quad \quad {\cataO}_{RR6} =  {\tempO}_{e d RR}^{V}~.
\end{align}

\noindent\textbf{For scalar operators:}
%
\begin{align}
    {\cataO}_{FY7} &={\tempO}_{e d RLRL}^{\prime S}~, &  {\cataO}_{ST9} &={\tempO}_{e u RLRL}^{\prime S}~,\label{CataS1S3}\\
    {\cataO}_{LR9} &= -2\,{\tempO}_{RLLR}^{S} - 2\,{\tempO}_{e d RLLR}^{S}~.& {\cataO}_{FY9} &={\tempO}_{e u RLLR}^{\prime S}~,\label{CataLR9} \\
     {\cataO}_{LR18} &=-2\,{\tempO}_{RLLR}^{S} + 2\,{\tempO}_{e d RLLR}^{S}~, & {\cataO}_{ST10} &= {\tempO}_{RLRL}^{\prime S}~.\label{CataLR18}
    %
\end{align}
Here $[{\tempO}_{e d(u) RLLR}^{\prime S}]^{\alpha\beta ij} = ([{\tempO}_{e d(u) RLLR}^{S}]^{\beta\alpha ji})^\dagger$ and $[{\tempO}_{RLRL}^{\prime S}]^{\alpha\beta ij} = ([{\tempO}_{RLRL}^{S}]^{\beta\alpha ji})^\dagger$. Note that the operators ${\cataO}_{LR9}$ and ${\cataO}_{LR18}$ are defined as products of vector currents in Table~\ref{heftlq}. However, they map to scalar operators after Fierz transformations, as can be seen from eqs.\,(\ref{CataLR9}--\ref{CataLR18}). 
 
In \cite{Buchalla:2012qq}, there is one more scalar operator,  ${\cataO}_{ST3}=\varepsilon_{ij} (\bar q^i u)(\bar l^j e)$ . This operator is not independent from the scalar operators appearing in Table~\ref{heftlq} and can be written as
\begin{align}
		{\cataO}_{ST3} & = {\cataO}_{ST9} - {\cataO}_{ST10}~.
\end{align}
Hence this operator has been omitted in our list.

\noindent\textbf{For tensor operators:}
%
\begin{align}
    {\cataO}_{FY8} &={\tempO}_{e d RLRL}^{\prime T}~, &  {\cataO}_{ST11} &={\tempO}_{e u RLRL}^{\prime T}~,\\
    {\cataO}_{ST13} &= {\tempO}_{e d RLLR}^{\prime T}~,& {\cataO}_{ST16} &= {\tempO}_{e u RLLR}^{\prime T}~,\\
    {\cataO}_{ST14} &= {\tempO}_{RLLR}^{T} & {\cataO}_{S12} &= ({\tempO}_{RLRL}^{\prime T})~,
\end{align}
where $[{\cataO}^\prime]^{\alpha\beta i j} \equiv ([{\cataO}]^{\beta\alpha j i})^\dagger$. The three tensor operators ${\cataO}_{ST13}$, ${\cataO}_{ST14}$ and ${\cataO}_{ST16}$ are absent in the list of HEFT operators presented in \cite{Buchalla:2012qq}. On the other hand, the operator ${\cataO}_{ST4}=\varepsilon_{ij}\, (\bar q^i \sigma_{\mu\nu}\,u)\, (\bar l^j \sigma^{\mu\nu}e)$ included in \cite{Buchalla:2012qq} is not an independent one. It can be written as 
\begin{align}
	{\cataO}_{ST4} & = {\cataO}_{ST11} - {\cataO}_{ST12}~,
\end{align}
and has been omitted in our list.

\begin{table}[b]
    \centering
    \renewcommand{\arraystretch}{1.2}
    \begin{tabular}{|l|l|}
        \hline
        \multicolumn{2}{|c|}{HEFT operators with $Z$, $W^{\pm}$ couplings}\\
        \hline
        ${\cataO}_{\psi V1}=\left(\bar q\gamma^\mu q\right) \l U^\dagger iD_\mu UT_3\r $ & ${\cataO}_{\psi V2}=\left(\bar q\gamma^\mu UT_3U^\dagger q\right) \l U^\dagger iD_\mu UT_3\r $ \\
        ${\cataO}_{\psi V3}=\left(\bar q\gamma^\mu U P_{12} U^\dagger q\right) \l U^\dagger iD_\mu U P_{21}\r  ~ + {\rm h.c.}$ & ${\cataO}_{\psi V4}=\left(\bar u\gamma^\mu u\right) \l U^\dagger iD_\mu UT_3\r  $ \\
        ${\cataO}_{\psi V5}=\left(\bar d\gamma^\mu d\right) \l U^\dagger iD_\mu UT_3\r  $ & ${\cataO}_{\psi V6}=\left(\bar u\gamma^\mu d\right) \l U^\dagger iD_\mu UP_{21}\r  ~ + {\rm h.c.}$ \\
        ${\cataO}_{\psi V7}=\left(\bar l\gamma^\mu l\right) \l U^\dagger iD_\mu UT_3\r  $ & ${\cataO}_{\psi V8}=\left(\bar l\gamma^\mu UT_3U^\dagger l\right) 
         \l U^\dagger iD_\mu UT_3\r $ \\
        ${\cataO}_{\psi V9}=\left(\bar l\gamma^\mu U P_{12} U^\dagger l\right)  \l U^\dagger iD_\mu U P_{21}\r  ~ + {\rm h.c.}$ & ${\cataO}_{\psi V10}=\left(\bar e\gamma^\mu e\right) \l U^\dagger iD_\mu UT_3\r  $ \\
        ${\cataO}_{\psi h1} = h \,\left(\bar q U P_{-}r\right)$ & ${\cataO}_{\psi h2} = h \,\left(\bar q U P_{+}r\right)$\\
        ${\cataO}_{\psi h3} = h \,\left(\bar l U P_{-}\eta\right)$ & \\
        \hline
    \end{tabular}
    \caption{\label{heftZWH}HEFT operators in \cite{Buchalla:2012qq} with $Z$, $W^{\pm}$ and $h$ couplings to fermions.}
\end{table}

For HEFT operators with BSM coupling of $Z$, $W^\pm$ to the fermions, we reproduce the list provided in \cite{Buchalla:2012qq} in Table~\ref{heftZWH}. In addition, we have also included the HEFT operators that modify the coupling of $h$ to fermions.   Once again there is a one-to-one mapping between the operators in Table~\ref{heftZWH} and Table~\ref{heftZW}:
\begin{align}
    {\cataO}_{\psi V1} & = -\frac{g}{2\cos{\theta}}\left( {\tempO}_{u_LZ} + {\tempO}_{d_LZ} \right)~,\quad {\cataO}_{\psi V2}  = -\frac{g}{2\cos{\theta}}\left( {\tempO}_{u_LZ} - {\tempO}_{d_LZ} \right)~,\\
    {\cataO}_{\psi V3} & = -\frac{g}{\sqrt{2}}{\tempO}_{ud_LW}~, \qquad\qquad\qquad ~{\cataO}_{\psi V4}  = -\frac{g}{2\cos{\theta}}{\tempO}_{u_RZ}~,\\
    {\cataO}_{\psi V5} & = -\frac{g}{2\cos{\theta}}{\tempO}_{d_RZ}~, \qquad\qquad\quad ~~{\cataO}_{\psi V6}  = -\frac{g}{\sqrt{2}}{\tempO}_{ud_RW}~,\\
    {\cataO}_{\psi V7} & = -\frac{g}{2\cos{\theta}}\left( {\tempO}_{\nu_LZ} + {\tempO}_{e_LZ} \right)~,\quad {\cataO}_{\psi V8}  = -\frac{g}{2\cos{\theta}}\left( {\tempO}_{\nu_LZ} - {\tempO}_{e_LZ} \right)~,\\
    {\cataO}_{\psi V9} & = -\frac{g}{\sqrt{2}}({\tempO}_{e\nu_LW})^\dagger~, \qquad\qquad~~\, {\cataO}_{\psi V10}  = -\frac{g}{2\cos{\theta}}{\tempO}_{e_RZ}~,\\
    {\cataO}_{\psi h1} &= {\hefto}_{dh}~,\qquad {\cataO}_{\psi h2} = {\hefto}_{uh}~, \qquad {\cataO}_{\psi h3} = {\hefto}_{eh}~.
\end{align}

Thus, we have explicitly demonstrated the one-to-one mapping between the HEFT operators in the $U(1)_{em}$ invariant language and the HEFT operators in $SU(2)_L\times SU(2)_R$ language in the unitary gauge.

\section{Details of the SMEFT basis used}\label{app: SMEFTbasis}

To obtain the SMEFT predictions, we have used the $\left(m_W,m_Z, \alpha_{EM}\right)$ input parameter scheme and the basis as proposed in Ref.~\cite{Masso:2014xra}. Note that this basis is different from the Warsaw basis that is conventionally used for SMEFT. In this appendix, we discuss the difference and the rationale for the choice of this basis.

In the Warsaw basis, the two operators\footnote{Note that in Ref.~\cite{Grzadkowski:2010es}  the operator ${\cal O}_T$ would actually appear  as a linear combination of the operators, $(H^\dagger\, D_\mu\,H)^*(H^\dagger\,D_\mu\,H)$ and $(H^\dagger H)\partial_\mu\partial^\mu(H^\dagger H)$. The combination of these operators orthogonal to ${\cal O}_T$, ${\cal O}_H =\frac{1}{2}(\partial_\mu |H|^2)^2 $, is part of the basis of Ref.~\cite{Masso:2014xra}. } 
\begin{align}
        {\cal O}_{WB} &= g g' H^\dagger \tau^I\,H\,W_{\mu\nu}^I\,B^{\mu\nu}~,\\
    {\cal O}_T &= (H^\dagger \, \overleftrightarrow{D} H)^2
\end{align}
would contribute to the couplings of gauge bosons to the fermions by affecting their mass and kinetic terms. One needs to carefully normalize the kinetic term to bring it to the canonical form and also take into account input parameter shifts. These subtleties become relevant when we try to write the $Z$ and $W^\pm$ coupling modifications in SMEFT as in eqs.\.(\ref{ZW1}-\ref{ZW3}).

Instead, in the basis of Ref.~\cite{Masso:2014xra}, the  operators  ${\cal O}_{WB}$ and ${\cal O}_T$ are  traded  for the following two operators:
\begin{align}
{\cal O}_{WB^\prime}&= {\cal O}_{WB}-2 ig^\prime  \left( H^\dagger  \overleftrightarrow{D}^\mu H \right )\partial^\nu  B_{\mu \nu},\nonumber\\
{\cal O}_{W'}&= \frac{ig}{2}\left( H^\dagger  \tau^a \overleftrightarrow {D}^\mu H \right )D^\nu  W_{\mu \nu}^a-  \frac{ig'}{2} \left( H^\dagger  \overleftrightarrow{D}^\mu H \right )\partial^\nu  B_{\mu \nu}.
\label{massoops}
\end{align}
This way of writing the operators eliminates their contributions to any mass or kinetic term ~\cite{Masso:2014xra}, thus allowing us to obtain the SMEFT predictions in a straightforward way.

Note that the final predictions we obtain should be independent of the basis being used and the Warsaw basis should also yield the same predictions, albeit with more complicated intermediate calculations. We show in the following that, in the Warsaw basis, the contributions of the two operators ${\cal O}_T$ and ${\cal O}_{WB}$ to the HEFT WCs associated with $Z$, $W^\pm$ couplings to the fermions cancel out in the final relations. In the Warsaw basis, these HEFT WCs receive the following SMEFT contributions~\cite{Efrati:2015eaa}:
\begin{align}
    [{\mc}_{u_LZ}]^{ij} &= \eta_{LZ} \, ([{\mC}_{Hq}^{(1)}]^{ij} - [{\mC}_{Hq}^{(3)}]^{ij}) +f(1/2,\,2/3)~,\label{heft-smeft-map1}\\
    [{\mc}_{d_LZ}]^{ij} &= \eta_{LZ} \, ([{\mC}_{Hq}^{(1)}]^{ij} + [{\mC}_{Hq}^{(3)}]^{ij}) +f(-1/2,\,-1/3)~,\\
    [{\mc}_{ud_LW}]^{ij} &= \eta_{LW} \, [{\mC}_{Hq}^{(3)}]^{ij} + f(1/2,\,2/3) - f(-1/2,\,-1/3)~,\label{heft-smeft-map3}
\end{align}
where $\eta_{LZ} \equiv -g/(2\,\cos\theta)$, $\eta_{LW} = g/(\sqrt{2})$ and the term $f(T^3, Q)$ is defined as~\cite{Efrati:2015eaa}
\begin{align}
    f(T^3, Q) &= \mathcal{I} \left[ -{\mC}_{WB}\,Q\,\frac{g^2\,g'^{2}}{g^2-g'^{2}} + ({\mC}_T - \delta v)\left(T^3 + Q\,\frac{g'^2}{g^2-g'^2} \right)\right]~,
\end{align}
with $[\delta v]^{ij} = ([{\mC}_{Hl}^{(3)}]^{11} + [{\mC}_{Hl}^{(3)}]^{22})/2 + [{\mC}_{{\ell}{\ell}}^{(1)}]^{1221}/4$. From eqs.\,(\ref{heft-smeft-map1}-\ref{heft-smeft-map3}), we obtain the SMEFT prediction
\begin{align}
    {\mc}_{ud_LW}= \frac{1}{\sqrt{2}} \cos\theta_w\, ({\mc}_{u_LZ} - \,{\mc}_{d_LZ})~,\label{ZWHcor1}
\end{align}
which is the same as the one in Table~\ref{CorrTable}. We see that in the prediction shown in eq.\,(\ref{ZWHcor1}), the function $f(T^3, Q)$ and the operators within do not appear. Similarly  for $Z$, $W^{\pm}$ coupling to leptons, we recover the prediction already presented in Table~\ref{CorrTable}, 
\begin{align}
    {\mc}_{e\nu_LW}= \frac{1}{\sqrt{2}} \cos\theta_w\, ({\mc}_{e_LZ} - \,{\mc}_{\nu_LZ})~.\label{ZWHcor2}
\end{align}
Thus, even in the Warsaw basis, the contributions to our relations from ${\cal O}_{WB}$ and ${\cal O}_T$ cancel out, confirming that the final relations among the HEFT WCs are independent of the choice of the basis for SMEFT. 

\section{Linear relations among LEFT and HEFT  operators}\label{app:left-heft}

In Sec.~\ref{sec: leftcor}, we have presented SMEFT predictions for LEFT WCs of the class $LLLL$. In this appendix, we provide a similar analysis for the other classes of LEFT operators.  We first write the matching of four-fermion semileptonic WCs between LEFT and HEFT.  We then substitute the HEFT WCs with the LEFT WCs for each of the analytic relations presented in Table~\ref{CorrTable}. As a result, we get relations among the LEFT WCs which also involve the BSM couplings of $Z, W^{\pm}$ and Higgs bosons to fermions. These relations for the vector operators are listed in Table~\ref{Copy1CorrTableLEFT}. For the scalar and the tensor operators, the relations are presented in Table~\ref{Copy2CorrTableLEFT}.

\begin{table}[h!]
	\begin{center}
		\small
		\begin{minipage}[t]{0.90\textwidth}
			\centering
			\renewcommand{\arraystretch}{1.7}
			\begin{tabular}[t]{|c| c | c|}
                \hline
                Class & Analytic relations for WCs of vector operators & Count\\
                \hline
                \multirow{1}{*}{\rotatebox{0}{$ LLLL $}} & \makecell*{$ \begin{aligned}
                &V_{ik}\left[[{\tempC}_{e dLL}^{V}]^{\alpha \beta k l}- \left( k_{e_L} \, [\hat{\mc}_{d_LZ}]^{kl} \, \delta_{\alpha\beta} + k_{d_L}\, [\hat{\mc}_{e_LZ}]^{\alpha\beta}\,\delta_{kl}\right)\right]V^\dagger_{{\ell} j}\nonumber\\
                &~~= U^\dagger_{\alpha \rho}\left[[{\tempC}_{\nu uLL}^{V}]^{\rho \sigma i j}-\chi\, \left(k_{\nu_L} \, [\hat{\mc}_{u_LZ}]^{ij} \, \delta_{\rho\sigma} + k_{u_L}\, [\hat{\mc}_{\nu_LZ}]^{\rho \sigma}\,\delta_{ij}\right)\right] U_{\sigma\beta}
                \end{aligned}$}
                & 81 (45)\\
                \cline{2-2}
                & \makecell*{$\begin{aligned}
                &V^\dagger_{ik}\left[[{\tempC}_{e uLL}^{V}]^{\alpha \beta k l}-\chi\, \left( k_{e_L} \, [{\mc}_{u_LZ}]^{kl} \, \delta_{\alpha\beta} + k_{u_L}\, [{\mc}_{e_LZ}]^{\alpha\beta}\,\delta_{kl}\right)\right]V_{{\ell} j} \nonumber\\
                &~= U^\dagger_{\alpha \rho}\left[[{\tempC}_{\nu dLL}^{V}]^{\rho \sigma i j}- \left(k_{\nu_L} \, [\hat{\mc}_{d_LZ}]^{ij} \, \delta_{\rho\sigma} + k_{d_L}\, [\hat{\mc}_{\nu_LZ}]^{\rho \sigma}\,\delta_{ij}\right)\right] U_{\sigma\beta}
                \end{aligned}$}
                & 81 (45)\\
                \cline{2-2}
                &\makecell*{$\begin{aligned} & V^\dagger_{ik} \, \left[[{\tempC}_{LL}^{V}]^{\alpha \beta k j} - \chi\, \left( k_{e\nu W} \, [\hat{\mc}_{ud_LW}]^{kj} \, \delta_{\alpha\beta} + [k_{ud W}]^{kj}\, [\hat{\mc}_{e\nu_LW}]^{\alpha\beta}\right)\right] \nonumber\\
                &=  \left[[{\tempC}_{e d LL}^{V}]^{\alpha \rho i j}-  \left( k_{e_L} \, [\hat{\mc}_{d_LZ}]^{ij} \, \delta_{\alpha\rho} + k_{d_L}\, [\hat{\mc}_{e_LZ}]^{\alpha\rho}\,\delta_{ij}\right)\right]\,U^\dagger_{\rho \beta}\nonumber\\
                & - U^\dagger_{\alpha \sigma} \, \left[[{\tempC}_{\nu d LL}^{V}]^{\sigma \beta i j}-  \left( k_{\nu_L} \, [\hat{\mc}_{d_LZ}]^{ij} \, \delta_{\sigma\beta} + k_{d_L}\, [\hat{\mc}_{\nu_LZ}]^{\sigma\beta}\,\delta_{ij}\right)\right]\end{aligned}$} & 162 (81)\\
                \hline\hline
                \multirow{1}{*}{\rotatebox{0}{$ RRRR $}} & No relations &\\
                \hline\hline
                \multirow{1}{*}{\rotatebox{0}{$ LLRR $}} & \makecell*{$\begin{aligned}&[{\tempC}_{e d LR}^{V}]^{\alpha \beta i j} -  \left(k_{e_L} \, [\hat{\mc}_{d_RZ}]^{ij} \, \delta_{\alpha\beta} + k_{d_R}\, [\hat{\mc}_{e_LZ}]^{\alpha\beta}\,\delta_{ij}\right)\nonumber\\
                &=U_{\alpha\rho}^\dagger \left[[{\tempC}_{\nu d LR}^{V}]^{\rho \sigma i j} -  \left( k_{\nu_L} \, [\hat{\mc}_{d_RZ}]^{ij} \, \delta_{\rho\sigma} + k_{d_R}\, [\hat{\mc}_{\nu_LZ}]^{\rho\sigma}\,\delta_{ij}\right)\right]\,U_{\sigma \beta} \end{aligned}$} & 81 (45)\\
                \cline{2-2}
                & \makecell*{$\begin{aligned}&[{\tempC}_{e u LR}^{V}]^{\alpha \beta i j} - \chi\, \left(k_{e_L} \, [\hat{\mc}_{u_RZ}]^{ij} \, \delta_{\alpha\beta} + k_{u_R}\, [\hat{\mc}_{e_LZ}]^{\alpha\beta}\,\delta_{ij}\right) \nonumber\\
                &= U^\dagger_{\alpha \rho} \, \left[[{\tempC}_{\nu u LR}^{V}]^{\rho \sigma i j} -\chi\,  \left(k_{\nu_L} \, [\hat{\mc}_{u_RZ}]^{ij} \, \delta_{\rho\sigma} + k_{u_R}\, [\hat{\mc}_{\nu_LZ}]^{\rho\sigma}\,\delta_{ij}\right)\right]\,U_{\sigma \beta}\end{aligned}$}  & 81 (45)\\
                \cline{2-2}
                &\makecell{$[\hat{\tempC}_{LR}^V]^{\alpha\beta i j} = \chi\,k_{e\nu W} \, [\hat{\mc}_{ud_RW}]^{ij}$}& 162 (81)\\
                \hline \hline
                \multirow{1}{*}{\rotatebox{0}{$ RRLL $}}& \makecell*{$\begin{aligned} & [{\tempC}_{e d RL}^{V}]^{\alpha \beta i j} -  \left(k_{e_R} \, [\hat{\mc}_{d_LZ}]^{ij} \, \delta_{\alpha\beta} + k_{d_L}\, [\hat{\mc}_{e_RZ}]^{\alpha\beta}\,\delta_{ij}\right) \nonumber\\
                &= V^\dagger_{ i k} \, \left[[{\tempC}_{e u RL}^{V}]^{\alpha \beta k l} - \chi\, \left( k_{e_R} \, [\hat{\mc}_{u_LZ}]^{kl} \, \delta_{\alpha\beta} + k_{u_L}\, [\hat{\mc}_{e_RZ}]^{\alpha\beta}\,\delta_{kl}\right)\right]\,V_{ lj}\end{aligned}$} & 81 (45)\\
                \hline
            \end{tabular}
		\end{minipage}
		\caption{\label{Copy1CorrTableLEFT} Relations among the LEFT semileptonic vector WCs and the BSM coupling of $Z, W^{\pm}$ and Higgs bosons to fermions, in the mass basis. Note that for the UV4f models, these BSM couplings vanish and this table becomes similar to Table~\ref{CorrTable}, but in terms of only LEFT WCs.}
	\end{center}
\end{table}

\begin{table}[h!]
    \begin{center}
    \small    
    \begin{minipage}[t]{0.90\textwidth}
        \centering
        \renewcommand{\arraystretch}{1.7}
        \begin{tabular}[t]{|c| c | c|}
            \hline
            Class & Analytic relations for WCs of scalar operators & Count\\
            \hline
            \multirow{2}{*}{\rotatebox{0}{\makecell{Scalar \\($d_R$)}}} & \makecell*{$\begin{aligned} &V_{ik}\, \left([{\tempC}_{edRLLR}^{S}]^{\alpha \beta k j} - [\hat{\mc}_{eh}]^{\beta\alpha\,*}{M}_{d}^{kj}\,k_{dh} -{M}_{e}^{\alpha\beta}\,k_{eh} [\hat{\mc}_{dh}]^{kj}\right.\nonumber\\
            &~ -{M}_{e}^{\alpha\rho}\,[\hat{\mc}_{e_LZ}]^{\rho\beta}\,{M}_{d}^{kj}y_d + [\hat{\mc}_{e_RZ}]^{\alpha\rho}{M}_{e}^{\rho\beta}\,{M}_{d}^{kj}y_d\nonumber\\
            &~\left.  - y_{e}{M}_{e}^{\alpha\beta}\,[\hat{\mc}_{d_LZ}]^{km}{M}_{d}^{mj} + y_{e}{M}_{e}^{\alpha\beta}\,{M}_{d}^{km}\,[\hat{\mc}_{d_RZ}]^{mj}  \right)\nonumber\\
            &= \left[[\hat {\tempC}_{RLLR}^{S}]^{\alpha \rho i j}  - \chi\left({M}_{e}^{\alpha\sigma} [\hat{\mc}_{e\nu_LW}]^{\sigma\rho} \, M_{d}^{kj}y_{ud} \right.\right.\nonumber\\
            &~\left.\left. +   y_{e\nu}M_{e}^{\alpha\rho}\, [\hat{\mc}_{ud_LW}]^{ik}{M}_{u}^{kj} -   y_{e\nu}M_e^{\alpha\rho}\, {M}_{u}^{kj} [\hat{\mc}_{ud_RW}]^{kj}\right)\right]\,U_{\rho \beta}\end{aligned}$} & 162 (81)\\
            \cline{2-2}
            & \makecell*{$\begin{aligned}&[{\tempC}_{edRLRL}^{S}]^{\alpha \beta i j} =  [\hat{\mc}_{eh}]^{\beta\alpha\,*}\,{M}_{d}^{ij}\,k_{dh} + k_{eh}{M}_{e}^{\alpha\beta}\,[\hat{\mc}_{dh}]^{ji\,*}\nonumber\\
            & -{M}_{e}^{\alpha\rho}[\hat{\mc}_{e_LZ}]^{\rho\beta}\,{M}_{d}^{ij}y_d  + [\hat{\mc}_{e_RZ}]^{\alpha\rho}{M}_{e}^{\rho\beta} \,{M}_{d}^{ij}y_{d}    ~\nonumber\\
            &~  -  y_e{M}_{e}^{\alpha\beta}\,{M}_{d}^{ik}[\hat{\mc}_{d_LZ}]^{kj} + y_{e}{M}_{e}^{\alpha\beta}\,[\hat{\mc}_{d_RZ}]^{ik}\,{M}_{d}^{kj}\end{aligned}$} & 162 (81)\\
            \hline\hline
            \multirow{2}{*}{\rotatebox{0}{\makecell{Scalar\\ ($u_R$)}}} & \makecell*{$\begin{aligned}&\left[[{\tempC}_{euRLRL}^{S}]^{\alpha \beta i k} -\chi\left([\hat{\mc}_{eh}]^{\beta\alpha\,*}\,{M}_{u}^{ik}\,k_{uh} + k_{eh}{M}_{e}^{\alpha\beta}\,[\hat{\mc}_{uh}]^{ki\,*}\right.\right.\nonumber\\
            &- {M}_{e}^{\alpha\rho}[\hat{\mc}_{e_LZ}]^{\rho\beta}\,{M}_{u}^{ik}y_u  + [\hat{\mc}_{e_RZ}]^{\alpha\rho}{M}_{e}^{\rho\beta} \,{M}_{u}^{ik}y_{u}    ~\nonumber\\
            &~\left.\left.  -  y_e {M}_{e}^{\alpha\beta}\,{M}_{u}^{im}[\hat{\mc}_{u_LZ}]^{mk} + y_{e}{M}_{e}^{\alpha\beta}\,[\hat{\mc}_{u_RZ}]^{im}\,{M}_{uRL}^{mk}\right)\right]\,V_{kj}\nonumber\\
            &=  -\left[[{\tempC}_{RLRL}^{S}]^{\alpha \rho i j}  - \chi\left({M}_{e}^{\alpha\sigma} [\hat{\mc}_{e\nu_LW}]^{\sigma\rho} \, M_{u}^{kj}y_{ud} \right.\right.\nonumber\\
            &\left.\left. -  y_{e\nu}M_e^{\alpha\rho}\, [\hat{\mc}_{ud_RW}]^{ik}{M}_{d}^{kj} +  y_{e\nu}M_e^{\alpha\rho}\, {M}_{u}^{kj} [\hat{\mc}_{ud_LW}]^{kj}\right)\right]\,U_{\rho \beta}\end{aligned}$ } & 162 (81)\\
            \cline{2-2}
            & \makecell*{$\begin{aligned}&[{\tempC}_{euRLLR}^{S}]^{\alpha \beta i j} = \chi\left([\hat{\mc}_{eh}]^{\beta\alpha\,*}{M}_{u}^{ij}\,k_{uh} + {M}_{e}^{\alpha\beta}\,k_{eh} [\hat{\mc}_{uh}]^{ij}\right.\nonumber\\
            &~ + {M}_{e}^{\alpha\rho}\,[\hat{\mc}_{e_LZ}]^{\rho\beta}\,{M}_{u}^{ij}y_u - [\hat{\mc}_{e_RZ}]^{\alpha\rho}{M}_{e}^{\rho\beta}\,{M}_{u}^{ij}y_u\nonumber\\
            &~\left.  + y_{e}{M}_{e}^{\alpha\beta}\,[\hat{\mc}_{u_LZ}]^{ik}{M}_{u}^{kj} - y_{e}{M}_{e}^{\alpha\beta}\,{M}_{u}^{ik}\,[\hat{\mc}_{u_RZ}]^{kj}  \right)\end{aligned}$} & 162 (81)\\
            \hline\hline
            \multirow{2}{*}{\rotatebox{0}{\makecell{Tensor \\($d_R$)}}} & $[ {\tempC}_{ed,\, \textrm{all}} ^{T}]^{\alpha \beta i j} = 0$ & 324 (162)\\
            & $[{\tempC}_{RLLR} ^{T}]^{\alpha \beta i j} = 0$ & 162 (81)\\
            \hline\hline
            \multirow{2}{*}{\rotatebox{0}{\makecell{Tensor \\($u_R$)}}} & $[ {\tempC}_{eu, RLRL}^{T, \alpha\beta i k} \, V_{kj}=  -[ {\tempC}_{RLRL}^{T}]^{\alpha \rho i j} \, U_{\rho \beta}$ & 162 (81)\\
            & $[{\tempC}_{e u,  LRRL}^{T}]^{\alpha \beta i j} = 0$ & 162 (81)\\
            \hline
	\end{tabular}
    \end{minipage}
    \caption{\label{Copy2CorrTableLEFT} Relations among the LEFT semileptonic scalar WCs and HEFT WCs for the BSM coupling of $Z, W^{\pm}$ and Higgs bosons to fermions, in the mass basis. Note that for the UV4f models, these BSM couplings vanish and this table becomes similar to Table~\ref{CorrTable}, but in terms of only LEFT WCs.}
    \end{center}
\end{table}
Now we present the matching relations among LEFT and HEFT operators in the flavor basis.

\noindent\textbf{For $LLLL$ vector operators:}
%
\begin{align}
	[\tilde{\tempC}_{\nu u LL}^{V}]^{\alpha \beta i j}&=\omega \,[{\mc}_{\nu u LL}^{V}]^{\alpha \beta i j} + \chi\,\left(k_{\nu_L} \, [{\mc}_{u_LZ}]^{ij} \, \delta_{\alpha\beta} + k_{u_L}\, [{\mc}_{\nu_LZ}]^{\alpha\beta}\,\delta_{ij}\right),\label{appleft-heftLL1}\\
	[\tilde{\tempC}_{e u LL}^{V}]^{\alpha \beta i j}&=\omega \,[{\mc}_{e u LL}^{V}]^{\alpha \beta i j} +\chi\,\left( k_{e_L} \, [{\mc}_{u_LZ}]^{ij} \, \delta_{\alpha\beta} + k_{u_L}\,[{\mc}_{e_LZ}]^{\alpha\beta}\,\delta_{ij}\right),\label{appleft-heftLL2}\\
	[\tilde{\tempC}_{\nu d LL}^{V}]^{\alpha \beta i j}&=\omega \,[{\mc}_{\nu d LL}^{V}]^{\alpha \beta i j} + k_{\nu_L} \, [{\mc}_{d_LZ}]^{ij} \, \delta_{\alpha\beta} + k_{d_L}\, [{\mc}_{\nu_LZ}]^{\alpha\beta}\,\delta_{ij},\label{appleft-heftLL3}\\
	[\tilde{\tempC}_{e d LL}^{V}]^{\alpha \beta i j}&=\omega \,[{\mc}_{e d LL}^{V}]^{\alpha \beta i j} + k_{e_L} \, [{\mc}_{d_LZ}]^{ij} \, \delta_{\alpha\beta} + k_{d_L}\, [{\mc}_{e_LZ}]^{\alpha\beta}\,\delta_{ij},\label{appleft-heftLL4}\\
	[\tilde{\tempC}_{LL}^{V}]^{\alpha \beta i j}&=\omega \,[{\mc}_{LL}^{V}]^{\alpha \beta i j} + \chi\,\left( k_{e\nu W} \, [{\mc}_{ud_LW}]^{ij} \, \delta_{\alpha\beta} + k_{ud W}\, [{\mc}_{e\nu_LW}]^{\alpha\beta}\,\delta_{ij}\right),\label{appleft-heftLLCC}
\end{align}
Where $l\in \{\nu,e\}$, $q\in \{u,d\}$, $\omega = v^2/(2\,\Lambda^2)$, and the $k$ coefficients are 
\begin{align}
    k_{f_L} &= \frac{2\cos\theta_w}{g}(T^3_f-Q_f \sin^2\theta_{w})~,\quad \textrm{and}~~k_{e\nu W} = k_{udW} = \frac{\sqrt{2}}{g}~,
\end{align}
where $f_L\in \{\nu_L, e_L, u_L, d_L\}$, as defined in Sec.~\ref{sec: leftcor}.

\noindent\textbf{For $LLRR$ vector operators:}
\begin{align}
	[\tilde{\tempC}_{\nu u LR}^{V}]^{\alpha \beta i j}&=\omega \,[{\mc}_{\nu u LL}^{V}]^{\alpha \beta i j} +\chi\,\left( k_{\nu_L} \, [{\mc}_{u_RZ}]^{ij} \, \delta_{\alpha\beta} + k_{u_R}\, {\mc}_{\nu_LZ}]^{\alpha\beta}\,\delta_{ij}\right),\label{left-heftLR1}\\
	[\tilde{\tempC}_{e u LR}^{V}]^{\alpha \beta i j}&=\omega \,[{\mc}_{e u LR}^{V}]^{\alpha \beta i j} + \chi\,\left(k_{e_L} \, [{\mc}_{u_RZ}]^{ij} \, \delta_{\alpha\beta} + k_{u_R}\, {\mc}_{e_LZ}]^{\alpha\beta}\,\delta_{ij}\right),\label{left-heftLR2}\\
	[\tilde{\tempC}_{\nu d LR}^{V}]^{\alpha \beta i j}&=\omega \,[{\mc}_{\nu d LR}^{V}]^{\alpha \beta i j} + k_{\nu_L} \, [{\mc}_{d_RZ}]^{ij} \, \delta_{\alpha\beta} + k_{d_R}\, [{\mc}_{\nu_LZ}]^{\alpha\beta}\,\delta_{ij},\label{left-heftLR3}\\
	[\tilde{\tempC}_{e d LR}^{V}]^{\alpha \beta i j}&=\omega \,[{\mc}_{e d LR}^{V}]^{\alpha \beta i j} + k_{e_L} \, [{\mc}_{d_RZ}]^{ij} \, \delta_{\alpha\beta} + k_{d_R}\, [{\mc}_{e_LZ}]^{\alpha\beta}\,\delta_{ij},\label{left-heftLR4}\\
	[\tilde{\tempC}_{LR}^{V}]^{\alpha \beta i j}&=\omega \,[{\mc}_{LR}^{V}]^{\alpha \beta i j} + \chi\,\left(k_{e\nu W} \, [{\mc}_{ud_RW}]^{ij} \, \delta_{\alpha\beta} \right).\label{left-heftLRCC}
\end{align}
Here,  $k_{e_R}$, $k_{u_R}$ and $k_{d_R}$ are 
\begin{align}
    k_{f_R} &= -\frac{2\cos\theta_w}{g}(Q_f \sin^2\theta_{w})~.
\end{align}

\noindent\textbf{For $RRRR$ vector operators:}
%
\begin{align}
    [\tilde{\tempC}_{e u RR}^{V}]^{\alpha \beta i j}&=\omega \,[{\mc}_{e u RR}^{V}]^{\alpha \beta i j} + \chi\,\left(k_{e_R} \, [{\mc}_{u_RZ}]^{ij} \, \delta_{\alpha\beta} + k_{u_R}\, {\mc}_{e_RZ}]^{\alpha\beta}\,\delta_{ij}\right)~,\label{left-heftLR2}\\
    [\tilde{\tempC}_{e d RR}^{V}]^{\alpha \beta i j}&=\omega \,[{\mc}_{e d RR}^{V}]^{\alpha \beta i j} + k_{e_R} \, [{\mc}_{d_RZ}]^{ij} \, \delta_{\alpha\beta} + k_{d_R}\, [{\mc}_{e_RZ}]^{\alpha\beta}\,\delta_{ij}~.\label{left-heftLR4}
\end{align}
In this category, the number of independent SMEFT and HEFT operators are the same. As a result, there is no prediction for the LEFT WCs in this category.

\noindent\textbf{For $RRLL$ vector operators:}
%
\begin{align}
	%
	[\tilde{\tempC}_{e u RL}^{V}]^{\alpha \beta i j}&=\omega \,[{\mc}_{e u RL}^{V}]^{\alpha \beta i j} +\chi\,\left( k_{e_R} \, [{\mc}_{u_LZ}]^{ij} \, \delta_{\alpha\beta} + k_{u_L}\, [{\mc}_{e_RZ}]^{\alpha\beta}\,\delta_{ij}\right),\label{left-heftRL2}\\
	%
	%
	[\tilde{\tempC}_{e d RL}^{V}]^{\alpha \beta i j}&=\omega \,[{\mc}_{e d RL}^{V}]^{\alpha \beta i j} + k_{e_R} \, [{\mc}_{d_LZ}]^{ij} \, \delta_{\alpha\beta} + k_{d_L}\, [{\mc}_{e_RZ}]^{\alpha\beta}\,\delta_{ij}~.\label{left-heftRL4}
\end{align}

\noindent\textbf{For scalar operators:}
%
\begin{align}
    [\tilde{\tempC}_{edRLLR}^{S}]^{\alpha \beta i j} &= \omega \,[ {\mc}_{edRLLR}^{S}]^{\alpha \beta i j} + [{\mc}_{eh}]^{\beta\alpha\,*}\tilde{M}_{dLR}^{ij}\,k_{dh} + \tilde{M}_{eRL}^{\alpha\beta}\,k_{eh} [{\mc}_{dh}]^{ij}\nonumber\\
    &~ + \tilde{M}_{eRL}^{\alpha\rho}\,[{\mc}_{e_LZ}]^{\rho\beta}\,\tilde{M}_{dLR}^{ij}y_d - [{\mc}_{e_RZ}]^{\alpha\rho}\tilde{M}_{eRL}^{\rho\beta}\,\tilde{M}_{dLR}^{ij}y_d\nonumber\\
    &~  + y_{e}\tilde{M}_{eRL}^{\alpha\beta}\,[{\mc}_{d_LZ}]^{ik}\tilde{M}_{dRL}^{kj} - y_{e}\tilde{M}_{eRL}^{\alpha\beta}\,\tilde{M}_{dLR}^{ik}\,[{\mc}_{d_RZ}]^{kj}~,   \\
    [\tilde{\tempC}_{edRLRL}^{S}]^{\alpha \beta i j} &= \omega \,[ {\mc}_{edRLRL}^{S}]^{\alpha \beta i j} + [{\mc}_{eh}]^{\beta\alpha\,*}\,\tilde{M}_{dRL}^{ij}\,k_{dh} + k_{eh}\tilde{M}_{eRL}^{\alpha\beta}\,[{\mc}_{dh}]^{ji\,*}\nonumber\\
    & - \tilde{M}_{eRL}^{\alpha\rho}[{\mc}_{e_LZ}]^{\rho\beta}\,\tilde{M}_{dRL}^{ij}y_d  + [{\mc}_{e_RZ}]^{\alpha\rho}\tilde{M}_{eLR}^{\rho\beta} \,\tilde{M}_{dRL}^{ij}y_{d}  ~\nonumber\\
    &~ -  y_e \tilde{M}_{eRL}^{\alpha\beta}\,\tilde{M}_{dRL}^{ik}[{\mc}_{d_LZ}]^{kj} + y_{e}\tilde{M}_{eRL}^{\alpha\beta}\,[{\mc}_{d_RZ}]^{ik}\,\tilde{M}_{dRL}^{kj}~,
\end{align}
\begin{align}
    [\tilde{\tempC}_{euRLLR}^{S}]^{\alpha \beta i j} &= \omega \,[ {\mc}_{euRLLR}^{S}]^{\alpha \beta i j} + \chi\left([{\mc}_{eh}]^{\beta\alpha\,*}\tilde{M}_{uLR}^{ij}\,k_{uh} + \tilde{M}_{eRL}^{\alpha\beta}\,k_{eh} [{\mc}_{uh}]^{ij}\right.\nonumber\\
    &~ + \tilde{M}_{eRL}^{\alpha\rho}\,[{\mc}_{e_LZ}]^{\rho\beta}\,\tilde{M}_{uLR}^{ij}y_u - [{\mc}_{e_RZ}]^{\alpha\rho}\tilde{M}_{eRL}^{\rho\beta}\,\tilde{M}_{uLR}^{ij}y_u \nonumber\\
    &~\left.  + y_{e}\tilde{M}_{eRL}^{\alpha\beta}\,[{\mc}_{u_LZ}]^{ik}\tilde{M}_{uRL}^{kj} - y_{e}\tilde{M}_{eRL}^{\alpha\beta}\,\tilde{M}_{uLR}^{ik}\,[{\mc}_{u_RZ}]^{kj}  \right)~, \\
    [\tilde{\tempC}_{euRLRL}^{S}]^{\alpha \beta i j} &= \omega \,[ {\mc}_{euRLRL}^{S}]^{\alpha \beta i j} + \chi\left([{\mc}_{eh}]^{\beta\alpha\,*}\,\tilde{M}_{uRL}^{ij}\,k_{uh} + k_{eh}\tilde{M}_{eRL}^{\alpha\beta}\,[{\mc}_{uh}]^{ji\,*}\right.\nonumber\\
    & - \tilde{M}_{eRL}^{\alpha\rho}[{\mc}_{e_LZ}]^{\rho\beta}\,\tilde{M}_{uRL}^{ij}y_u  + [{\mc}_{e_RZ}]^{\alpha\rho}\tilde{M}_{eLR}^{\rho\beta} \,\tilde{M}_{uRL}^{ij}y_{u} ~\nonumber\\
    &~\left. -  y_e \tilde{M}_{eRL}^{\alpha\beta}\,\tilde{M}_{uRL}^{ik}[{\mc}_{u_LZ}]^{kj} + y_{e}\tilde{M}_{eRL}^{\alpha\beta}\,[{\mc}_{u_RZ}]^{ik}\,\tilde{M}_{uRL}^{kj}\right)~, 
\end{align}
\begin{align}
    [\tilde{\tempC}_{RLLR}^S]^{\alpha \beta i j} &= \omega \,[{\mc}_{RLLR}^S]^{\alpha \beta i j}      +  \chi\left(     \tilde{M}_{eRL}^{\alpha\rho} [{\mc}_{e\nu_LW}]^{\rho\beta} \,\tilde{M}_{dLR}^{ij}y_{ud}  + y_{e\nu}\tilde{M}_{eRL}^{\alpha\beta}\, [{\mc}_{ud_LW}]^{ik}\tilde{M}_{dLR}^{kj}\right.\nonumber\\
    &~\left.  ~ -  y_{e\nu}\tilde{M}_{eRL}^{\alpha\beta}\, \tilde{M}_{uLR}^{ik} [{\mc}_{ud_RW}]^{kj}                  \right)~,\\
    [\tilde{\tempC}_{RLRL}^S]^{\alpha \beta i j} &= \omega \,[{\mc}_{RLRL}^S]^{\alpha \beta i j} + \chi\left( \tilde{M}_{eRL}^{\alpha\rho} [{\mc}_{e\nu_LW}]^{\rho\beta} \,\tilde{M}_{uRL}^{ij}y_{ud}  - y_{e\nu}\tilde{M}_{eRL}^{\alpha\beta}\, [{\mc}_{ud_RW}]^{ik}\tilde{M}_{dRL}^{kj}\right.\nonumber\\
    &~\left. ~ +  y_{e\nu}\tilde{M}_{eRL}^{\alpha\beta}\, \tilde{M}_{uRL}^{ik} [{\mc}_{ud_LW}]^{kj} \right)~,
\end{align}
where
\begin{align}
    \tilde{M}_{uLR} &= S^{u}_L\,M_u\,S^{u\dagger}_R,\quad \tilde{M}_{dLR} = S^{d}_L\,M_d\,S^{d\dagger}_R, \quad\tilde{M}_{eLR} = K^{e}_L\,M_e\,K^{e\dagger}_R~,
    \label{MLR}
\end{align}
with $M_e = {\rm diag}(m_e,~ m_\mu,~ m_\tau)$, $M_u = {\rm diag}(m_u,~ m_c,~ m_t)$, and $M_d = {\rm diag}(m_d,~ m_s,~ m_b)$. For $\tilde{M}_{fRL}$, $L$ and $R$ are interchanged compared to the $\tilde{M}_{fLR}$ expressions in eq.\,(\ref{MLR}).
The $k$-coefficients are
\begin{align}
    k_{eh} &=  k_{uh} =   k_{dh} = \frac{v}{2m_h^2}~.
\end{align}
Note that the BSM couplings of   $W^{\pm}$ and $Z$ bosons can also contribute to scalar LEFT operators because of the second term in the unitary gauge propagator for massive vectors. In the matching relations above for the $Z$ and $W^{\pm}$ couplings to fermions, the corresponding coefficients are
\begin{align}
    y_f &= \frac{1}{m_Z^2}(k_{f_LZ} - k_{f_RZ})~,\quad y_{ud} =  y_{e\nu} = \frac{k_{udW}}{m_W^2}~.
\end{align}

Note that in the UV4f scenario, the vanishing of HEFT WCs ${\mc}_{edLRLR}^{S}$ and ${\mc}_{euLRRL}^{S}$ results in the vanishing of the LEFT WCs ${\tempC}_{edLRLR}^{S}$ and ${\tempC}_{euLRRL}^{S}$. In conventional LEFT notation~\cite{Aebischer:2015fzz}, these identities are presented as $C_S =  - C_P$ and $C_S^\prime = C_P^\prime$ in both of these categories.

\noindent\textbf{For tensor operators:}
%
\begin{align}
    [\tilde{\tempC}_{edLRRL}^{T}]^{\alpha \beta i j} &= \omega \,[ {\mc}_{edLRRL}^{T}]^{\alpha \beta i j}~,\quad [\tilde{\tempC}_{edLRLR}^{T}]^{\alpha \beta i j} = \omega \, [ {\mc}_{edLRRL}^{T}]^{\alpha \beta i j} ~,\\
    [\tilde{\tempC}_{euLRRL}^{T}]^{\alpha \beta i j} &= \omega \,[ {\mc}_{euLRRL}^{T}]^{\alpha \beta i j}~,\quad [\tilde{\tempC}_{euLRLR}^{T}]^{\alpha \beta i j} = \omega \, [ {\mc}_{euLRRL}^{T}]^{\alpha \beta i j} ~,\\
    [\tilde{\tempC}_{RLLR}^T]^{\alpha \beta i j} &= \omega \,[{\mc}_{RLLR}^T]^{\alpha \beta i j}\quad \textrm{, and}\quad [\tilde{\tempC}_{RLRL}^T]^{\alpha \beta i j} = \omega \,[{\mc}_{RLRL}^T]^{\alpha \beta i j}~.
\end{align}
The relations among the tensor LEFT WCs remain the same as for the HEFT WCs. 

Note that, while writing the relations in Tables~\ref{Copy1CorrTableLEFT} and \ref{Copy2CorrTableLEFT}, we have replaced the four-fermion HEFT WCs with the corresponding LEFT ones using the matching relations. However, when the high energy bounds for the four-fermion WCs are stronger than the low energy bounds, it may be more efficient to use some of the original HEFT WCs instead of replacing them with the LEFT ones.

\section{Tables for direct and indirect bounds on the WCs}\label{app: numbers}

In this section, we have listed the $90\%$ C.L. bounds on the WCs (on both their real and the imaginary parts if they are complex ) obtained in Sec.~\ref{sec: indirect_bounds}. The actual bounds especially between real and imaginary parts of the WCs are correlated and the figures shown in sec. 3 are a better representation of the bounds. However, for the sake of simplicity, we present the $90\%$ C.L. bounds for individual quantities.

In Tables~\ref{tab: nud} to \ref{tab: nuu}, we present the direct bounds derived from low-energy observables (e.g. meson decays, neutrino oscillation), high-energy observables (e.g. high-$p_T$ dimuon searches, production and decays of top quark) and indirect bounds derived using eqs.\,(\ref{corLFV1}) and (\ref{corLFV2}). In Table~\ref{tab: charged}, we present the direct bounds from high-$p_T$ monomuon searches and the indirect bounds derived from eq.\,(\ref{corLFV3}). Note that the high-energy bounds directly constrain only the HEFT WCs, $[\hat{\mc}]^{\alpha \beta i j}$, but here we work
in the context of the UV4f scenario, which allows us to provide high-energy bounds on LEFT WCs.

\begin{table}[h!]
    \centering
    \renewcommand{\arraystretch}{1.3}
    \begin{tabular}{|c|c|c|c|}
        \hline
        WC & Low energy & High energy & Indirect\\
        \hline
        $[{\tempC}_{\nu d LL}^{V}]^{2 2 1 1}$ & [-0.042, 0.049] & - & [-0.00021, 0.00046] \\
        \hline
        $\textrm{Re}\left([{\tempC}_{\nu d LL}^{V}]^{2 2 1 2}\right)$ & [-0.00001, 0.00001] & - & - \\
        $\textrm{Im}\left([{\tempC}_{\nu d LL}^{V}]^{2 2 1 2}\right)$ & [-0.00001, 0.00001] & - & - \\
        \hline
        $\textrm{Re}\left([{\tempC}_{\nu d LL}^{V}]^{2 2 1 3}\right)$ & [-0.0010, 0.0012] & - & - \\
        $\textrm{Im}\left([{\tempC}_{\nu d LL}^{V}]^{2 2 1 3}\right)$ & [-0.0012, 0.0011] & - & - \\
        \hline
        $[{\tempC}_{\nu d LL}^{V}]^{2 2 2 2}$ & - & - & [-0.0029, 0.0032] \\
        \hline
        $\textrm{Re}\left([{\tempC}_{\nu d LL}^{V}]^{2 2 2 3}\right)$ & [-0.0014, 0.00078] & - & - \\
        $\textrm{Im}\left([{\tempC}_{\nu d LL}^{V}]^{2 2 2 3}\right)$ & [-0.0011, 0.0011] & - & - \\
        \hline
        $[{\tempC}_{\nu d LL}^{V}]^{2 2 3 3}$ & - & - & [-4.34, 4.44] \\
        \hline
    \end{tabular}
    \caption{$90\%$ C.L. bounds for the WCs of the type ${\tempC}_{\nu dLL}^V$. The low-energy bounds (second column) correspond to bounds from rare decays of $K$ and $B$ mesons. For $[{\tempC}_{\nu dLL}^V]^{2211}$, the low-energy bound is from atmospheric and accelerator neutrino experiments. There is no high-energy bound on these WCs. The fourth column shows the indirect bounds derived using eq.\,(\ref{corLFV1}).}
    \label{tab: nud}
\end{table}

\begin{table}[h!]
    \centering
    \renewcommand{\arraystretch}{1.3}
    \begin{tabular}{|c|c|c|c|}
         \hline
         WC & Low energy & High energy & Indirect\\
         \hline
          $[{\tempC}_{e u LL}^{V}]^{2 2 1 1}$& -  & [-0.000077, 0.00033] & - \\
          \hline
          $\textrm{Re}\left([{\tempC}_{e u LL}^{V}]^{2 2 1 2}\right)$& [-0.00056, 0.00056]  & [-0.00091, 0.00091] & - \\
          $\textrm{Im}\left([{\tempC}_{e u LL}^{V}]^{2 2 1 2}\right)$& [-0.00056, 0.00056]  & [-0.00091, 0.00091] & - \\
          \hline
          $\textrm{Re}\left([{\tempC}_{e u LL}^{V}]^{2 2 1 3}\right)$& -  & [-0.0025, 0.0025] & [-0.0078, 0.0077] \\
          $\textrm{Im}\left([{\tempC}_{e u LL}^{V}]^{2 2 1 3}\right)$& - & [-0.0025, 0.0025]  & [-0.013, 0.013] \\
          \hline
          $[{\tempC}_{e u LL}^{V}]^{2 2 2 2}$& -  & [-0.0039, 0.00093] & [-0.0067, 0.0071] \\
          \hline
          $\textrm{Re}\left([{\tempC}_{e u LL}^{V}]^{2 2 2 3}\right)$& -  & [-0.0208, 0.0208] & [-0.18, 0.19] \\
          $\textrm{Im}\left([{\tempC}_{\nu d LL}^{V}]^{2 2 2 3}\right)$& -  & [-0.021, 0.021] & [-0.0040, 0.0041] \\
          \hline
          $[{\tempC}_{e u LL}^{V}]^{2 2 3 3}$& -  & [-0.20, 0.20] & [-4.34, 4.44] \\
          \hline
    \end{tabular}
    \caption{$90\%$ C.L. bounds for the WCs of the type ${\tempC}_{e uLL}^V$. The low-energy bounds (second column) correspond to bounds from rare $D$ meson decays. The high-energy bounds (third column) correspond to bounds derived from high-$p_T$ dimuon searches and from top production and decays, particularly for $[{\tempC}_{e uLL}^V]^{2213}$, $[{\tempC}_{e uLL}^V]^{2223}$ and $[{\tempC}_{e uLL}^V]^{2233}$. The fourth column shows the indirect bounds derived using eq.\,(\ref{corLFV1}).}
    \label{tab: eu}
\end{table}

\begin{table}[h!]
    \centering
    \renewcommand{\arraystretch}{1.3}
    \begin{tabular}{|c|c|c|c|}
         \hline
         WC & Low energy & High energy & Indirect\\
         \hline
          $[{\tempC}_{e d LL}^{V}]^{2 2 1 1}$& -  & [-0.0010, 0.00014] & - \\
          \hline
          $\textrm{Re}\left([{\tempC}_{e d LL}^{V}]^{2 2 1 2}\right)$&   [-0.00001, 0.00001]& [-0.0011, 0.0011] & - \\
          $\textrm{Im}\left([{\tempC}_{e d LL}^{V}]^{2 2 1 2}\right)$& [-0.00001, 0.00001]& [-0.0011, 0.0011] & - \\
          \hline
          $\textrm{Re}\left([{\tempC}_{e d LL}^{V}]^{2 2 1 3}\right)$& [-0.00003, 0.00001]  & [-0.0017, 0.0017] & - \\
          $\textrm{Im}\left([{\tempC}_{e d LL}^{V}]^{2 2 1 3}\right)$& [-0.00002, 0.00003]  & [-0.0017, 0.0017] & - \\
          \hline
          $[{\tempC}_{e d LL}^{V}]^{2 2 2 2}$& -  & [-0.0043, 0.0014] & -\\
          \hline
          $\textrm{Re}\left([{\tempC}_{e d LL}^{V}]^{2 2 2 3}\right)$& [-0.00013, 0.00010]  & [-0.0041, 0.0041] & - \\
          $\textrm{Im}\left([{\tempC}_{e d LL}^{V}]^{2 2 2 3}\right)$& [-0.00017, 0.00017]  & [-0.0041, 0.0041] & - \\
          \hline
          $[{\tempC}_{e d LL}^{V}]^{2 2 3 3}$& -  & [-0.0096, 0.0054] & - \\
          \hline
    \end{tabular}
    \caption{$90\%$ C.L. bounds for the WCs of the type ${\tempC}_{e dLL}^V$. The low-energy bounds (second column) correspond to bounds from rare decays of $K$ and $B$ mesons. High-energy bounds (third column) correspond to bounds derived from high-$p_T$ dimuon searches. The indirect bounds are absent as we have used these WCs as inputs to eq.\,(\ref{corLFV2}).}
    \label{tab: ed}
\end{table}

\begin{table}[h!]
    \centering
    \renewcommand{\arraystretch}{1.3}
    \begin{tabular}{|c|c|c|c|}
         \hline
         WC & Low energy & High energy & Indirect\\
         \hline
          $[{\tempC}_{\nu u LL}^{V}]^{2 2 1 1}$& [-0.075, 0.052]  & - & [-0.0012, 0.00021] \\
          \hline
          $\textrm{Re}\left([{\tempC}_{\nu u LL}^{V}]^{2 2 1 2}\right)$& - & - & [-0.00098, 0.00053] \\
          $\textrm{Im}\left([{\tempC}_{\nu u LL}^{V}]^{2 2 1 2}\right)$& - & - & [-0.000012, 0.000013] \\
          \hline
          $\textrm{Re}\left([{\tempC}_{\nu u LL}^{V}]^{2 2 1 3}\right)$& -  & - & [-0.000057, 0.000074] \\
          $\textrm{Im}\left([{\tempC}_{\nu u LL}^{V}]^{2 2 1 3}\right)$& -  & - & [-0.000070, 0.000092] \\
          \hline
          $[{\tempC}_{\nu u LL}^{V}]^{2 2 2 2}$& -  & - & [-0.0042, 0.0013] \\
          \hline
          $\textrm{Re}\left([{\tempC}_{\nu u LL}^{V}]^{2 2 2 3}\right)$& -  & - & [-0.00044, 0.00049] \\
          $\textrm{Im}\left([{\tempC}_{\nu u LL}^{V}]^{2 2 2 3}\right)$& -  & - & [-0.00017, 0.00017] \\
          \hline
          $[{\tempC}_{\nu u LL}^{V}]^{2 2 3 3}$& -  & - & [-0.0096, 0.0054] \\
          \hline
    \end{tabular}
    \caption{$90\%$ C.L. bounds for the WCs of the type ${\tempC}_{\nu uLL}^V$.The low-energy bound (second column) for $[{\tempC}_{\nu uLL}^V]^{2211}$ is obtained from atmospheric  and accelerator neutrino experiments. There is no high-energy bound on these WCs. The fourth column shows the indirect bounds derived using eq.\,(\ref{corLFV2}).}
    \label{tab: nuu}
\end{table}

\begin{table}[h!]
    \centering
    \renewcommand{\arraystretch}{1.3}
    \begin{tabular}{|cc|c|c|}
         \hline
         WC &  & High energy & Indirect\\
         \hline
          $\textrm{Re}\left([{\tempC}_{LL}^{V}]^{2 2 1 1}\right)$&  & [-0.0006, 0.0002] & [-0.0012, 0.00028] \\
          $\textrm{Im}\left([{\tempC}_{LL}^{V}]^{2 2 1 1}\right)$&  & [-0.0014, 0.0014] & [-0.0000070, 0.0000078] \\
          \hline
          $\textrm{Re}\left([{\tempC}_{LL}^{V}]^{2 2 1 2}\right)$&  & [-0.0016, 0.0009] & [-0.0015, 0.00084] \\
          $\textrm{Im}\left([{\tempC}_{LL}^{V}]^{2 2 1 2}\right)$&  & [-0.0013, 0.0013] & [-0.000024, 0.000019] \\
          \hline
          $\textrm{Re}\left([{\tempC}_{LL}^{V}]^{2 2 1 3}\right)$&  & [-0.0017, 0.0013] & [-0.0015, 0.0015] \\
          $\textrm{Im}\left([{\tempC}_{LL}^{V}]^{2 2 1 3}\right)$&  & [-0.0015, 0.0016] & [-0.0018, 0.0018] \\
          \hline
          $\textrm{Re}\left([{\tempC}_{LL}^{V}]^{2 2 2 1}\right)$&  & [-0.0014, 0.0022] & [-0.00010, 0.00033] \\
          $\textrm{Im}\left([{\tempC}_{LL}^{V}]^{2 2 2 1}\right)$&  & [-0.0018, 0.0018] & [-0.000056, 0.000059] \\
          \hline
          $\textrm{Re}\left([{\tempC}_{LL}^{V}]^{2 2 2 2}\right)$&  & [-0.006, 0.002] & [-0.0066, 0.0037] \\
          $\textrm{Im}\left([{\tempC}_{LL}^{V}]^{2 2 2 2}\right)$&  & [-0.0043, 0.0043] & [-0.000052, 0.000054] \\
          \hline
          $\textrm{Re}\left([{\tempC}_{LL}^{V}]^{2 2 2 3}\right)$&  & [-0.005, 0.005] & [-0.0095, 0.0098] \\
          $\textrm{Im}\left([{\tempC}_{LL}^{V}]^{2 2 2 3}\right)$&  & [-0.005, 0.005] & [-0.0014, 0.0014] \\
          \hline
          $\textrm{Re}\left([{\tempC}_{LL}^{V}]^{2 2 3 1}\right)$&  & - & [-0.0012, 0.0010] \\
          $\textrm{Im}\left([{\tempC}_{LL}^{V}]^{2 2 3 1}\right)$&  & - & [-0.0011, 0.0011] \\
          \hline
          $\textrm{Re}\left([{\tempC}_{LL}^{V}]^{2 2 3 2}\right)$&  & - & [-0.00087, 0.0017] \\
          $\textrm{Im}\left([{\tempC}_{LL}^{V}]^{2 2 3 2}\right)$&  & - & [-0.0012, 0.0012] \\
          \hline
          $\textrm{Re}\left([{\tempC}_{LL}^{V}]^{2 2 3 3}\right)$&  & - & [-0.21, 0.20] \\
          $\textrm{Im}\left([{\tempC}_{LL}^{V}]^{2 2 3 3}\right)$&  & - & [-0.000059, 0.000059] \\
          \hline         
    \end{tabular}
    \caption{$90\%$ C.L. bounds for the charged-current WCs of the type ${\tempC}_{LL}^V$. The high-energy bounds (second column) are derived from high-$p_T$ mono-muon searches. The third column shows the indirect bounds derived using eq.\,(\ref{corLFV3}).}
    \label{tab: charged}
\end{table}

\clearpage


\providecommand{\href}[2]{#2}\begingroup\raggedright\endgroup



\begin{thebibliography}{10}

\bibitem{Buchmuller:1985jz}
W.~Buchmuller and D.~Wyler, \emph{{Effective Lagrangian Analysis of New
Interactions and Flavor Conservation}},
\href{https://doi.org/10.1016/0550-3213(86)90262-2}{\emph{Nucl. Phys. B}
{\bfseries 268} (1986) 621}.

\bibitem{Grzadkowski:2010es}
B.~Grzadkowski, M.~Iskrzynski, M.~Misiak and J.~Rosiek, \emph{{Dimension-Six
Terms in the Standard Model Lagrangian}},
\href{https://doi.org/10.1007/JHEP10(2010)085}{\emph{JHEP} {\bfseries 10}
(2010) 085} [\href{https://arxiv.org/abs/1008.4884}{{\tt arXiv:1008.4884}}].

\bibitem{Jenkins:2013zja}
E.E.~Jenkins, A.V.~Manohar and M.~Trott, \emph{{Renormalization Group Evolution
of the Standard Model Dimension Six Operators I: Formalism and lambda
Dependence}}, \href{https://doi.org/10.1007/JHEP10(2013)087}{\emph{JHEP}
{\bfseries 10} (2013) 087} [\href{https://arxiv.org/abs/1308.2627}{{\tt
arXiv:1308.2627}}].

\bibitem{Isidori:2023pyp}
G.~Isidori, F.~Wilsch and D.~Wyler, \emph{{The Standard Model effective field
theory at work}},  [\href{https://arxiv.org/abs/2303.16922}{{\tt
arXiv:2303.16922}}].

\bibitem{Jenkins:2017jig}
E.E.~Jenkins, A.V.~Manohar and P.~Stoffer, \emph{{Low-Energy Effective Field
Theory below the Electroweak Scale: Operators and Matching}},
\href{https://doi.org/10.1007/JHEP03(2018)016}{\emph{JHEP} {\bfseries 03}
(2018) 016} [\href{https://arxiv.org/abs/1709.04486}{{\tt
arXiv:1709.04486}}].

\bibitem{Aebischer:2017gaw}
J.~Aebischer, M.~Fael, C.~Greub and J.~Virto, \emph{{B physics Beyond the
Standard Model at One Loop: Complete Renormalization Group Evolution below
the Electroweak Scale}},
\href{https://doi.org/10.1007/JHEP09(2017)158}{\emph{JHEP} {\bfseries 09}
(2017) 158} [\href{https://arxiv.org/abs/1704.06639}{{\tt
arXiv:1704.06639}}].

\bibitem{Aebischer:2017ugx}
J.~Aebischer et~al., \emph{{WCxf: an exchange format for Wilson coefficients
beyond the Standard Model}},
\href{https://doi.org/10.1016/j.cpc.2018.05.022}{\emph{Comput. Phys. Commun.}
{\bfseries 232} (2018) 71} [\href{https://arxiv.org/abs/1712.05298}{{\tt
arXiv:1712.05298}}].

\bibitem{London:2021lfn}
D.~London and J.~Matias, \emph{{$B$ Flavour Anomalies: 2021 Theoretical Status
Report}},
\href{https://doi.org/10.1146/annurev-nucl-102020-090209}{\emph{Ann. Rev.
Nucl. Part. Sci.} {\bfseries 72} (2022) 37}
[\href{https://arxiv.org/abs/2110.13270}{{\tt arXiv:2110.13270}}].

\bibitem{Buchalla:1995vs}
G.~Buchalla, A.J.~Buras and M.E.~Lautenbacher, \emph{{Weak decays beyond
leading logarithms}},
\href{https://doi.org/10.1103/RevModPhys.68.1125}{\emph{Rev. Mod. Phys.}
{\bfseries 68} (1996) 1125} [\href{https://arxiv.org/abs/hep-ph/9512380}{{\tt
hep-ph/9512380}}].

\bibitem{Alonso:2012px}
R.~Alonso, M.B.~Gavela, L.~Merlo, S.~Rigolin and J.~Yepes, \emph{{The Effective
Chiral Lagrangian for a Light Dynamical ''Higgs Particle''}},
\href{https://doi.org/10.1016/j.physletb.2013.04.037}{\emph{Phys. Lett. B}
{\bfseries 722} (2013) 330} [\href{https://arxiv.org/abs/1212.3305}{{\tt
arXiv:1212.3305}}].

\bibitem{Buchalla:2013rka}
G.~Buchalla, O.~Cat\`a and C.~Krause, \emph{{Complete Electroweak Chiral
Lagrangian with a Light Higgs at NLO}},
\href{https://doi.org/10.1016/j.nuclphysb.2014.01.018}{\emph{Nucl. Phys. B}
{\bfseries 880} (2014) 552} [\href{https://arxiv.org/abs/1307.5017}{{\tt
arXiv:1307.5017}}].

\bibitem{Pich:2016lew}
A.~Pich, I.~Rosell, J.~Santos and J.J.~Sanz-Cillero, \emph{{Fingerprints of
heavy scales in electroweak effective Lagrangians}},
\href{https://doi.org/10.1007/JHEP04(2017)012}{\emph{JHEP} {\bfseries 04}
(2017) 012} [\href{https://arxiv.org/abs/1609.06659}{{\tt
arXiv:1609.06659}}].

\bibitem{Alonso:2014csa}
R.~Alonso, B.~Grinstein and J.~Martin~Camalich, \emph{{$SU(2)\times U(1)$ gauge
invariance and the shape of new physics in rare $B$ decays}},
\href{https://doi.org/10.1103/PhysRevLett.113.241802}{\emph{Phys. Rev. Lett.}
{\bfseries 113} (2014) 241802} [\href{https://arxiv.org/abs/1407.7044}{{\tt
arXiv:1407.7044}}].

\bibitem{Cata:2015lta}
O.~Cat\`a and M.~Jung, \emph{{Signatures of a nonstandard Higgs boson from
flavor physics}},
\href{https://doi.org/10.1103/PhysRevD.92.055018}{\emph{Phys. Rev. D}
{\bfseries 92} (2015) 055018} [\href{https://arxiv.org/abs/1505.05804}{{\tt
arXiv:1505.05804}}].

\bibitem{Fuentes-Martin:2020lea}
J.~Fuentes-Martin, A.~Greljo, J.~Martin~Camalich and J.D.~Ruiz-Alvarez,
\emph{{Charm physics confronts high-p$_{T}$ lepton tails}},
\href{https://doi.org/10.1007/JHEP11(2020)080}{\emph{JHEP} {\bfseries 11}
(2020) 080} [\href{https://arxiv.org/abs/2003.12421}{{\tt
arXiv:2003.12421}}].

\bibitem{Bause:2020auq}
R.~Bause, H.~Gisbert, M.~Golz and G.~Hiller, \emph{{Lepton universality and
lepton flavor conservation tests with dineutrino modes}},
\href{https://doi.org/10.1140/epjc/s10052-022-10113-6}{\emph{Eur. Phys. J. C}
{\bfseries 82} (2022) 164} [\href{https://arxiv.org/abs/2007.05001}{{\tt
arXiv:2007.05001}}].

\bibitem{Bause:2020xzj}
R.~Bause, H.~Gisbert, M.~Golz and G.~Hiller, \emph{{Rare charm
$\boldsymbol{c\to u\,\nu\bar{\nu}}$ dineutrino null tests for $e^+e^-$
machines}}, \href{https://doi.org/10.1103/PhysRevD.103.015033}{\emph{Phys.
Rev. D} {\bfseries 103} (2021) 015033}
[\href{https://arxiv.org/abs/2010.02225}{{\tt arXiv:2010.02225}}].

\bibitem{Bissmann:2020mfi}
S.~Bi\ss{}mann, C.~Grunwald, G.~Hiller and K.~Kr\"oninger, \emph{{Top and
Beauty synergies in SMEFT-fits at present and future colliders}},
\href{https://doi.org/10.1007/JHEP06(2021)010}{\emph{JHEP} {\bfseries 06}
(2021) 010} [\href{https://arxiv.org/abs/2012.10456}{{\tt
arXiv:2012.10456}}].

\bibitem{Bause:2021cna}
R.~Bause, H.~Gisbert, M.~Golz and G.~Hiller, \emph{{Interplay of dineutrino
modes with semileptonic rare B-decays}},
\href{https://doi.org/10.1007/JHEP12(2021)061}{\emph{JHEP} {\bfseries 12}
(2021) 061} [\href{https://arxiv.org/abs/2109.01675}{{\tt
arXiv:2109.01675}}].

\bibitem{Bause:2021ihn}
R.~Bause, H.~Gisbert-Mullor, M.~Golz and G.~Hiller, \emph{{Dineutrino modes
probing lepton flavor violation}},
\href{https://doi.org/10.22323/1.398.0563}{\emph{PoS} {\bfseries EPS-HEP2021}
(2022) 563} [\href{https://arxiv.org/abs/2110.08795}{{\tt
arXiv:2110.08795}}].

\bibitem{Bause:2022rrs}
R.~Bause, H.~Gisbert, M.~Golz and G.~Hiller, \emph{{Model-independent analysis
of $b \rightarrow d$ processes}},
\href{https://doi.org/10.1140/epjc/s10052-023-11586-9}{\emph{Eur. Phys. J. C}
{\bfseries 83} (2023) 419} [\href{https://arxiv.org/abs/2209.04457}{{\tt
arXiv:2209.04457}}].

\bibitem{Sun:2023cuf}
S.~Sun, Q.-S.~Yan, X.~Zhao and Z.~Zhao, \emph{{Constraining rare B decays by
\ensuremath{\mu}+\ensuremath{\mu}-\textrightarrow{}tc at future lepton
colliders}}, \href{https://doi.org/10.1103/PhysRevD.108.075016}{\emph{Phys.
Rev. D} {\bfseries 108} (2023) 075016}
[\href{https://arxiv.org/abs/2302.01143}{{\tt arXiv:2302.01143}}].

\bibitem{Grunwald:2023nli}
C.~Grunwald, G.~Hiller, K.~Kr\"oninger and L.~Nollen, \emph{{More synergies
from beauty, top, Z and Drell-Yan measurements in SMEFT}},
\href{https://doi.org/10.1007/JHEP11(2023)110}{\emph{JHEP} {\bfseries 11}
(2023) 110} [\href{https://arxiv.org/abs/2304.12837}{{\tt
arXiv:2304.12837}}].

\bibitem{Greljo:2023bab}
A.~Greljo, J.~Salko, A.~Smolkovi\v{c} and P.~Stangl, \emph{{SMEFT restrictions
on exclusive b \textrightarrow{} u\ensuremath{\ell}\ensuremath{\nu} decays}},
\href{https://doi.org/10.1007/JHEP11(2023)023}{\emph{JHEP} {\bfseries 11}
(2023) 023} [\href{https://arxiv.org/abs/2306.09401}{{\tt
arXiv:2306.09401}}].

\bibitem{Fajfer:2012vx}
S.~Fajfer, J.F.~Kamenik and I.~Nisandzic, \emph{{On the $B \to D^* \tau \bar
\nu_{\tau}$ Sensitivity to New Physics}},
\href{https://doi.org/10.1103/PhysRevD.85.094025}{\emph{Phys. Rev. D}
{\bfseries 85} (2012) 094025} [\href{https://arxiv.org/abs/1203.2654}{{\tt
arXiv:1203.2654}}].

\bibitem{Bause:2023mfe}
R.~Bause, H.~Gisbert and G.~Hiller, \emph{{Implications of an enhanced
B\textrightarrow{}K\ensuremath{\nu}\ensuremath{\nu}\textasciimacron{}
branching ratio}},
\href{https://doi.org/10.1103/PhysRevD.109.015006}{\emph{Phys. Rev. D}
{\bfseries 109} (2024) 015006} [\href{https://arxiv.org/abs/2309.00075}{{\tt
arXiv:2309.00075}}].

\bibitem{Chen:2024jlj}
F.-Z.~Chen, Q.~Wen and F.~Xu, \emph{{Correlating $B\to K^{(\ast)} \nu\bar{\nu}$
and flavor anomalies in SMEFT}},
[\href{https://arxiv.org/abs/2401.11552}{{\tt arXiv:2401.11552}}].

\bibitem{gupta}
R.S.~Gupta, A.~Pomarol and F.~Riva, \emph{{BSM Primary Effects}},
\href{https://doi.org/10.1103/PhysRevD.91.035001}{\emph{Phys. Rev. D}
{\bfseries 91} (2015) 035001} [\href{https://arxiv.org/abs/1405.0181}{{\tt
arXiv:1405.0181}}].

\bibitem{LHCHiggs}
{\scshape LHC Higgs Cross Section Working Group} collaboration, \emph{{Handbook
of LHC Higgs Cross Sections: 4. Deciphering the Nature of the Higgs Sector}},
[\href{https://arxiv.org/abs/1610.07922}{{\tt arXiv:1610.07922}}].

\bibitem{Hiller:2014yaa}
G.~Hiller and M.~Schmaltz, \emph{{$R_K$ and future $b \to s \ell \ell$ physics
beyond the standard model opportunities}},
\href{https://doi.org/10.1103/PhysRevD.90.054014}{\emph{Phys. Rev. D}
{\bfseries 90} (2014) 054014} [\href{https://arxiv.org/abs/1408.1627}{{\tt
arXiv:1408.1627}}].

\bibitem{Gripaios:2014tna}
B.~Gripaios, M.~Nardecchia and S.A.~Renner, \emph{{Composite leptoquarks and
anomalies in $B$-meson decays}},
\href{https://doi.org/10.1007/JHEP05(2015)006}{\emph{JHEP} {\bfseries 05}
(2015) 006} [\href{https://arxiv.org/abs/1412.1791}{{\tt arXiv:1412.1791}}].

\bibitem{deMedeirosVarzielas:2015yxm}
I.~de~Medeiros~Varzielas and G.~Hiller, \emph{{Clues for flavor from rare
lepton and quark decays}},
\href{https://doi.org/10.1007/JHEP06(2015)072}{\emph{JHEP} {\bfseries 06}
(2015) 072} [\href{https://arxiv.org/abs/1503.01084}{{\tt
arXiv:1503.01084}}].

\bibitem{Sahoo:2015wya}
S.~Sahoo and R.~Mohanta, \emph{{Scalar leptoquarks and the rare $B$ meson
decays}}, \href{https://doi.org/10.1103/PhysRevD.91.094019}{\emph{Phys. Rev.
D} {\bfseries 91} (2015) 094019}
[\href{https://arxiv.org/abs/1501.05193}{{\tt arXiv:1501.05193}}].

\bibitem{Altmannshofer:2014cfa}
W.~Altmannshofer, S.~Gori, M.~Pospelov and I.~Yavin, \emph{{Quark flavor
transitions in $L_\mu-L_\tau$ models}},
\href{https://doi.org/10.1103/PhysRevD.89.095033}{\emph{Phys. Rev. D}
{\bfseries 89} (2014) 095033} [\href{https://arxiv.org/abs/1403.1269}{{\tt
arXiv:1403.1269}}].

\bibitem{Bonilla:2017lsq}
C.~Bonilla, T.~Modak, R.~Srivastava and J.W.F.~Valle,
\emph{{$U(1)_{B_3-3L_\mu}$ gauge symmetry as a simple description of $b\to s$
anomalies}}, \href{https://doi.org/10.1103/PhysRevD.98.095002}{\emph{Phys.
Rev. D} {\bfseries 98} (2018) 095002}
[\href{https://arxiv.org/abs/1705.00915}{{\tt arXiv:1705.00915}}].

\bibitem{Bian:2017rpg}
L.~Bian, S.-M.~Choi, Y.-J.~Kang and H.M.~Lee, \emph{{A minimal flavored $U(1)'$
for $B$-meson anomalies}},
\href{https://doi.org/10.1103/PhysRevD.96.075038}{\emph{Phys. Rev. D}
{\bfseries 96} (2017) 075038} [\href{https://arxiv.org/abs/1707.04811}{{\tt
arXiv:1707.04811}}].

\bibitem{Alonso:2017uky}
R.~Alonso, P.~Cox, C.~Han and T.T.~Yanagida, \emph{{Flavoured $B-L$ local
symmetry and anomalous rare $B$ decays}},
\href{https://doi.org/10.1016/j.physletb.2017.10.027}{\emph{Phys. Lett. B}
{\bfseries 774} (2017) 643} [\href{https://arxiv.org/abs/1705.03858}{{\tt
arXiv:1705.03858}}].

\bibitem{Cline:2017ihf}
J.M.~Cline and J.~Martin~Camalich, \emph{{$B$ decay anomalies from nonabelian
local horizontal symmetry}},
\href{https://doi.org/10.1103/PhysRevD.96.055036}{\emph{Phys. Rev. D}
{\bfseries 96} (2017) 055036} [\href{https://arxiv.org/abs/1706.08510}{{\tt
arXiv:1706.08510}}].

\bibitem{Buchalla:2012qq}
G.~Buchalla and O.~Cata, \emph{{Effective Theory of a Dynamically Broken
Electroweak Standard Model at NLO}},
\href{https://doi.org/10.1007/JHEP07(2012)101}{\emph{JHEP} {\bfseries 07}
(2012) 101} [\href{https://arxiv.org/abs/1203.6510}{{\tt arXiv:1203.6510}}].

\bibitem{Aebischer:2015fzz}
J.~Aebischer, A.~Crivellin, M.~Fael and C.~Greub, \emph{{Matching of gauge
invariant dimension-six operators for $b\to s$ and $b\to c$ transitions}},
\href{https://doi.org/10.1007/JHEP05(2016)037}{\emph{JHEP} {\bfseries 05}
(2016) 037} [\href{https://arxiv.org/abs/1512.02830}{{\tt
arXiv:1512.02830}}].

\bibitem{Burgess:2021ylu}
C.P.~Burgess, S.~Hamoudou, J.~Kumar and D.~London, \emph{{Beyond the standard
model effective field theory with $B \rightarrow c \tau^- \overline{\nu}$}},
\href{https://doi.org/10.1103/PhysRevD.105.073008}{\emph{Phys. Rev. D}
{\bfseries 105} (2022) 073008} [\href{https://arxiv.org/abs/2111.07421}{{\tt
arXiv:2111.07421}}].

\bibitem{Efrati:2015eaa}
A.~Efrati, A.~Falkowski and Y.~Soreq, \emph{{Electroweak constraints on
flavorful effective theories}},
\href{https://doi.org/10.1007/JHEP07(2015)018}{\emph{JHEP} {\bfseries 07}
(2015) 018} [\href{https://arxiv.org/abs/1503.07872}{{\tt
arXiv:1503.07872}}].

\bibitem{NA62:2022qes}
{\scshape NA62} collaboration, \emph{{A measurement of the
K$^{+}$\textrightarrow{}
\ensuremath{\pi}$^{+}$\ensuremath{\mu}$^{+}$\ensuremath{\mu}$^{-}$ decay}},
\href{https://doi.org/10.1007/JHEP06(2023)040}{\emph{JHEP} {\bfseries 11}
(2022) 011} [\href{https://arxiv.org/abs/2209.05076}{{\tt
arXiv:2209.05076}}].

\bibitem{ParticleDataGroup:2022pth}
{\scshape Particle Data Group} collaboration, \emph{{Review of Particle
Physics}}, \href{https://doi.org/10.1093/ptep/ptac097}{\emph{PTEP} {\bfseries
2022} (2022) 083C01}.

\bibitem{Belle-II:2023esi}
{\scshape Belle-II} collaboration, \emph{{Evidence for $B^{+}\to
K^{+}\nu\bar{\nu}$ Decays}},  [\href{https://arxiv.org/abs/2311.14647}{{\tt
arXiv:2311.14647}}].

\bibitem{Straub:2018kue}
D.M.~Straub, \emph{{flavio: a Python package for flavour and precision
phenomenology in the Standard Model and beyond}},
[\href{https://arxiv.org/abs/1810.08132}{{\tt arXiv:1810.08132}}].

\bibitem{Farzan:2017xzy}
Y.~Farzan and M.~Tortola, \emph{{Neutrino oscillations and Non-Standard
Interactions}}, \href{https://doi.org/10.3389/fphy.2018.00010}{\emph{Front.
in Phys.} {\bfseries 6} (2018) 10}
[\href{https://arxiv.org/abs/1710.09360}{{\tt arXiv:1710.09360}}].

\bibitem{Escrihuela:2011cf}
F.J.~Escrihuela, M.~Tortola, J.W.F.~Valle and O.G.~Miranda, \emph{{Global
constraints on muon-neutrino non-standard interactions}},
\href{https://doi.org/10.1103/PhysRevD.83.093002}{\emph{Phys. Rev. D}
{\bfseries 83} (2011) 093002} [\href{https://arxiv.org/abs/1103.1366}{{\tt
arXiv:1103.1366}}].

\bibitem{CMS:2021ctt}
{\scshape CMS} collaboration, \emph{{Search for resonant and nonresonant new
phenomena in high-mass dilepton final states at $ \sqrt{s} $ = 13 TeV}},
\href{https://doi.org/10.1007/JHEP07(2021)208}{\emph{JHEP} {\bfseries 07}
(2021) 208} [\href{https://arxiv.org/abs/2103.02708}{{\tt
arXiv:2103.02708}}].

\bibitem{Allwicher:2022mcg}
L.~Allwicher, D.A.~Faroughy, F.~Jaffredo, O.~Sumensari and F.~Wilsch,
\emph{{HighPT: A tool for~ high-$p_T$ Drell-Yan tails beyond the standard
model}}, \href{https://doi.org/10.1016/j.cpc.2023.108749}{\emph{Comput. Phys.
Commun.} {\bfseries 289} (2023) 108749}
[\href{https://arxiv.org/abs/2207.10756}{{\tt arXiv:2207.10756}}].

\bibitem{Allwicher:2022gkm}
L.~Allwicher, D.A.~Faroughy, F.~Jaffredo, O.~Sumensari and F.~Wilsch,
\emph{{Drell-Yan tails beyond the Standard Model}},
\href{https://doi.org/10.1007/JHEP03(2023)064}{\emph{JHEP} {\bfseries 03}
(2023) 064} [\href{https://arxiv.org/abs/2207.10714}{{\tt
arXiv:2207.10714}}].

\bibitem{Afik:2021jjh}
Y.~Afik, S.~Bar-Shalom, A.~Soni and J.~Wudka, \emph{{New flavor physics in di-
and trilepton events from single-top production at the LHC and beyond}},
\href{https://doi.org/10.1103/PhysRevD.103.075031}{\emph{Phys. Rev. D}
{\bfseries 103} (2021) 075031} [\href{https://arxiv.org/abs/2101.05286}{{\tt
arXiv:2101.05286}}].

\bibitem{Alwall:2014hca}
J.~Alwall, R.~Frederix, S.~Frixione, V.~Hirschi, F.~Maltoni, O.~Mattelaer
et~al., \emph{{The automated computation of tree-level and next-to-leading
order differential cross sections, and their matching to parton shower
simulations}}, \href{https://doi.org/10.1007/JHEP07(2014)079}{\emph{JHEP}
{\bfseries 07} (2014) 079} [\href{https://arxiv.org/abs/1405.0301}{{\tt
arXiv:1405.0301}}].

\bibitem{ATLAS:2019lsy}
{\scshape ATLAS} collaboration, \emph{{Search for a heavy charged boson in
events with a charged lepton and missing transverse momentum from $pp$
collisions at $\sqrt{s} = 13$ TeV with the ATLAS detector}},
\href{https://doi.org/10.1103/PhysRevD.100.052013}{\emph{Phys. Rev. D}
{\bfseries 100} (2019) 052013} [\href{https://arxiv.org/abs/1906.05609}{{\tt
arXiv:1906.05609}}].

\bibitem{Banta:2021dek}
I.~Banta, T.~Cohen, N.~Craig, X.~Lu and D.~Sutherland, \emph{{Non-decoupling
new particles}}, \href{https://doi.org/10.1007/JHEP02(2022)029}{\emph{JHEP}
{\bfseries 02} (2022) 029} [\href{https://arxiv.org/abs/2110.02967}{{\tt
arXiv:2110.02967}}].

\bibitem{Cohen:2020xca}
T.~Cohen, N.~Craig, X.~Lu and D.~Sutherland, \emph{{Is SMEFT Enough?}},
\href{https://doi.org/10.1007/JHEP03(2021)237}{\emph{JHEP} {\bfseries 03}
(2021) 237} [\href{https://arxiv.org/abs/2008.08597}{{\tt
arXiv:2008.08597}}].

\bibitem{Falkowski:2019tft}
A.~Falkowski and R.~Rattazzi, \emph{{Which EFT}},
\href{https://doi.org/10.1007/JHEP10(2019)255}{\emph{JHEP} {\bfseries 10}
(2019) 255} [\href{https://arxiv.org/abs/1902.05936}{{\tt
arXiv:1902.05936}}].

\bibitem{Proceedings:2019qno}
\emph{{Neutrino Non-Standard Interactions: A Status Report}}, vol.~2, 2019.
\newblock 10.21468/SciPostPhysProc.2.001.

\bibitem{Ardu:2022pzk}
M.~Ardu, S.~Davidson and M.~Gorbahn, \emph{{Sensitivity of
\ensuremath{\mu}\textrightarrow{}e processes to \ensuremath{\tau} flavor
change}}, \href{https://doi.org/10.1103/PhysRevD.105.096040}{\emph{Phys. Rev.
D} {\bfseries 105} (2022) 096040}
[\href{https://arxiv.org/abs/2202.09246}{{\tt arXiv:2202.09246}}].

\bibitem{Farzan:2019xor}
Y.~Farzan, \emph{{A model for lepton flavor violating non-standard neutrino
interactions}},
\href{https://doi.org/10.1016/j.physletb.2020.135349}{\emph{Phys. Lett. B}
{\bfseries 803} (2020) 135349} [\href{https://arxiv.org/abs/1912.09408}{{\tt
arXiv:1912.09408}}].

\bibitem{murayama}
L.~Gr\'af, B.~Henning, X.~Lu, T.~Melia and H.~Murayama, \emph{{Hilbert series,
the Higgs mechanism, and HEFT}},
\href{https://doi.org/10.1007/JHEP02(2023)064}{\emph{JHEP} {\bfseries 02}
(2023) 064} [\href{https://arxiv.org/abs/2211.06275}{{\tt
arXiv:2211.06275}}].

\bibitem{Masso:2014xra}
E.~Masso, \emph{{An Effective Guide to Beyond the Standard Model Physics}},
\href{https://doi.org/10.1007/JHEP10(2014)128}{\emph{JHEP} {\bfseries 10}
(2014) 128} [\href{https://arxiv.org/abs/1406.6376}{{\tt arXiv:1406.6376}}].

\end{thebibliography}
\end{document}